\documentclass[twocolumn]{aastex62}
\pdfoutput=1 
\usepackage{amssymb}
\usepackage{graphicx}
\usepackage{amsmath}
\usepackage{pbox}
\usepackage{pifont}
\usepackage{xspace}
\usepackage{array}
\usepackage[caption=false]{subfig}
\usepackage{apjfonts} 
\usepackage{fixltx2e}
\usepackage{color}
\usepackage{mathtools}
\usepackage{tabularx}

\begin{document}

\author{David K. Khatami}
\affil{Department of Astronomy, University of California, Berkeley, CA 94720}
\affil{Lawrence Livermore National Laboratory, Livermore, CA 94550}
\author{Daniel N. Kasen}
\affil{Department of Astronomy, University of California, Berkeley, CA 94720}
\affil{Nuclear Science Division, Lawrence Berkeley National Laboratory, 1 Cyclotron Road, Berkeley, CA 94720}
\title{The Landscape of Thermal Transients from Supernova Interacting with a Circumstellar Medium}
\shorttitle{Thermal Transients from CSM Interaction}
\shortauthors{Khatami and Kasen}

\begin{abstract}
    The interaction of supernova ejecta with a surrounding circumstellar medium (CSM) generates a strong shock which can convert the ejecta kinetic energy into observable radiation. Given the diversity of potential CSM structures (arising from diverse mass loss processes such as late-stage stellar outbursts, binary interaction, and winds), the resulting transients can display a wide range of light curve morphologies. We provide a framework for classifying the transients arising from interaction with a spherical CSM shell. The light curves are decomposed into five consecutive phases, starting from the onset of interaction and extending through shock breakout and subsequent shock cooling. The relative prominence of each phase in the light curve is determined by two dimensionless quantities representing the CSM-to-ejecta mass ratio $\eta$, and a breakout parameter $\xi$. These two parameters define four light curve morphology classes, where each class is characterized by the location of shock breakout and the degree of deceleration as the shock sweeps up the CSM. We compile analytic scaling relations connecting the luminosity and duration of each light curve phase to the physical parameters. We then run a grid of radiation hydrodynamics simulations for a wide range of ejecta and CSM parameters to numerically explore the landscape of interaction light curves, and to calibrate and confirm the analytic scalings. We connect our theoretical framework to several case studies of observed transients, highlighting the relevance in explaining slow-rising and superluminous supernovae, fast blue optical transients, and double-peaked light curves.

\end{abstract}

\section{Introduction}

The light curves of typical supernovae are generally understood to be radiation diffusing from the hot stellar debris produced in the explosion blastwave and often further heated by the radioactive decay of $^{56}$Ni \citep{1966ApJ...143..626C,1982ApJ...253..785A, 2002RvMP...74.1015W,2007PhR...442...38J}. 
Diversity in the ejecta/nickel masses and explosion energies can produce a wide range of light curve durations and luminosities \citep{2017hsn..book..403S,2017suex.book.....B}. Recent all-sky observations have enlarged the domain of transient types (e.g. \cite{2009PASP..121.1334R,2019PASP..131g8001G,2021arXiv210508811H}), uncovering highly luminous events outside of the realm of ``typical'' supernovae. These events occur on timescales as short as a day \citep{2014ApJ...794...23D,2019ApJ...887..169H,2019ApJ...872...18M}, to as long as several months \citep{2007ApJ...666.1116S,2013ApJ...770..128I}. Their extreme brightness and gamut of timescales pose a challenge to usual explanations of luminous transients \citep{2012Sci...337..927G}.

In typical core-collapse supernovae, roughly half of the explosion energy is converted into thermal energy from the passage of a strong neutrino-driven shock \citep{1986ARA&A..24..205W,2017hsn..book.1095J}. Due to the high optical depths of stellar interiors, most of this energy is lost to adiabatic expansion of the ejecta \citep{1980ApJ...237..541A,1982ApJ...253..785A}. The bulk of  the explosion energy is then stored in a reservoir of kinetic energy of order $\sim 10^{51}$ ergs \citep{2016ApJ...821...38S}. If this prodigious store of energy can be tapped into and converted into observable electromagnetic radiation, it can  power some of the most energetic events in the transient sky.

Interaction of the expanding supernova ejecta with a surrounding medium results in shocks that convert kinetic energy into internal energy of the gas \citep{1967pswh.book.....Z} which can be radiated in a light curve;
If the shock is optically thin, a \textit{collisionless shock} forms and most of the kinetic energy remains as internal gas energy \citep{1982ApJ...259..302C}. This is typically the case for supernova remnants \citep{1986ApJ...301..790W,1988ARA&A..26..295W}. While such sites are expected to be efficient sources of energetic cosmic rays \citep{1978MNRAS.182..147B,1987PhR...154....1B,1995Natur.378..255K} and non-thermal radio/X-ray emission \citep{2017hsn..book..875C}, they are incapable of powering the luminous \textit{optical} transients that are being discovered (e.g. \cite{2014ApJ...794...23D}). Instead, these events require the formation of a \textit{radiative} shock \citep{2005Ap&SS.298...49D,2017hsn..book..967W}. 

The formation of radiative shocks requires the presence of a dense circumstellar medium (CSM) that is optically thick and moving slowly relative to the ejecta velocity. 
Supernova progenitor stars typically lose significant mass to stellar winds over their lifetime
\citep{2001A&A...369..574V,2012ARA&A..50..107L,2014ARA&A..52..487S,2021ApJ...913..145W}.
Gradual mass loss in winds will disperse into the interstellar medium. 
To produce a dense, local CSM, requires episodes of extreme mass loss that
occur shortly before the supernova explosion.
Such mass loss events are often referred to as stellar outbursts, and numerous explanations have been proposed regarding their origin, such as binary interaction \citep{2012Sci...337..444S,2015MNRAS.451.2123T,2022ApJ...940L..27W} and wave-driven mass loss from e.g. unstable nuclear burning \citep{2012MNRAS.423L..92Q,2017MNRAS.470.1642F,2021ApJ...906....3W}.

Observations of late-stage stellar outbursts \citep{1994PASP..106.1025H,1997ARA&A..35....1D,2007ARA&A..45..177C} and the presence of narrow lines in supernova spectra \citep{1997ARA&A..35..309F} lend credence to the CSM interaction model as a viable explanation for at least some of the transients \citep{2017hsn..book..403S,2017ApJ...849...70V,2020A&A...637A..73N,2020ApJ...899...56S}. Given the diversity of mass loss rates, they are an appealing mechanism for atypical supernovae, including superluminous events \citep{2007ApJ...666.1116S,2012Sci...337..927G,2013ApJ...773...76C,2015MNRAS.449.4304D,2018MNRAS.475.1046I} and the recently emerging class of so-called fast blue optical transients, or FBOTs \citep{2014ApJ...794...23D,2018NatAs...2..307R,2018ApJ...865L...3P,2019ApJ...887..169H,2019ApJ...872...18M,2022ApJ...926..125P}. 

The physics of CSM interaction has been extensively researched in the literature \citep[e.g.,][]{1982ApJ...258..790C,1994ApJ...420..268C,2013MNRAS.428.1020M,2015MNRAS.449.4304D,2017hsn..book..875C,2022ApJ...928..122M} including both numerical and analytical works that predict the light curve and spectra of CSM interaction \citep{2011ApJ...729L...6C,2012ApJ...757..178G,2017ApJ...838...28M,2020ApJ...899...56S,2022ApJ...932...84M}, as well as models to explain specific events \citep{2010ApJ...724.1396O,2013MNRAS.428.1020M,2020ApJ...903...66L,2021ApJ...915...80L}. Different theoretical models, however, may make different physical assumptions and derive divergent expressions for how the light curve luminosity and duration depend upon physical parameters. The regions of applicability of such models is not always clear, and the degeneracy in parameter estimation when fitting observations with numerical models is often unconstrained. The same observed light curve, for example, may be fit with ``shock breakout'' \citep{2010ApJ...724.1396O} or ``shock cooling'' models \citep{2015ApJ...808L..51P,2021ApJ...909..209P}, leading to different inferences as to the nature of the event.

In this work, we outline a theoretical framework to help clarify the categorization  of interaction light curves. We discuss how the physical parameters describing the configuration of supernova ejecta plus CSM shell can be reduced to two dimensionless parameters that primarily determine the  light curve morphology. The values of these two quantities naturally partition the parameter space of interaction light curves into four classes. We compile analytic relations that express how the luminosity and duration of the light curve scale with physical parameters, and clarify their regimes of applicability.  We then run a comprehensive set of spherically symmetric radiation-hydrodynamical simulations of interacting supernovae and explore the landscape of  optical light curves. The numerical models are used to confirm the analytic relations and highlight the break in scaling relations that occurs when transitioning from one light curve class to the next.  

The numerical models presented here aim to provide an expansive library of bolometric light curves for interacting supernova that can aid in the interpretation of observed events.
Follow-up work will explore spectroscopic properties of the models and possible non-thermal emission mechanisms. 
In Section \ref{sec:configuration} we give a qualitative overview of CSM interaction, and the basic physics that controls each phase of the light curve. We give a more quantitative analysis in Section \ref{sec:analytics}, including useful scaling relations for each phase, which we compare with numerical simulations in Section \ref{sec:numerical}. Finally, in Section \ref{sec:discussion}  we show how the results can be used to infer properties of the CSM mass and radius, and discuss the relevance of different interaction classes to observed classes of transient phenomena. For clarity of presentation, we provide a more complete description of the numerics and supplementary equations in the Appendix.

\section{Qualitative Picture}
\label{sec:configuration}

We provide in this section a qualitative picture of the dynamics of interacting supernovae and the context of radiation emission. This is used to define the possible morphologies of the resulting light curves. 

\subsection{System Configuration} 

We consider supernovae interacting with a single CSM shell of mass $M_{\rm csm}$. Such a configuration roughly approximates the structure of material ejected in a presupernova outburst. The key dimensional parameters of the system are 
\begin{itemize}
\item $M_{\rm ej}$: ejecta mass
\item $E_{\rm sn}$: ejecta kinetic energy 
\item $M_{\rm csm}$: circumstellar mass
\item $R_{\rm csm}$: outer radius of circumstellar material
\item $\kappa$: opacity
\end{itemize}
 The density profile within the CSM shell is taken to be a power-law  $\rho(r) \propto r^{-s}$ which transitions to a steep power law cutoff at the outer edge at $R_{\rm csm}$. Usually, we take $s=2$ (i.e. a wind-like CSM) but select models  explore different density profiles.  The CSM velocity is assumed to be much less than that of the ejecta and so set to zero.

The ejecta is assumed to be in homologous expansion with a broken power-law density profile,  $\rho_{\rm ej}\propto r^{-n}$ (where typically $n\approx 7-10$ in the outer layers of ejecta \citep{1989ApJ...341..867C}).
The ejecta is taken to be 
cold (thermal energy $\ll E_{\rm sn}$)  with a characteristic velocity $v_{\rm ej}\equiv \sqrt{2 E_{\rm sn}/M_{\rm ej}}$. These assumptions apply when the radius of the progenitor star is much less than $R_{\rm csm}$, such that the ejecta are able to expand, cool and reach homology before interaction begins. Inclusion of initial thermal energy or heating due to radioactivity are unlikely to influence the interaction dynamics, but  could contribute additional luminosity to the light curve. Finally, we assume that the inner CSM edge is much less than the outer CSM radius, $R_*\ll R_{\rm csm}$.

It is helpful to combine the above five physical quantities into three dimensionless parameters which determine the morphology of the  light curve; and two dimensional parameters which set the overall luminosity and time scale. The dimensionless parameters are
\begin{itemize}
\item $\eta \equiv M_{\rm csm}/M_{\rm ej}$: ratio of CSM to ejecta mass
\item $\beta_0 \equiv v_{\rm ej}/c$ : ejecta velocity relative to the speed of light
\item $\tau_0 \equiv \kappa M_{\rm csm}/4\pi R_{\rm csm}^2$: characteristic CSM optical depth 
\end{itemize}

The dimensional scale parameters  of the light curve are
\begin{itemize}
\item $L_0\equiv  M_{\rm csm}v_{\rm ej}^3/R_{\rm csm}$:  luminosity scale
\item $t_0\equiv R_{\rm csm}/v_{\rm ej}$:  temporal scale
\end{itemize}

A combination of the dimensionless parameters that will be critical to understanding the light curve behavior is the \textbf{breakout parameter}
\begin{align}\label{eq:bo_param}
\boxed{
\xi\equiv \tau_0\beta_0\eta^{-\alpha}},
\end{align}
where the factor  $\eta^{-\alpha}$ accounts for how shock propagation through the CSM modifies the velocity scale $\beta_0$ of the shock. Here $\alpha$ is an order-unity exponent that depends on the mass ratio $\eta$ and the power-law exponent, $n$, of the ejecta density profile in the outer layers, whose expression is given by Eq. (\ref{eq:alpha_eta}) and derived in Appendix B.
 In terms of the physical quantities, the breakout parameter is
\begin{align}
\xi \approx 10\,\,\kappa\,M_{\rm csm,\odot} v_9 R_4^{-2}\eta^{-\alpha}
\end{align}
where $\kappa\approx 0.34$ cm$^{2}$ g$^{-1}$ for solar electron scattering, $M_{\rm csm,\odot}=M_{\rm csm}/M_\odot$, $v_9=v_{\rm ej}/10^9$ cm s$^{-1}$, and $R_4 = R_{\rm csm}/10^4 R_\odot$.

\begin{figure}
    \includegraphics[width=0.45\textwidth]{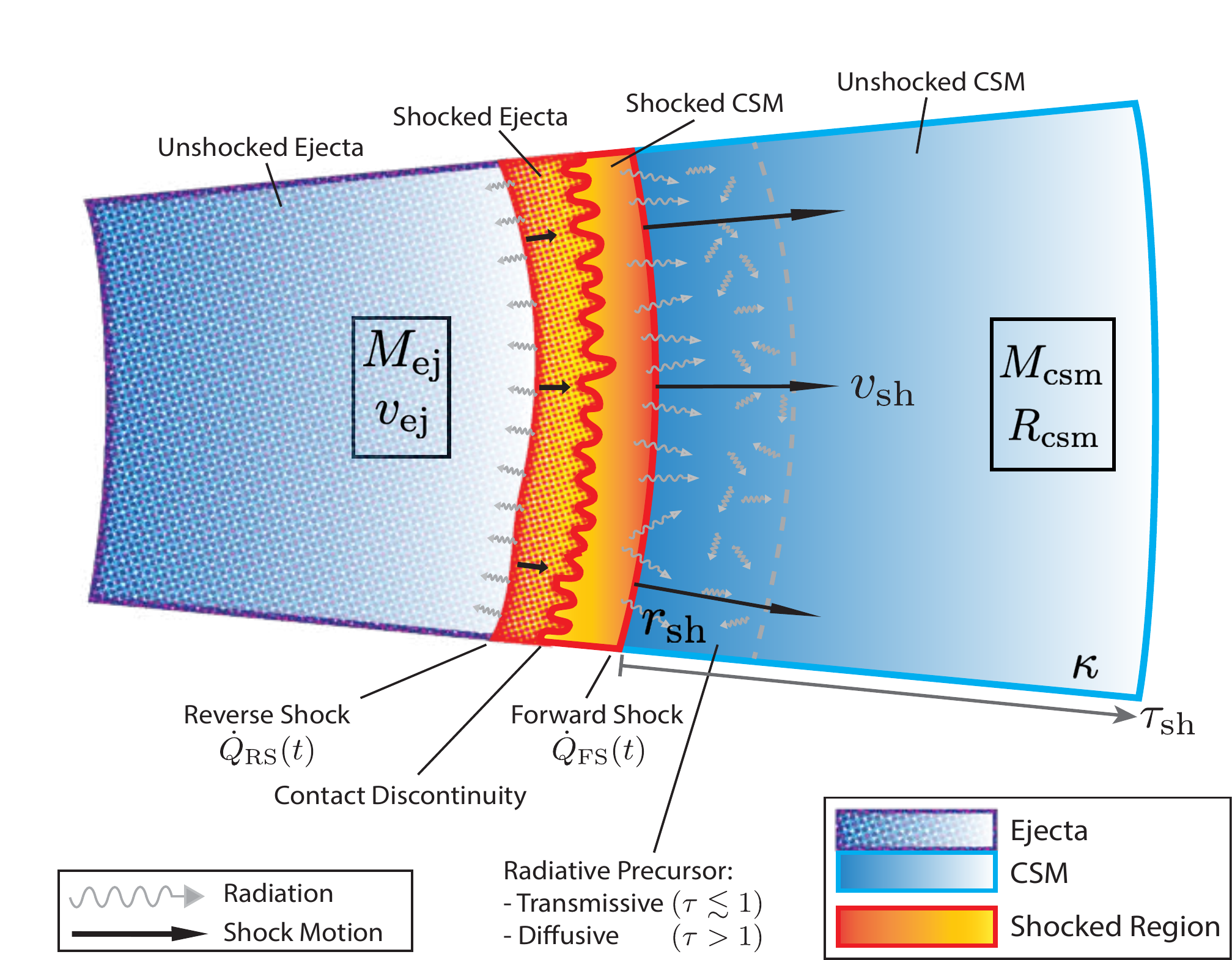}
    \caption{Illustration of the radiative shock structure during ejecta-CSM interaction, with the different shock features labeled.}
    \label{fig:shock_structure}
\end{figure}

\subsection{Interaction Dynamics} 
\label{subsec:dynamics}

Fig. \ref{fig:shock_structure} illustrates the structure of a generic interaction shock, with a forward shock propagating into the CSM and a reverse shock decelerating the ejecta. 
The properties of interaction light curves depend critically on the forward shock velocity, $v_{\rm sh}$, which initially is characteristic of the fast outermost ejecta layers, $v_{\rm sh}\gtrsim v_{\rm ej}$, but decelerates as the shock progressively sweeps up the CSM. The degree to which the shock decelerates depends on the relative masses of the ejecta and CSM, $\eta$. If the shock evolves as a power-law in time $r_{\rm sh}\propto t^\lambda$, then we can derive the shock velocity in terms of radius as \citep{1988RvMP...60....1O}
\begin{align}\label{eq:shock_velocity}
v_{\rm sh}(r_{\rm sh}) \approx v_{\rm ej} \eta^{-\alpha}\left(\frac{r_{\rm sh}}{R_{\rm csm}}\right)^{(\lambda-1)/\lambda}
\end{align}
where the factor $\eta^{-\alpha}$ accounts for the shock deceleration, and the order-unity shock exponent $\alpha$ depends on the density structure of the ejecta and CSM (see Appendix \ref{sec:shock_derivation}). For $M_{\rm csm}< M_{\rm ej}$, the shock may decelerate only the outermost layers of ejecta and $v_{\rm sh}\approx v_{\rm ej}$. However, for $M_{\rm csm}\gtrsim M_{\rm ej}$ the shock velocity will be substantially lower than $v_{\rm ej}$. The shock decelerates only so long as the density profile is shallower than $s<3$. Acceleration of the shock for $s > 3$ \citep{1999ApJ...510..379M} must be accounted for shocks that reach the outer edge of the CSM, above which the density drops off steeply.

The properties of the shock can be influenced by radiative diffusion. From Fig. \ref{fig:shock_structure}, the shock front is located a distance $\Delta r = R_{\rm csm}-r_{\rm sh}$ from the outer edge of the CSM. The timescale for photons to diffuse out ahead of the shock and escape is $t_{\rm esc}\sim \tau_{\rm sh} \Delta r/c$, where $\tau_{\rm sh}\sim \kappa \rho \Delta r$ is the radial optical depth from the shock to the CSM surface. We can compare this timescale to the dynamical timescale of the shock, $t_{\rm sh}\sim \Delta r/v_{\rm sh}$ for the shock of speed $v_{\rm sh}$ to traverse the same distance $\Delta r$. The ratio of these two timescales is
\begin{align}\label{eq:timescale_ratio}
\frac{t_{\rm esc}}{t_{\rm sh}} \approx \tau_{\rm sh}\frac{v_{\rm sh}}{c}
\end{align}
When $\tau_{\rm sh} \gtrsim c/v_{\rm sh}$, radiation is trapped at the shock front and advected with the flow. Radiation pressure mediates the shock, and assuming gas and radiation are in equilibrium the shock temperature is found by setting the ram pressure $\frac{1}{2}\rho v_{\rm sh}^2$ equal to the radiation pressure $\frac{1}{3} a T_{\rm eq}^4$, giving
\begin{align}\label{eq:temp_eq}
T_{\rm sh} = T_{\rm eq} \approx 10^5\,\,\rho_{-12}^{1/4}v_9^{1/2}\,\,{\rm K}\,\,\,\,\,(\tau_{\rm sh}\gtrsim c/v_{\rm sh})
\end{align}
where $\rho_{-12}=\rho/10^{-12}$ g cm$^{-3}$, and $v_9=v_{\rm sh}/10^9$ cm s$^{-1}$. The trapped radiation collects into a reservoir behind the shock front until it is able to escape at a later time, either due to the shock reaching the edge of the CSM or due to the shock decelerating sufficiently that the photon diffusion speed $\sim c/\tau_{\rm sh}$ exceeds  $v_{\rm sh}$. When  radiation remains trapped in the expanding medium, photons adiabatically degrade, converting the internal shock energy back into kinetic energy and decreasing the radiative throughput of the interaction.

When $\tau_{\rm sh} < c/v_{\rm sh}$, photons are able to escape ahead of the shock and power the light curve. If gas and radiation are not in equilibrium, the immediate post-shock temperature is determined by equating the ram pressure with the gas pressure $P_{\rm g}=\rho k_b T/\mu m_p$, giving
\begin{align}\label{eq:temp_shock}
T_{\rm sh}\approx 10^{9}\,\,v_9^2\,\,{\rm K}\,\,\,\,\,\,\,\,\,\,(\tau_{\rm sh}<c/v_{\rm sh})
\end{align}
which is much hotter than $T_{\rm eq}$ by several orders of magnitude. As photons are not trapped in the $\tau_{\rm sh}<c/v_{\rm sh}$ regime, we also need to determine how efficiently the shock can cool.
The thermal radiative cooling timescale is given by
\begin{align}\label{eq:cooling_time}
t_{\rm cool} = \frac{1}{\epsilon c \kappa\rho}\frac{n k_b T / (\gamma-1)}{a T^4}\approx 10^{-4}\,\,{\rm s}\,\,\epsilon^{-1}T_5^{-3}
\end{align}
where $\epsilon = \chi_{\rm abs}/(\chi_{\rm abs}+\chi_{\rm sc})$ is the ratio of absorptive to total (absoprtive plus scattering) extinction and we take the primary opacity source as electron scattering $\kappa\rho\approx \chi_{\rm sc}=n_e\sigma_T$, $n_e$ is the electron number density, and $\sigma_T$ is the Thomson cross-section. Thermal free-free emission \citep{1979rpa..book.....R} is important in cooling the radiative shocks discussed here, where
\begin{align}\label{eq:eps_freefree}
\epsilon_{\rm ff}\approx \frac{\chi_{\rm ff}}{n_e\sigma _T}\approx 10^{-6}\,\,T_5^{-7/2}\rho_{-12}
\end{align}
The free-free cooling time will therefore increase with shock temperature and density as $t_{\rm cool,ff}\propto T_{\rm sh}^{1/2}\rho^{-1}$. For high enough shock temperatures (or low enough densities), the gas will not be able to radiatively cool faster than the shock dynamical timescale. More specifically, free-free cooling will be efficient so long as $t_{\rm cool}<t_{\rm sh}$, which holds for shock optical depths greater than \citep{2022ApJ...928..122M}
\begin{align}\label{eq:tau_cool}
\tau_{\rm sh} \gtrsim 0.3v_9
\end{align}

For an optical depth less than $0.3v_9$, the shock inefficiently cools and is adiabatic. In this regime, non-thermal emission will become important. Here, we limit our focus to CSM optical depths where $\tau_0 \gtrsim 1$, and additionally assume non-relativistic shock velocities $v_{\rm ej}\lesssim 0.1c$ such that Eq. \ref{eq:tau_cool} more readily holds across the shock's evolution. Note that other processes which may aid in radiative cooling of the shock include lines and bound-free absorption, increasing the effective $\epsilon$ in Eq. \ref{eq:cooling_time}. In particular, from Eq. \ref{eq:eps_freefree} we see that free-free thermalization becomes less efficient at high temperatures (i.e. faster shocks). \cite{2022ApJ...928..122M} show that at these higher shock temperatures, inverse Compton scattering becomes the dominant thermalization process, which expands the $(\tau_{\rm sh},v_{\rm sh})$ parameter space in which the shock can efficiently radiate.

The kinetic luminosity of the forward shock (in the strong shock limit) is approximately the kinetic energy density $\rho v_{\rm sh}^2/2$ times the flux $4 \pi r_{\rm sh}^2 v_{\rm sh}$ through the shock front,
\begin{align}\label{eq:shock_lum}
L_{\rm sh}\approx 2\pi r_{\rm sh}^2\rho(r_{\rm sh})v_{\rm sh}^3.
\end{align}
A detailed analysis of how the shock heating evolves and eventually escapes to power the light curve is given in  \S \ref{sec:analytics}.

\begin{figure}
    \centering
    \includegraphics[width=0.5\textwidth]{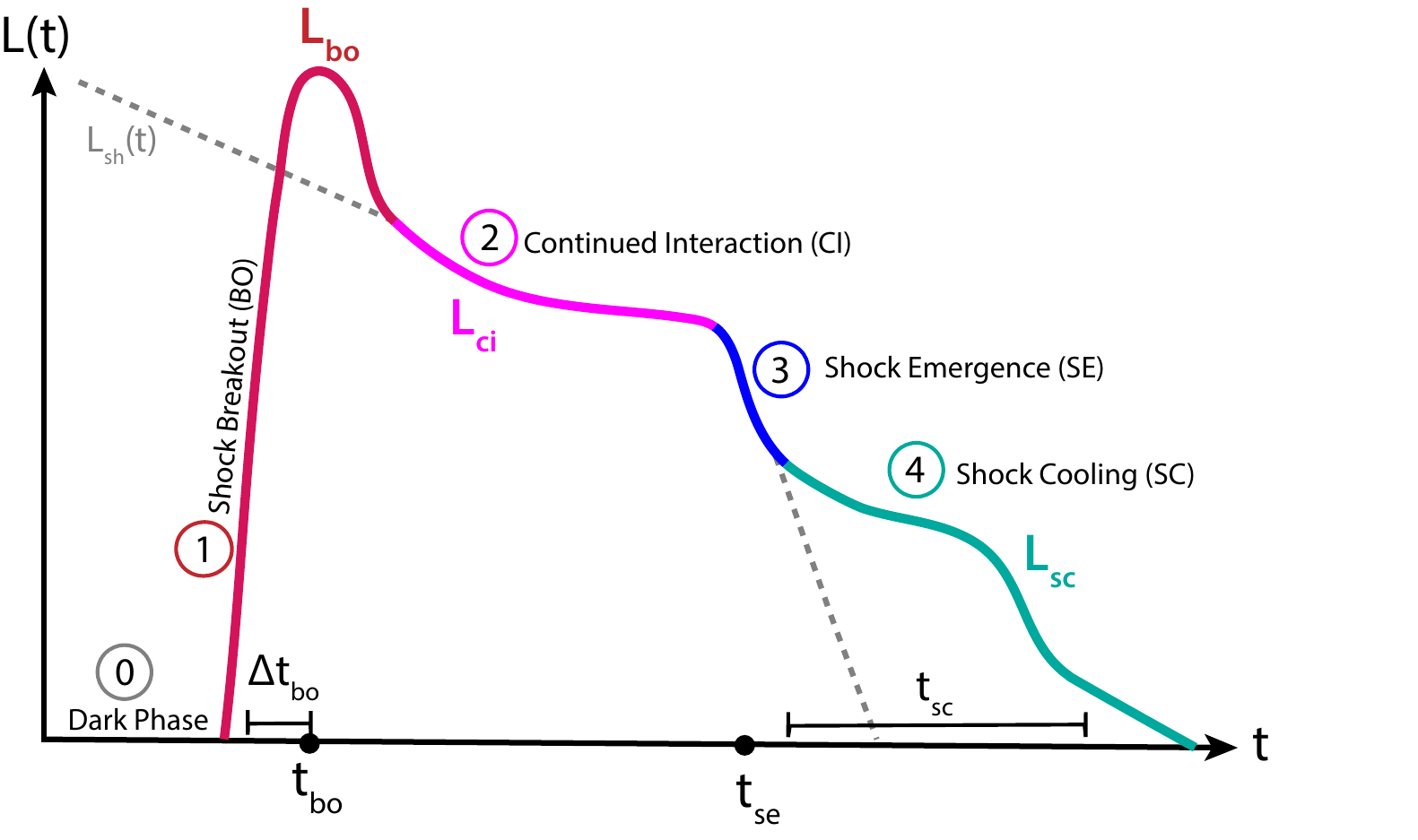}
    \caption{Schematic diagram (not to scale) of a CSM interaction light curve and the distinct phases that appear as the shock evolves in time. Also indicated are the characteristic luminosity and timescale of each phase, appearing in \S\ref{sec:analytics} as the boxed equations.}
    \label{fig:lc_phases}
\end{figure}    

\begin{figure*}
    \centering
        \includegraphics[width=1.0\textwidth]{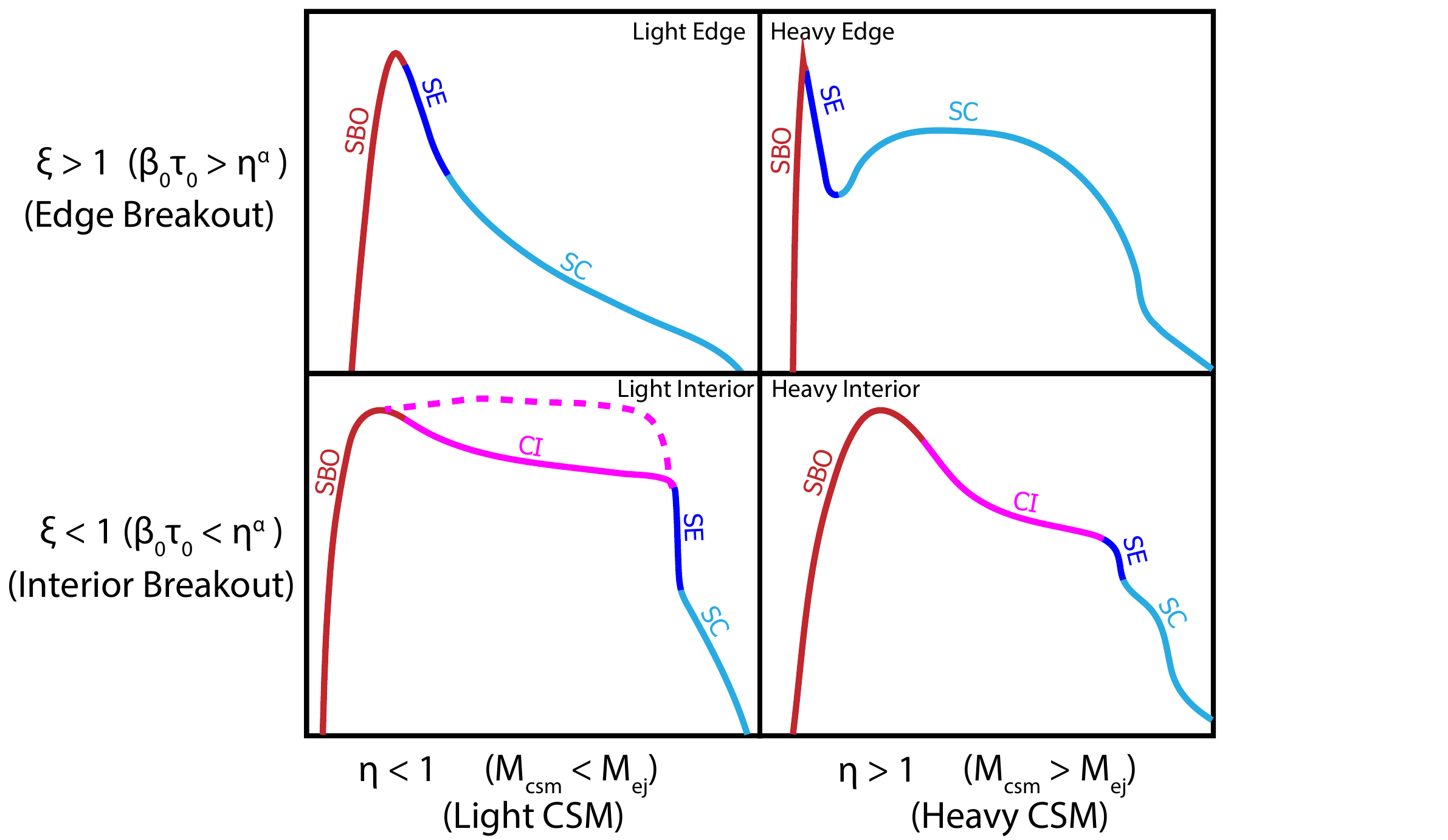}
    \caption{The four general classes of CSM interaction light curves, and the color-coded phases observable in each class. (\textit{SBO} = shock breakout (Phase 1); \textit{CI} = continued interaction (Phase 2); \textit{SE} = shock emergence (Phase 3); \textit{SC} = shock cooling (Phase 4)).}
    \label{fig:lc_classes}
\end{figure*}

\subsection{Light Curve Phases and Morphology}
\label{subsec:phases}
We can conceptually decompose the light curve arising from interaction into five phases, as illustrated in Figure~2.

\begin{enumerate}
\item[(0)] \textbf{Dark Phase:} The shock is propagating through the CSM, but photons are unable to escape ($\tau_{\rm sh} \gg c/v_s$) and remain trapped at the shock front. The interaction therefore produces no observable signal.

\item[(1)]\textbf{Shock Breakout:} The forward shock front reaches a low enough optical depth ($\tau_{\rm sh} \sim c/v_s$) that photons can diffuse ahead of the shock front, and the light curve rises to a peak.

\item[(2)]\textbf{Continued Interaction:} The forward shock continues propagating through the CSM and photons efficiently escape, such that the luminosity tracks the instantaneous energy deposition rate of the shock. Additionally, the reverse shock propagates inwards (in mass), generating additional heating of the ejecta.

\item[(3)] \textbf{Shock Emergence:} The forward shock reaches the outer edge of the CSM resulting in a sharp drop in luminosity as the shock heating abates.

\item[(4)] \textbf{Shock Cooling:} Photons produced at earlier times in deeper shock-heated regions continue to escape and power the light curve. Continued heating from the reverse shock or other sources (e.g. radioactivity) may also contribute. 
\end{enumerate}

We emphasize that in this terminology "shock breakout" refers to the escape of photons from the shock and not the exiting of shock from the system (which we instead label "shock emergence"). In some scenarios, "breakout" and "emergence" occur almost simultaneously at the CSM edge and this distinction is not be significant, but in extended CSM it is essential to consider the case where breakout occurs interior to the CSM edge well before emergence \citep{2011ApJ...729L...6C}. 

\begin{figure}
    \centering
    \includegraphics[width=0.5\textwidth]{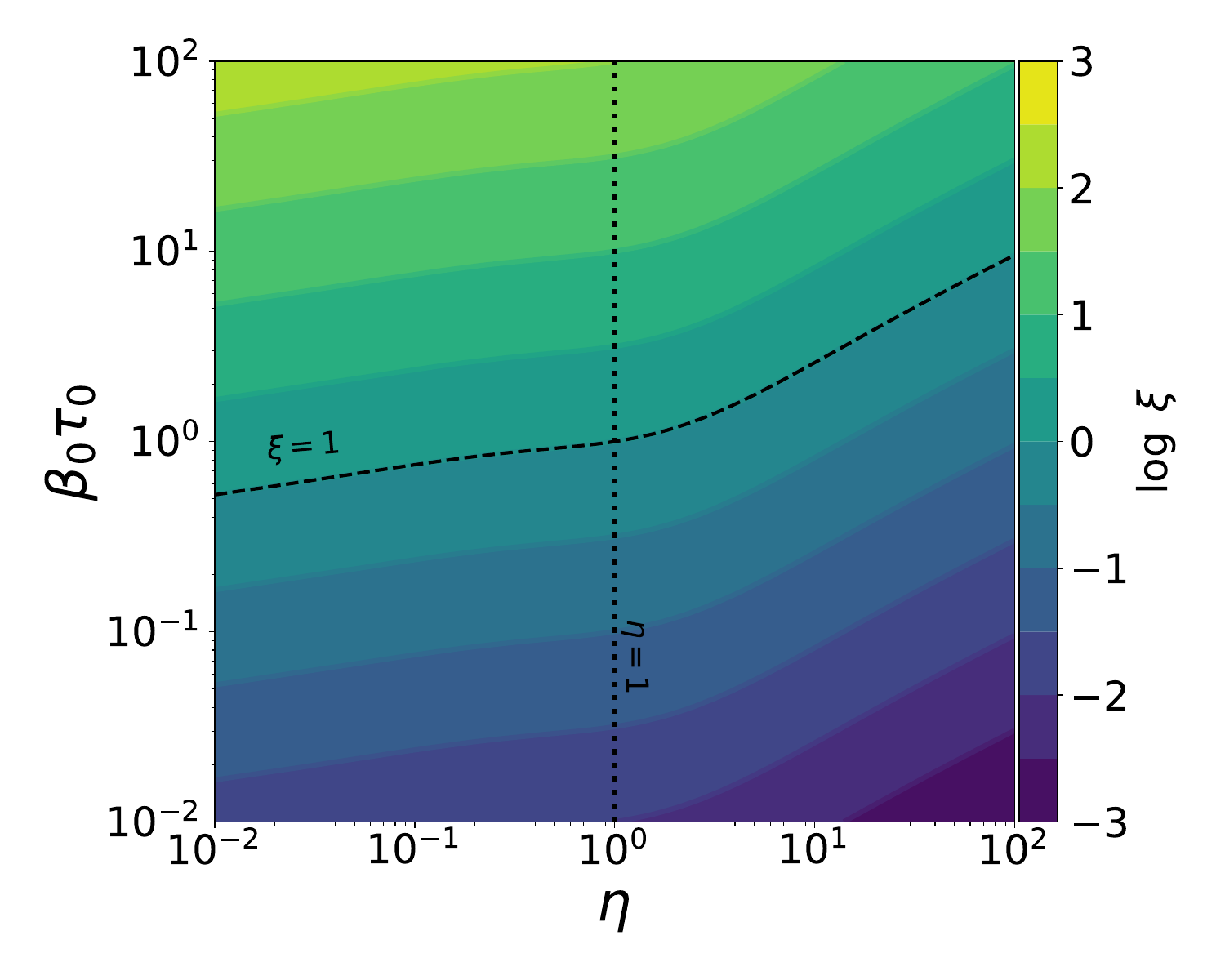}
    \caption{The breakout parameter $\xi$ in the $\eta,\beta_0\tau_0$ space. Dashed and dotted lines denote $\xi=1$ and $\eta=1$, respectively, separating the four classes in Fig. \ref{fig:lc_classes}. }
    \label{fig:xibo}
\end{figure}

The relative prominence of each of the above phases will depend on the parameters of the CSM/ejecta configuration.  As a result, interaction is capable of producing a  diversity of light curve behaviors. By taking $\tau_{\rm sh}\sim \tau_0$ and $v_{\rm sh}\sim v_{\rm ej}\eta^{-\alpha}$ in Eq. (\ref{eq:timescale_ratio}), we get $t_{\rm esc}/t_{\rm sh}\sim \xi$, the breakout parameter defined in Eq. \ref{eq:bo_param}. If $\xi>1$, shock breakout occurs at the CSM edge; if $\xi<1$, breakout will instead occur within the CSM. We can then define  four light curve classes (see Fig. \ref{fig:xibo}) based on whether the CSM significantly decelerates the shock ($\eta > 1$, "heavy CSM") or not ($\eta < 1$, "light CSM") and whether breakout occurs at the CSM edge ($\xi > 1$) or in its interior ($\xi < 1$).

\bigskip \noindent {\bf Edge-breakout; Light CSM} ($\xi \gg 1$, $\eta \ll 1$):  
In this scenario, the CSM is so optically thick that shock breakout (phase 2) occurs in the steep density profile just outside the CSM edge at $R_{\rm csm}$ \citep{2010ApJ...724.1396O,2017hsn..book..967W}. The breakout of radiation  is almost immediately followed by shock emergence (phase 4), with essentially no phase of continued interaction (phase 3). This leads to a relatively sharply rising and falling light curve. Shock cooling (phase 5) after the shock has emerged results in an extended light curve tail after breakout \citep{2015ApJ...808L..51P,2021ApJ...909..209P}.

\bigskip \noindent {\bf Edge-breakout; Heavy CSM} ($\xi \gg 1$, $\eta \gtrsim 1$):  
As with the previous scenario, breakout and emergence happen at the edge of the CSM, producing a shapr breakout peak in the light curve. The subsequent shock cooling phase, however, is more prominent, as the high CSM mass leads to a deceleration and thermalization of the bulk of the ejecta kinetic energy. Given the higher mass and lower velocity of the shocked gas, the cooling emission diffuses out on a longer timescale, leading to a distinct second "shock-cooling" bump in the emergent light curve.


\bigskip \noindent {\bf Interior-breakout; Light CSM} ($\xi \lesssim  1; \eta \ll 1$):  
In this scenario, shock breakout occurs well before the forward shock has reached the CSM edge (see e.g., \cite{2011ApJ...729L...6C,2019ApJ...884...87T}). The  peak in the light curve associated with breakout is followed by an extended continued interaction phase, where the luminosity tracks the shock kinetic luminosity Eq. \ref{eq:shock_lum}. The slope of the light curve in the continued interaction phase thus depends directly on the density profile of the CSM. Once the shock reaches the outer edge of the CSM shell and emerges, the shock luminosity drops rapidly, leading to a sharp decline in the light curve followed by a shock cooling tail. 

\bigskip \noindent {\bf Interior-breakout; Heavy CSM} ($\xi \lesssim  1; \eta \gtrsim 1$):  As with the previous scenario, shock breakout occurs before the forward shock has reached the CSM edge (see \cite{2012ApJ...757..178G,2015MNRAS.449.4304D}). Given the large CSM mass, the shock velocity is significantly decelerated as it sweeps up the CSM, such that the breakout condition $\tau_{\rm sh} \sim c/v_{\rm sh}$ is reached within the CSM, resulting in a more gradual rise to a breakout peak. Following a phase of continued interaction, shock emergence leads to a modest drop in luminosity as the light curve transitions to shock cooling emission.

\section{Analytic Scalings}
\label{sec:analytics}

Numerous previous works have considered analytical models and scaling relations for interaction light curves (see e.g. \cite{2010ApJ...724.1396O,2011ApJ...729L...6C,2012ApJ...746..121C,2012ApJ...757..178G,2015ApJ...808L..51P,2017hsn..book..967W,2019ApJ...884...87T}). These various results often contradict each other, usually due to different assumptions made in the derivation which renders particular results valid only in specific regions of parameter space. Here we present scaling relations for the light curve luminosity and duration, clarifying the regimes of applicability within the  $(\xi,\eta)$ parameter space. \S \ref{sec:numerical} validates these relations with numerical radiation-hydrodynamics simulations. Equations that correspond to quantities that appear in Fig. \ref{fig:lc_phases} are boxed for clarity. Correction factors for the scaling relations, calibrated to the numerical simulations, are provided in Appendix \ref{sec:physical_scalings}.

 Consider a shell of shocked material at radius $r$ and of thickness $\Delta r$.  After a shock has passed through the shell, the post-shock thermal energy is roughly
\begin{equation}\label{eq:shock_energy}
E_{\rm sh} = 4\pi r^2\Delta r \rho_{\rm s} \frac{v_{\rm sh}^2}{2}
\end{equation}
Initially the shock velocity will be of order the ejecta velocity scale, $v_{\rm ej} = (2E_{\rm sn}/M_{\rm ej})^{1/2}$. However, as the shock sweeps up material in the CSM, it is decelerated. From Eq. \ref{eq:shock_velocity}, by assuming the shock radius evolves as a power-law $r_{\rm sh}\propto t^{\lambda}$, we have that the shock velocity evolves as
\begin{align*}
v_{\rm sh}(r_{\rm sh}) \approx v_{\rm ej}\eta^{-\alpha}\left(\frac{r_{\rm sh}}{R_{\rm csm}}\right)^{(\lambda-1)/\lambda}
\end{align*}
where the factor of $\eta^{-\alpha}$ accounts for the slowing down of the shock, and $\eta=M_{\rm csm}/M_{\rm ej}$. A full derivation of $v_{\rm sh}$, as well as the shock exponents $\alpha$ and $\lambda$, is provided in Appendix \ref{sec:shock_derivation}.

If $\eta \gtrsim 1$, then the bulk of the ejecta kinetic energy is tapped by the interaction, and $\alpha=1/2$ from energy conservation. On the other hand, if $\eta\ll 1$, then only a fraction of $E_{\rm sn}$ will be thermalized. The amount of deceleration that occurs will therefore depend on the outer density profile of the ejecta in the $\eta<1$ case, and so
\begin{align}\label{eq:alpha_eta}
\alpha = \left\{\begin{array}{lr}
    1/2, & (\eta\gtrsim 1) \\
    1/(n-3), & (\eta\ll1)
\end{array}\right.
\end{align}
where $n$ is the power-law exponent of the outer ejecta, $\rho_{\rm ej}\propto r^{-n}$ with $n\approx 7-10$ \citep{2016ApJ...821...36K}. The mass ratio between the inner and outer ejecta is equal to $(3-\delta)/(n-3)$ \citep{1989ApJ...341..867C}. For $\delta = 1$ and $n=10$, the shock will transition between the inner and outer portions of the ejecta for $\eta\gtrsim 0.3$.

The time it takes for the shock to reach the outermost shell at $r = R_{\rm csm}$, accounting for the shock deceleration, is the \textit{shock emergence} timescale,
\begin{align}\label{eq:t_se}
\boxed{t_{\rm se} \approx \eta^{\alpha}t_0},
\end{align}
where $t_0=R_{\rm csm}/v_{\rm ej}$. Depending on the optical depth of the CSM and how fast the shock is moving, shock breakout may occur at a deeper shell than the one located at $R_{\rm csm}$, and must be accounted for. In this case, the light curve begins rising at a time $t<t_{\rm se}$. We now separate our analysis into these two breakout regimes.


\bigskip \noindent \underline{{\bf  Scenario 1: Shock Breakout at CSM Edge ($\xi\gg 1$):}} 
\medskip

In this regime, the CSM is sufficiently optically thick  that the condition $\tau_{\rm sh} \approx c/v_{\rm sh}$ is not reached until the shock has traversed the entire CSM and begun accelerating down the steep outer edge. This scenario resembles stellar surface shock breakout in several ways \citep{1999ApJ...510..379M,2012ApJ...747..147K}, and so we proceed along a similar analysis.

Breakout happens at a radius $r_{\rm bo}$, where the photons contained in a shell of width $\Delta r_{\rm bo}$ escape. The post-shock energy in the shell is
\begin{equation}
\Delta E_{\rm bo} \approx 4\pi r_{\rm bo}^2\Delta r_{\rm bo} \left( \frac{1}{2} \rho_{\rm bo} v_{\rm bo}^2 \right)
\label{eq:deltaE_shell}
\end{equation}
where $v_{\rm bo}$ and $\rho_{\rm bo}$ are the shock velocity and CSM density at the breakout location. At breakout, $t_{\rm bo}\approx t_{\rm se}\approx \eta^\alpha t_0$, radiation escapes from the shell on a timescale comparable to the dynamical timescale, $\Delta t_{\rm bo} \sim \Delta r/v$, giving a luminosity of
\begin{equation}
L_{\rm bo} \approx \frac{\Delta E_{\rm bo}}{\Delta t_{\rm bo}} \approx 2 \pi r_{\rm bo}^2 \rho_{\rm bo} v_{\rm bo}^3
\label{eq:breakout_L_outeredge}
\end{equation}
When the shock just reaches the CSM edge its velocity is
$v \sim v_{\rm ej} \eta^{-\alpha}$, where the factor $\eta^{-\alpha}$ accounts for interaction with the bulk of the CSM (see Appendix \ref{sec:shock_derivation}). Once the shock passes $R_{\rm csm}$ it begins accelerating  down the steeply dropping outer density profile, which we take to be a power law
\begin{equation}
\rho(r)\approx \rho_0 x^{-p}
\label{eq:rho_outer_edge}
\end{equation}
where $x = r/R_{\rm csm}$ and the exact value of $p$ will not matter in the limit $p \gg 1$. We account for shock acceleration using Sakurai's law  $v \propto \rho^{-\hat{\delta}}$, \citep{https://doi.org/10.1002/cpa.3160130303}, where $\hat{\delta} \approx 0.2$ for a strong shock \citep{2017hsn..book..967W}. The shock velocity in the steep outer region is then
\begin{equation}
v_{\rm sh} \approx v_{\rm ej} \eta^{-\alpha} x^{p \hat{\delta}} 
\label{eq:vsh_Sakurai}
\end{equation}
To find the point $x_{\rm bo}$ where the shock reaches optical depth $\tau\sim c/v_s$ we integrate the density profile 
\begin{equation}
\tau = \int_x^\infty \rho(r) \kappa dr
\approx \tau_0 x^{-p}
\end{equation}
where $\tau_0=\kappa M_{\rm csm}/4\pi R_{\rm csm}^2$ and we have assumed $p \gg 1$. 
Setting this $\tau$ equal to $c/v_{\rm sh}$ where $v_{\rm sh}$ is given by Eq.~\ref{eq:vsh_Sakurai}, we can solve for the radius
where breakout occurs
\begin{equation}
x_{\rm bo} = \left[ \tau_0 \beta_0 \eta^{-\alpha} \right]^{1/p(1-\hat{\delta})} = \xi^{1/p(1-\hat{\delta})}
\label{eq:breakout_xbo}
\end{equation}
where $\beta_0 = v_{\rm ej}/c$ and $\xi=\beta_0\tau_0\eta^{-\alpha}$.  For $p \gg 1$ we have $x_{\rm bo} \approx 1$, but it is important to use Eq.~\ref{eq:breakout_xbo} to evaluate the breakout velocity, $v_{\rm bo}$ (from Eq.~\ref{eq:vsh_Sakurai}) and the breakout density (from Eq.~\ref{eq:rho_outer_edge}). 
Using these in the expression for the breakout luminosity Eq.~\ref{eq:breakout_L_outeredge} and choosing $\hat{\delta}=0.2$ gives
\begin{equation}\label{eq:breakout_lbo}
\boxed{L_{\rm bo} \approx \eta^{-3\alpha}\xi^{-1/2}L_0}
\end{equation}
where $L_0 =  M_{\rm csm}v_{\rm ej}^3/R_{\rm csm}$ is the characteristic luminosity scale defined in  \S\ref{sec:configuration}.

The duration of this breakout emission is $\Delta r_{\rm bo}/v_{\rm bo}$.  Given that the optical depth through the breakout layer
$\tau \approx \kappa \rho_{\rm bo} \Delta r_{\rm bo} $ is roughly equal to $c/v_{\rm bo}$, we have $\Delta r_{\rm bo} = c/\kappa \rho_{\rm bo} v_{\rm bo}$ so the timescale is \citep{2017hsn..book..967W}
\begin{equation}
\Delta t_{\rm bo} \approx \frac{\Delta r_{\rm bo}}{v_{\rm bo}} \approx 
\frac{c}{\rho_{\rm bo} \kappa v_{\rm bo}^2}
\approx \frac{R}{c} \frac{1}{\tau_0 \beta_0^2} \eta^{2\alpha} x_{\rm bo}^{p(1 - 2\hat{\delta})}
\end{equation}
Plugging in $x_{\rm bo}$ from Eq.~\ref{eq:breakout_xbo} and taking $\delta=0.2$ gives 
\begin{equation}\label{eq:breakout_dtbo}
\boxed{\Delta t_{\rm bo}
\approx  \eta^{\alpha}\xi^{-1/4}t_0}
\end{equation}

\bigskip
{\noindent\underline{\textit{\textbf{Post-Breakout Cooling Emission:}}}} 

\medskip \noindent Following shock  breakout, energy deposited by the shock at earlier times will begin to diffuse out from deeper layers. The total  energy deposited will be 
\begin{equation}
\Delta E \approx \frac{1}{2} M_{\rm csm} (v_{\rm ej} \eta^{-\alpha} )^2
\end{equation}
To derive scaling relations we treat the system in a one zone approximation \citep{2015ApJ...808L..51P,2022ApJ...933..238M}. Assuming that the remnant expands on a ballistic trajectory after shock breakout with speed $v_{\rm ej} \eta^{-\alpha}$, the radius increases in time as $r(t) \approx  R_{\rm csm} + v_{\rm ej}\eta^{-\alpha} t$. We can consider two limits. When the diffusion time is much faster than the expansion time $t_e = R_{\rm csm}/v_{\rm ej} \eta^{-\alpha}$, the remnant can be considered quasi-static with fixed radius $R_{\rm csm}$. The diffusion time is then $\Delta t = \tau_0 R_{\rm csm}/c$, and the shock cooling luminosity $L_{\rm sc} \sim \Delta E/\Delta t$ scales as 
\begin{equation}\label{eq:L_cool}
\boxed{L_{\rm sc} \approx \eta^{-3\alpha}\xi^{-1}L_0}
\end{equation}

In the other limit where the diffusion time is much longer than the expansion time, the remnant will expand by a significant factor before radiating and we approximate the radius as $r(t) \approx v_{\rm ej}\eta^{-\alpha} t$.  Radiation will escape when the remnant has expanded to the point that $\tau \sim c/v_{\rm ej}\eta^{-\alpha}$ which occurs at a time 
\begin{align}\label{eq:t_cool}
\boxed{t_{\rm sc} \approx \eta^{\alpha}\xi^{1/2}t_0}
\end{align}
The shock deposited energy will be reduced due to expansion, such that the thermal energy remaining when radiation can diffuse out is 
\begin{equation}
\Delta E \approx \frac{1}{2} M_{\rm csm} (v_{\rm ej} \eta^{-\alpha} )^2
\left( \frac{R_{\rm csm}}{v_{\rm ej} \eta^{-\alpha} t_{\rm sc}} \right)
\end{equation}
where the factor in parentheses accounts for the losses due to adiabatic expansion of a radiation dominated gas from an initial radius $R_{\rm csm}$ to a final one at 
$v_{\rm ej} \eta^{-\alpha} t_{\rm sc}$. The peak luminosity in this shock cooling phase will then scale as $L_{\rm sc} \sim \Delta E/t_{\rm sc}$ which results
in a similar expression for the luminosity as Eq. \ref{eq:L_cool}.

Note that Eqs. \ref{eq:L_cool} and \ref{eq:t_cool} are identical to the expressions found in \cite{2015ApJ...808L..51P} for the case of $\alpha=0.15$.
A more detailed analysis of the shock cooling emission is presented in \cite{2022ApJ...933..238M}.

\bigskip \noindent \underline{{\bf Scenario 2: Shock Breakout in CSM Interior ($\xi\lesssim 1$):}}
\medskip
\medskip\noindent

If the breakout shell is located within the CSM, then we need to account for the time-dependent evolution of the shock. In this case, the shock will propagate and be continuously decelerated within the CSM. 

We assume the shock evolves in time as a power law, $r_{\rm sh}\propto t^{\lambda}$, and so at $x_{\rm bo}=r_{\rm bo}/R<1$

\begin{align}\label{eq:interior_vbo}
v_{\rm bo}\approx v_{\rm ej}\eta^{-\alpha} x_{\rm bo}^{1-1/\lambda}
\end{align}

The shock exponent $\lambda$ will depend on the CSM density profile for $\eta\gtrsim 1$ and, for $\eta \ll 1$, the ejecta density profile as well (\cite{1963idp..book.....P,1982ApJ...258..790C,1988RvMP...60....1O}; see also Appendix \ref{sec:shock_derivation}),

\begin{align}\label{eq:shock_lambda}
\lambda=\left\{\begin{array}{cc}
   2/(5-s)  &  (\eta \gtrsim 1)\\
   (n-3)/(n-s)  & (\eta\ll 1) 
\end{array}\right.
\end{align}

The breakout luminosity will be roughly equal to the shock luminosity at breakout (Eq.~\ref{eq:shock_lum}),
\begin{align}
L_{\rm bo} \approx L_0\eta^{-3\alpha}x_{\rm bo}^{5-s-3/\lambda}
\end{align}
where we have assumed a power-law CSM density profile $\rho_{\rm csm}\propto r^{-s}$.

In order to calculate the breakout time, we must integrate the CSM density profile to the breakout optical depth $\tau_{\rm bo}=c/v_{\rm bo}$ and set $v_{\rm bo}$ equal to Eq. \ref{eq:interior_vbo}, which results in a nonlinear equation that must be solved numerically (see Appendix \ref{sec:shock_derivation}). Instead, we here approximate the breakout radius with the expression
\begin{align}\label{eq:interior_xbo}
x_{\rm bo}\approx \left[\tau_0\beta_0\eta^{-\alpha}\right]^{\lambda k_0} = \xi^{\lambda k_0}
\end{align}
where $0 \leq k_0\leq 1$. For $k_0\approx 0$, $x_{\rm bo}\approx 1$ i.e. breakout occurs at the CSM edge; this is the case discussed in the previous section. Interior breakout $x_{\rm bo}<1$ thus requires $k_0>0$ and $\xi<1$. Note that $k_0=1$ gives $t_{\rm bo}\approx \kappa M_{\rm csm}/4\pi R_{\rm csm}c$, which is simply the static diffusion timescale. This corresponds to the breakout time used in \cite{2011ApJ...729L...6C} and \cite{2012ApJ...757..178G} up to a constant prefactor.

In general, $k_0$ will depend on the configuration of the ejecta and CSM parameters, which can be viewed as a weighted average of the shock emergence and static diffusion timescales. If radiation is able to immediately escape the CSM at the onset of interaction, then a choice of $k_0=1$ is more appropriate. On the other hand, if radiation escapes only once the shock nears the edge, then $k_0\approx 0$. Here we adopt an intermediate value of $k_0\approx 0.6$ based on fits to numerical simulations presented in \S \ref{sec:numerical}, which is appropriate in the regime where the shock sweeps up a fraction of the CSM before being able to radiate ahead of the shock.

Using this approximation for $x_{\rm bo}$, the breakout time is then given by
\begin{align}\label{eq:interior_tbo}
\boxed{t_{\rm bo} \approx \eta^{\alpha}\xi^{k_0}t_0}
\end{align}

The breakout duration $\Delta t_{\rm bo}$ is proportional to the breakout time; comparison with numerical simulations of \S \ref{sec:numerical} show that $\Delta t_{\rm bo}\approx t_{\rm bo}/2$. Finally, to find the breakout luminosity, we use Eq.\ref{eq:interior_xbo} for $x_{\rm bo}$ to get
 \begin{align}\label{eq:interior_Lbo}
\boxed{L_{\rm bo}\approx   \eta^{-3\alpha}\xi^{\sigma_s}L_0}
\end{align}
where $\sigma_s = \left[\left(5-s\right)\lambda-3\right]k_0 $; the shock exponent $\lambda$ is given by Eq. \ref{eq:shock_lambda} ; and $k_0$ is the same as in Eq. \ref{eq:interior_xbo}. For the case of $n=10$ and $s=2$, this gives $\sigma_s\approx -0.23$ for $\eta<1$ and $k_0\approx 0.6$. For $\eta>1$, $\sigma_s = -k_0$, independent of the density profiles. Note that Eqs. \ref{eq:interior_tbo} and \ref{eq:interior_Lbo} are equivalent to that derived in \cite{2011ApJ...729L...6C} for the choice of $k_0=1$, $\lambda=4/5$, and $\alpha=1/4$ (corresponding to an outer ejecta density profile $\rho_{\rm ej}\propto r^{-7}$ and CSM profile $\rho_{\rm csm}\propto r^{-2}$), where they implicitly work in the $\eta<1$ regime.

\begin{figure*}
    \centering
    \includegraphics[width=0.85\textwidth]{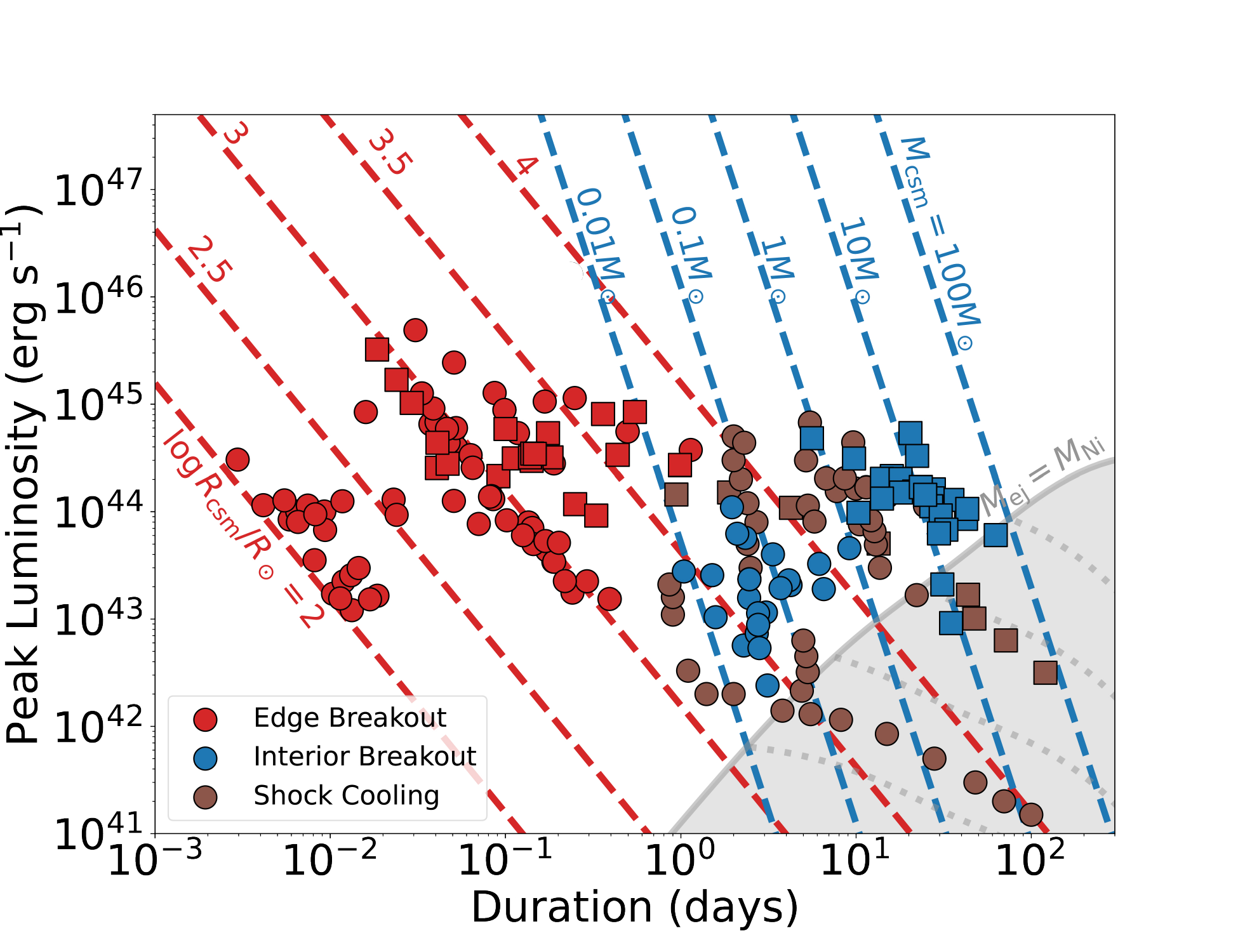}
    \caption{Numerical model results in terms of their breakout duration and luminosity for edge (red) and interior (blue) breakout events separated by light CSM (circles) and heavy CSM (squares). We also show the shock cooling emission for a subset of edge breakout events (brown points). Dashed blue and red lines denote $M_{\rm csm}$ and $R_{\rm csm}$ contours from Eqs. \ref{eq:mcsm_bo} and \ref{eq:rcsm_bo}, respectively. The space of radioactive-powered transients is shown as a shaded grey region. Here, the models cover ejecta mass/energy in the range $0.1M_\odot\leq M_{\rm ej}\leq 10M_\odot$ and $10^{49}\leq E_{\rm sn}\leq 10^{52}$ ergs, respectively.}
    \label{fig:dlps}
\end{figure*}

\medskip\noindent {\underline{\textit{\textbf{ Post-Breakout Continued Interaction:}}}} 

\medskip 
After shock breakout within the CSM, the light curve will continue to be powered by an additional supply of unshocked CSM, in addition to the reverse shock propagating inwards through the ejecta \citep{2011ApJ...729L...6C,2012ApJ...746..121C,2019ApJ...884...87T}. As the shock photons are able to efficiently radiate after breakout, the light curve tracks the instantaneous shock energy deposition rate Eq.~\ref{eq:shock_lum}.
Using the power-law forms of the shock evolution Eq. \ref{eq:shock_velocity} and assuming the forward shock dominates, the continued interaction luminosity becomes
\begin{align}\label{eq:continued_Lsh}
\boxed{L_{\rm ci}(t) \approx  L_0\eta^{-3\alpha}\left[\frac{t}{\eta^\alpha t_0}\right]^{(5-s)\lambda-3}}
\end{align}

For $\eta \ll 1$ the exponents $\alpha$ and $\lambda$ are identical to those provided in Eqs. \ref{eq:alpha_eta} and \ref{eq:shock_lambda}, respectively, as these exponents hold for both energy and momentum-conserving shocks \citep{1982ApJ...258..790C}. However, for $\eta \gtrsim 1$, the \cite{1959sdmm.book.....S} exponents no longer hold, as the blastwave transitions to a momentum-conserving snowplow whose exponents are given by \citep{1988RvMP...60....1O}
\begin{align}\label{eq:snowplow_exp}
\alpha = 1,\,\,\lambda = \frac{1}{(4-s)}~~~~(\eta>1)
\end{align}

\begin{figure*}
\plottwo{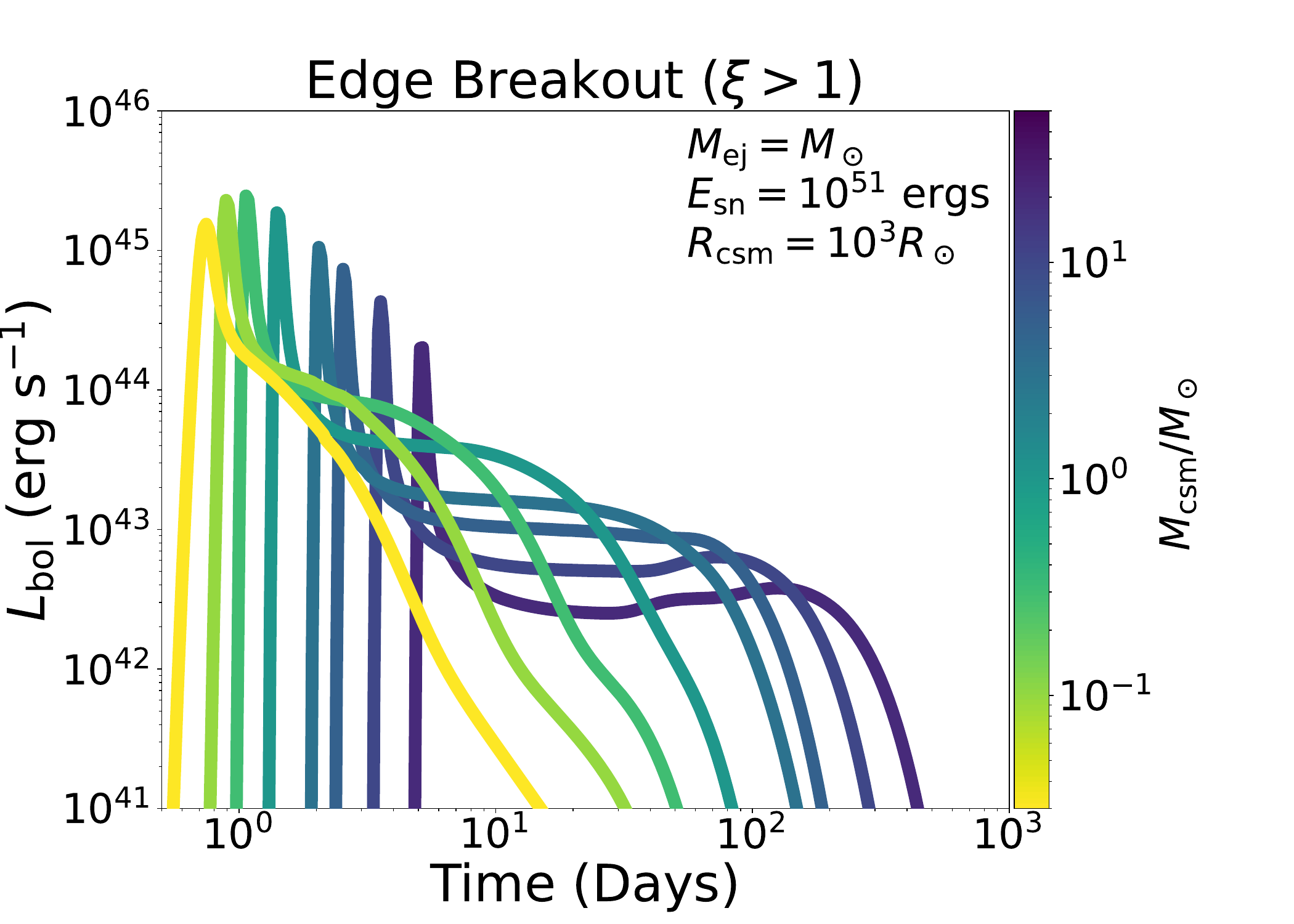}{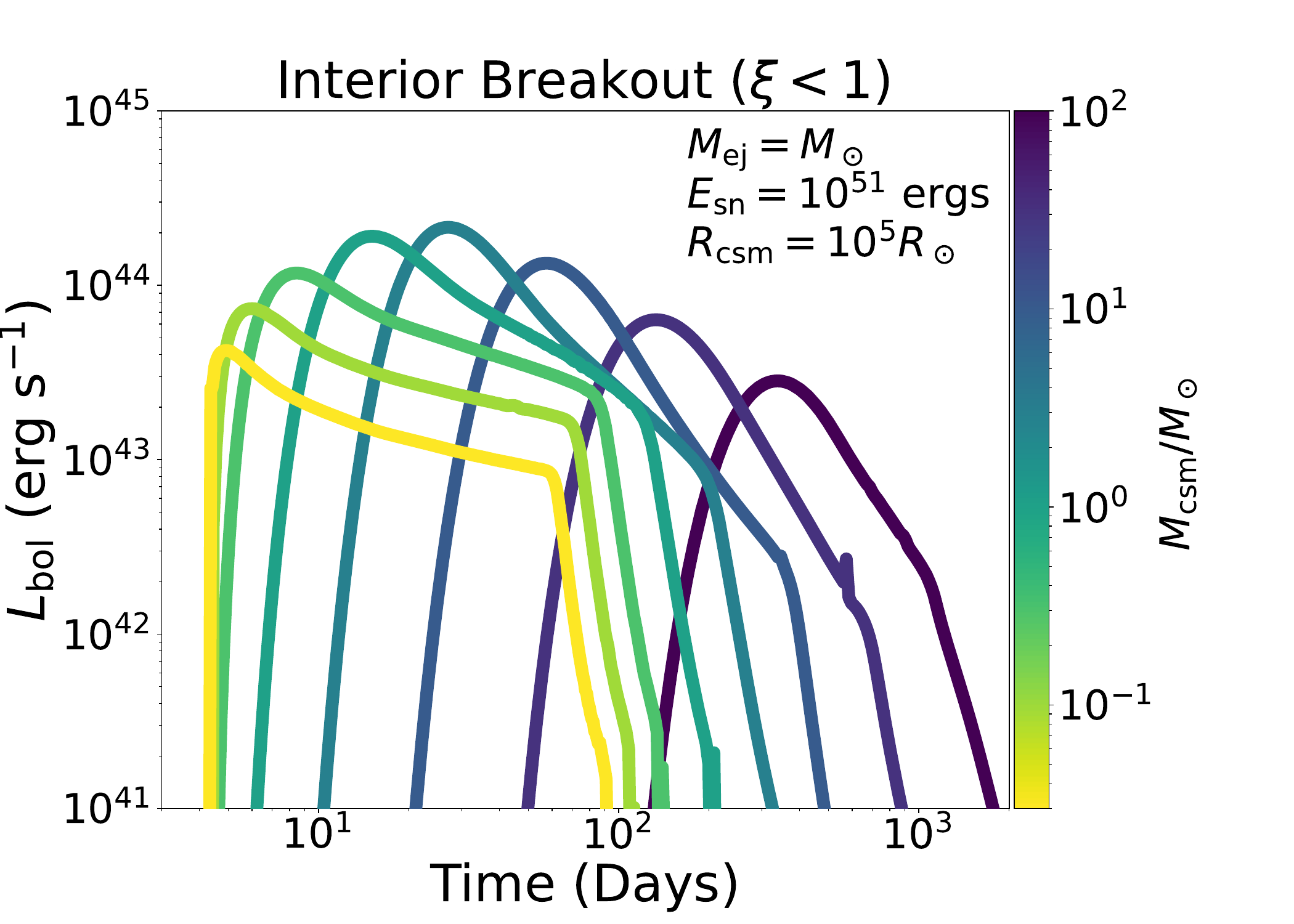}
\plottwo{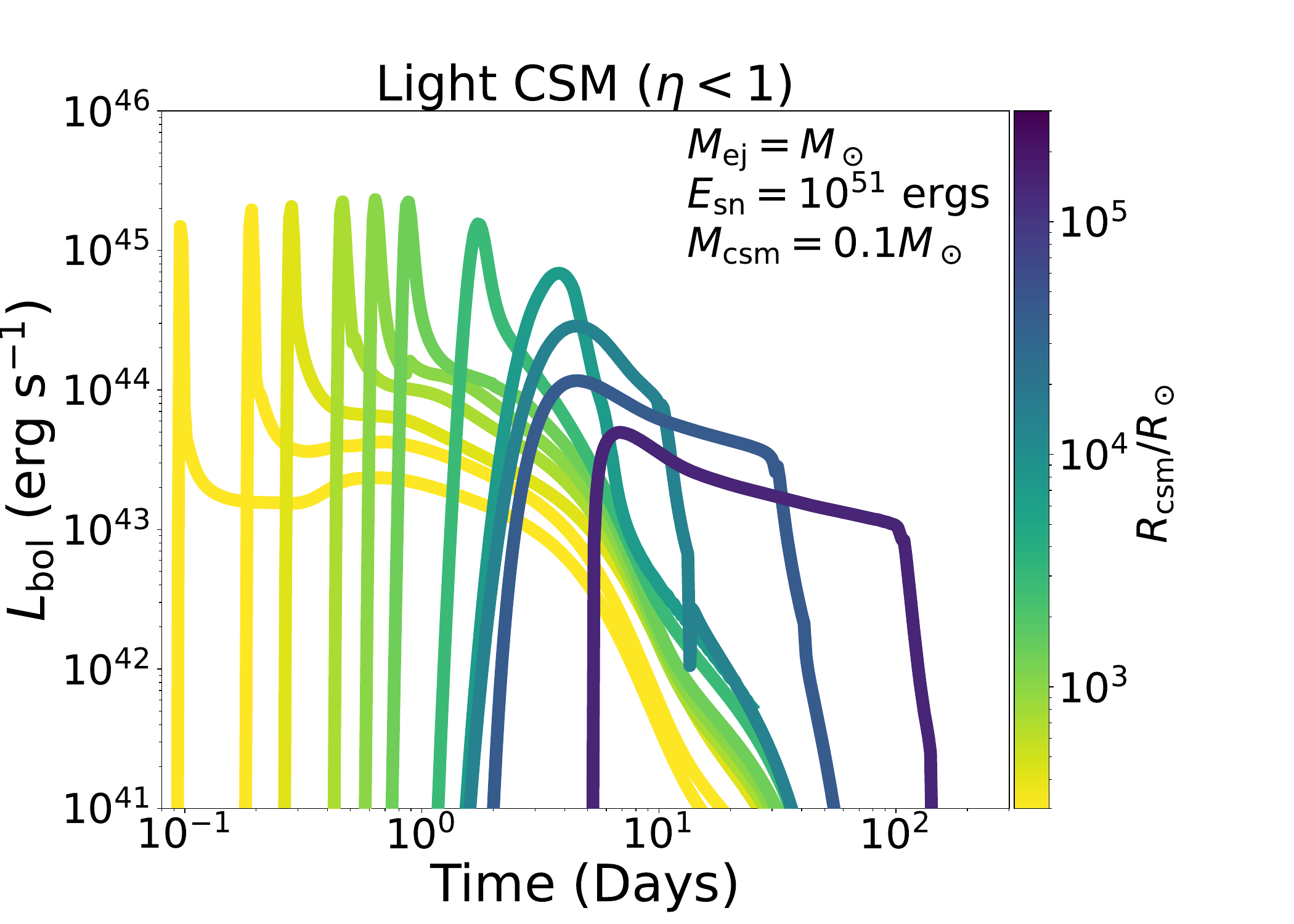}{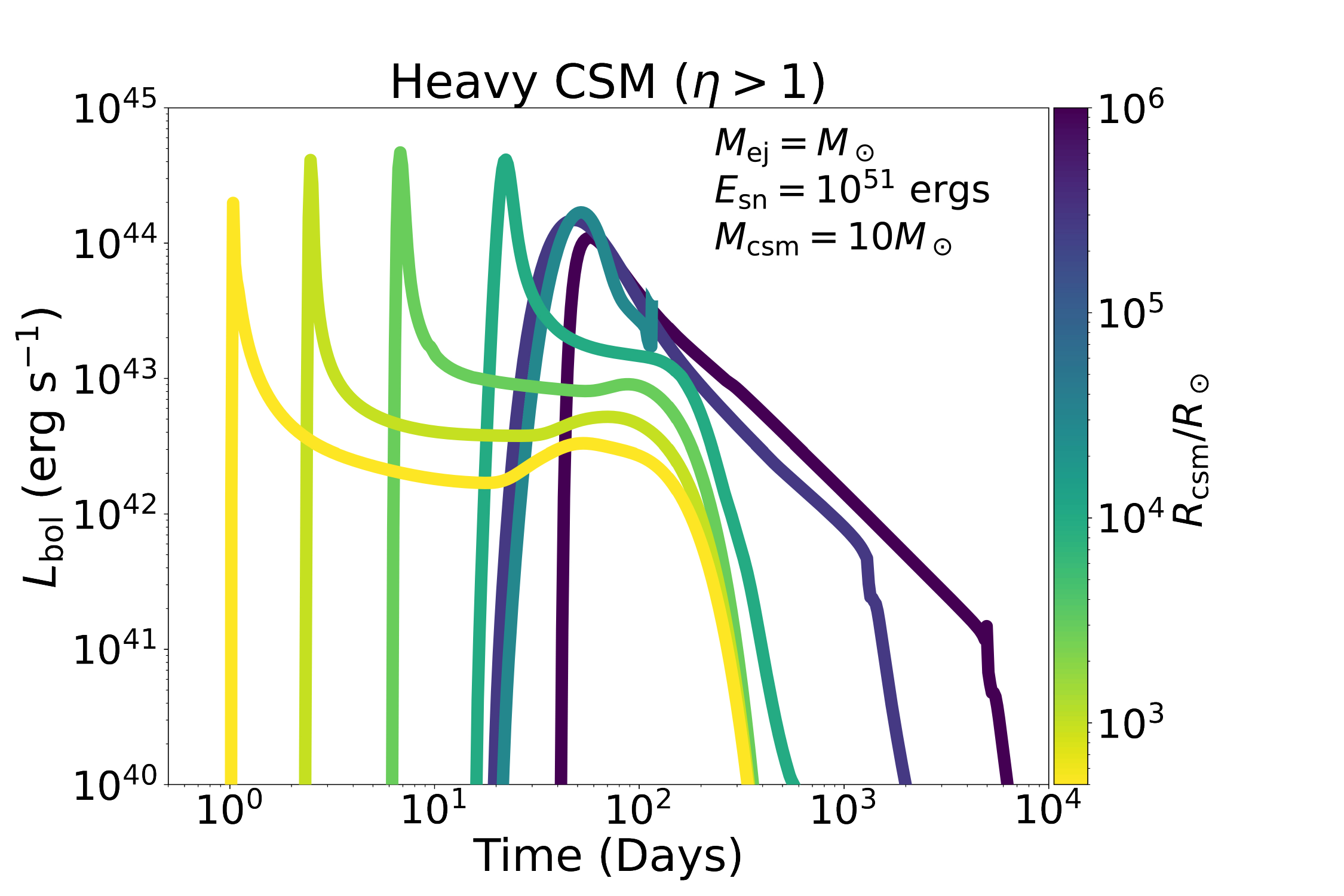}
\caption{Numerical light curves of CSM interaction for different CSM parameters, assuming fixed ejecta properties $M_{\rm ej}=M_\odot$ and $E_{\rm sn}=10^{51}$ ergs. \textit{Top row}: varying $M_{\rm csm}$ for two fiducial choices of $R_{\rm csm}$ in the edge (\textit{left}) and interior (\textit{right}) regimes. \textit{Bottom row}: varying $R_{\rm csm}$ for two fiducial choices of $M_{\rm csm}$ in the light (\textit{left}) and heavy (\textit{right}) CSM regimes. }
\label{fig:csm_lightcurves}
\end{figure*}



\begin{figure*}
\plottwo{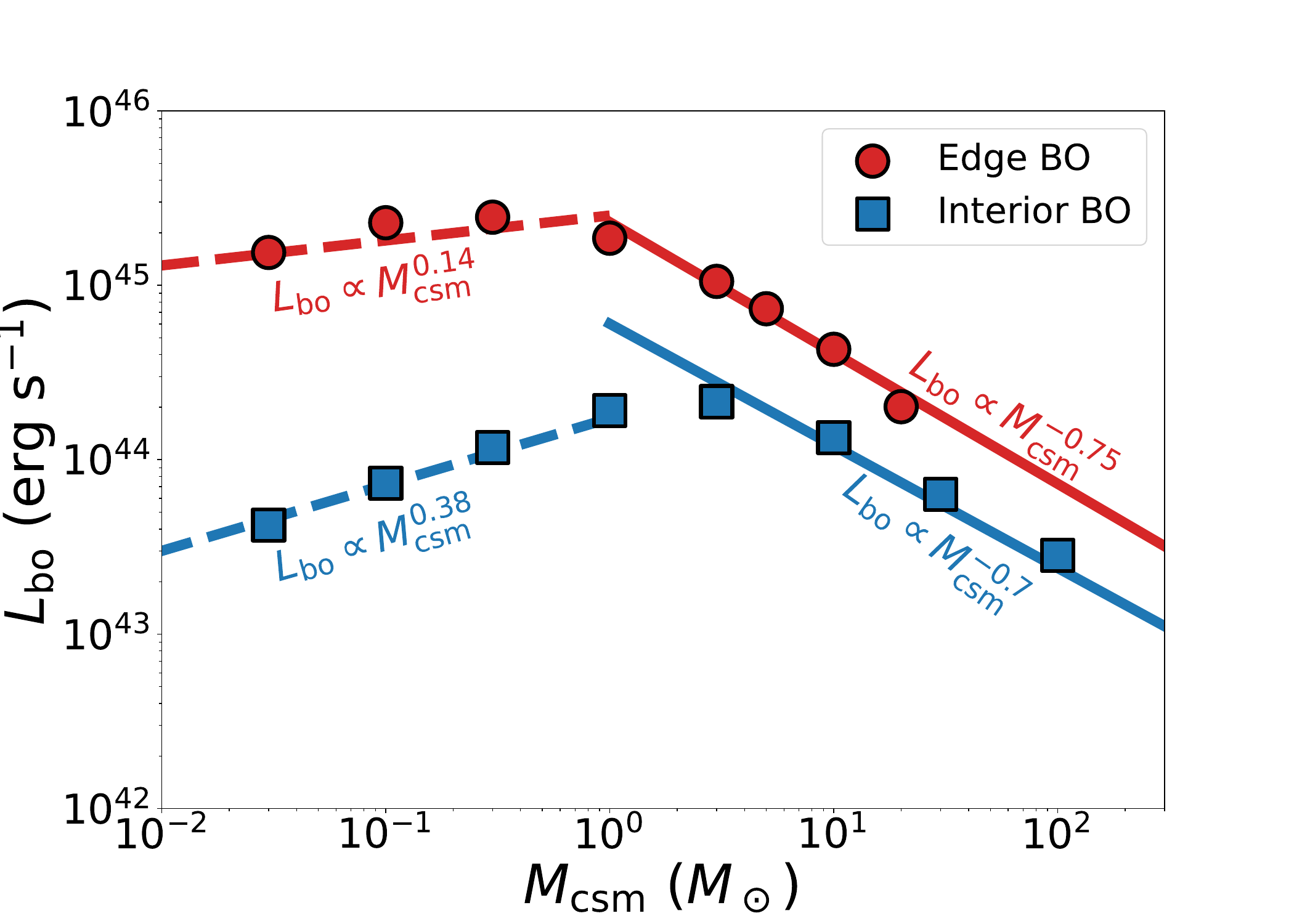}{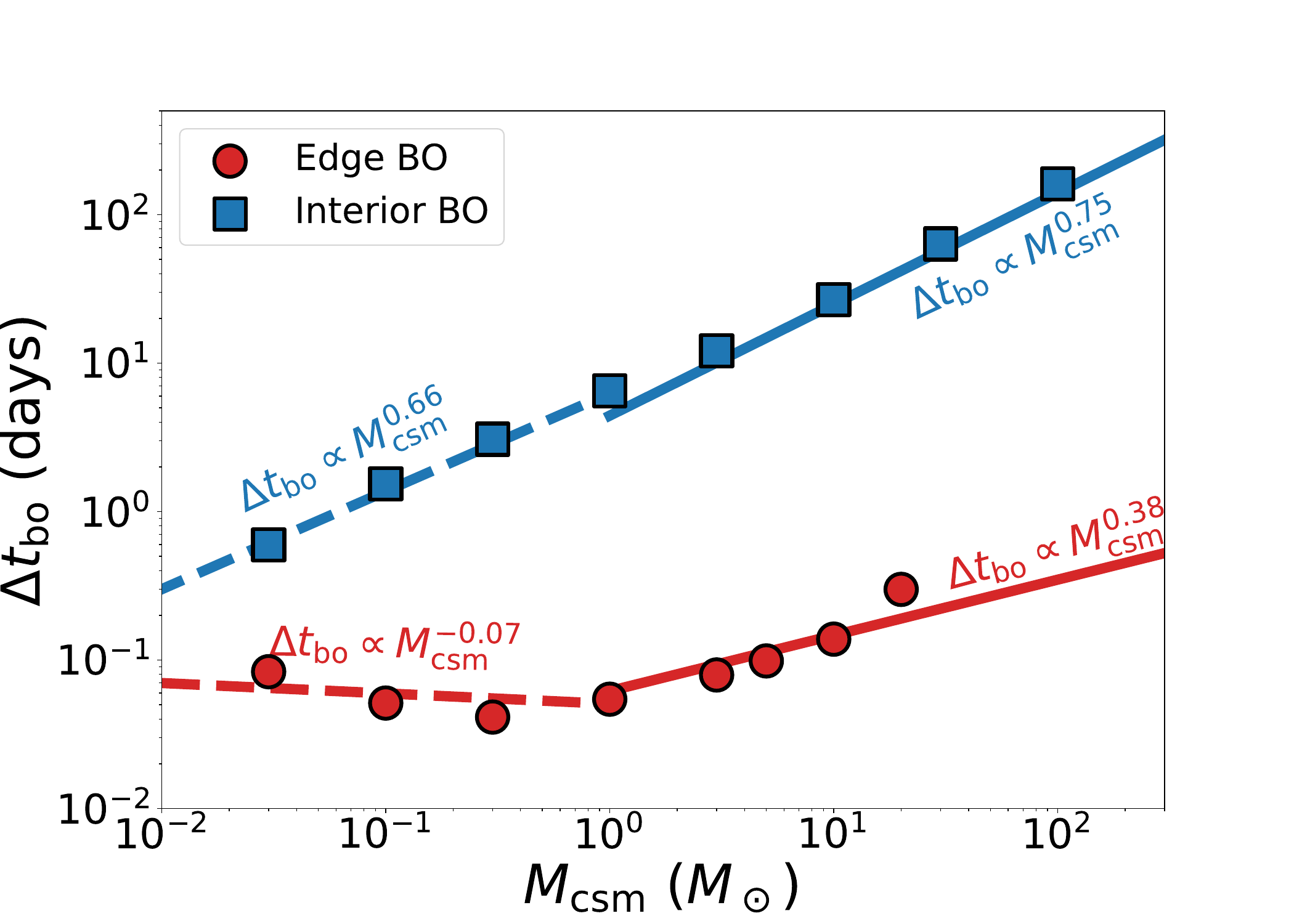}
\plottwo{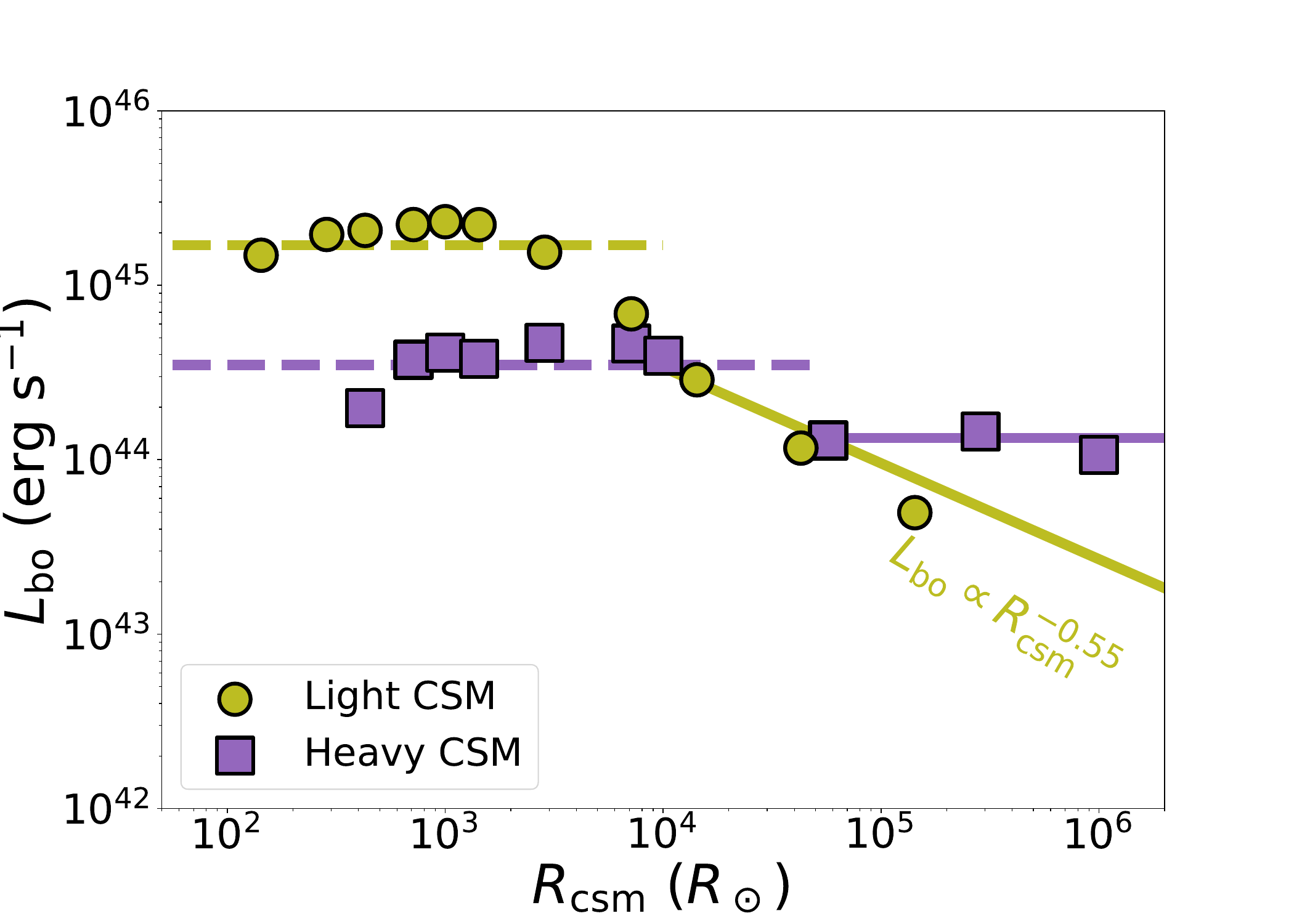}{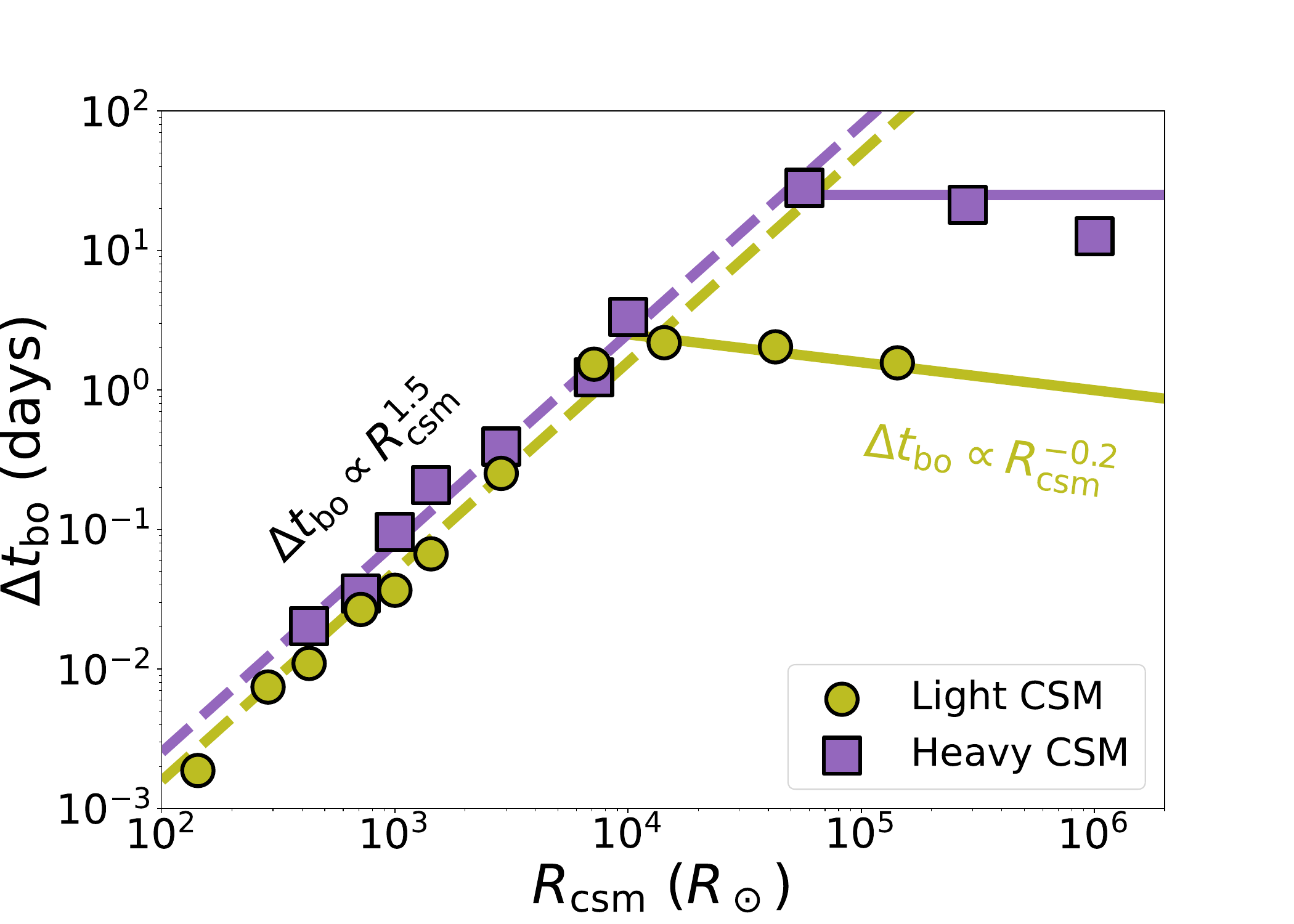}
\caption{Correlation of the CSM properties with the breakout luminosity $L_{\rm bo}$ (\textit{left column}) and duration $\Delta t_{\rm bo}$ (\textit{right column}). Points correspond to model light curves in Fig. \ref{fig:csm_lightcurves}, while lines give the analytic scalings of \S \ref{sec:analytics} with specific formulae provided in Appendix A. \textit{Top row}: dependence of $L_{\rm bo}$ and $\Delta t_{\rm bo}$ on $M_{\rm csm}$, in the edge $\xi>1$ \textit{(red)} and interior $\xi<1$ \textit{(blue)} breakout regimes. The break in behavior around $M_{\rm csm}=M_\odot$ corresponds to the transition from light ($\eta\ll 1$) to heavy ($\eta\gtrsim 1$) mass regimes. \textit{Bottom row}: dependence of $L_{\rm bo}$ and $\Delta t_{\rm bo}$ on $R_{\rm csm}$, in the light \textit{(yellow)} and heavy \textit{(purple)} CSM regimes. The break in the scaling behavior denotes the transition from edge to interior breakout. }
\label{fig:csm_scalings}
\end{figure*}

One interesting property of continued interaction emission is its direct dependence on the CSM density structure. In particular, if $s<5-3/\lambda$ then the continued interaction phase will \textit{rise} in time. For $\eta \ll 1$ and ejecta density $n\approx 10$, this requires a CSM shallower than $s< 5/4$.Note also that for a steady wind-like CSM profile $s=2$, the light curve will decrease in time irrespective of the ejecta denisty profile.

The CSM density power-law index, $s$,  does not significantly affect the time of shock breakout or shock emergence, but it does affect the overall luminosity (see Fig. \ref{fig:density_profile_lightcurves}). Breakout luminosity is more luminous for steeper CSM, while the luminosity at shock emergence $L_{\rm ci}(t_{\rm se})$ will be more luminous for shallower CSM profiles. Accounting for the density profile effects in the CSM (Appendix \ref{sec:numeric_setup}), the characteristic shock luminosity scales with $s$ as
\begin{align}\label{eq:Lsh_s}
L_{\rm sh,0} = \frac{3-s}{4\pi}\left[1-\left(\frac{R_*}{R_{\rm csm}}\right)^{3-s}\right]^{-1}L_0
\end{align}
where $R_*$ is the inner edge of the CSM and we have assumed $s<3$. 

Once the shock reaches the outer edge, the light curve will drop with only a residual amount of shock cooling emission, as nearly all of the shock energy had already been radiated away through continued interaction.


\section{Numerical Light Curves}
\label{sec:numerical}

We perform one-dimensional radiation hydrodynamics simulations of CSM interaction. We couple the finite-volume moving mesh method of \cite{2016ApJ...821...76D} to a gray flux-limited diffusion solver based on \cite{2003JCoPh.184...53H} and \cite{2011ApJS..196...20Z}. 
A full description of the numerical method is given in Appendix \ref{sec:numeric_setup}.  We adopt the same assumptions and parameters described in \S \ref{sec:analytics}. Namely, the ejecta is described by a mass $M_{\rm ej}$ and kinetic energy $E_{\rm sn}$ undergoing homologous expansion, whose density structure is given by the broken power-law form of \cite{1989ApJ...341..867C}. We assume fiducial values of $\delta=1$ and $n=10$ for the inner and outer ejecta density profiles, respectively. 

The CSM extends from an inner edge $R_*$ to an outer radius $R_{\rm csm}$ with mass $M_{\rm csm}$, with a power-law density profile $s$. We adopt a fiducial wind-like $s=2$ for most runs, unless otherwise stated, and take $R_*=10^{-2}R_{\rm csm}$ (i.e. $R_*\ll R_{\rm csm}$). Additionally, we attach a steep cutoff layer at the outer edge of the CSM described by a power-law $r^{-p}$, which we take $p=30$ as fiducial. 
As discussed in Sec. \S \ref{sec:analytics}, the exact numerical choice does not matter so long as $p\gg 1$. The CSM and ejecta are both initially cold, with $T_0=10^2$ K, and described by the same uniform gray opacity $\kappa=0.34$ cm$^{2}$ g$^{-1}$. We use a thermalization fraction of $\epsilon= 10^{-3}$. Finally, we initialize the setup at a time $t=10^3  $ seconds after explosion. For a more detailed description of the problem setup and relevant equations, see Appendix \ref{sec:numeric_setup}. 

We consider a range of ejecta and CSM properties to cover the diversity of light curves expected from the different regimes outlined in \S \ref{sec:configuration}. Specifically, we use ejecta masses and energies between $0.1M_\odot\leq M_{\rm ej}\leq 100M_\odot$ and $10^{49}\leq E_{\rm sn}\leq 10^{52}$ ergs. For the CSM, we consider mass and radii in the range $0.01M_\odot \leq M_{\rm csm}\leq 100M_\odot$ and $10^2 R_\odot\leq R_{\rm csm}\leq 10^6 R_\odot$, respectively.
In total, we ran approximately $100$ different ejecta-CSM interaction scenarios within the numerical parameter space.

For each run, we measure the breakout peak luminosity $L_{\rm bo}$ and time $t_{\rm bo}$, as well as the duration $\Delta t_{\rm bo}$ which we take to be the time to rise to peak by one order of magnitude. For the edge breakout events which feature two light curve peaks, we measure the secondary peak to determine $L_{\rm sc}$ and $\Delta t_{\rm sc}$. We also fit a power-law to the continued interaction tail of interior breakout events to compare with Eq.\ref{eq:continued_Lsh}.

We use the grid to also construct numerical scalings for each phase of each interaction type. In particular, we adopt fitting formulae for luminosity and time of the $i$-th phase as
\begin{align}\label{eq:fit_equations}
L_i &= a_i \eta^{-3\alpha_i}\xi^{k_i}L_0 \\
t_i &=  b_i \eta^{\alpha_i}\xi^{c_i}t_0
\end{align}
where $(\alpha_i,k_i,c_i)$ are fitting exponents, and $(a_i,b_i)$ are normalization factors to account for numerical differences compared to the analytic scalings. The results of the numerical fits for the different phases and classes, as well as correction factors for the analytic scalings of \S \ref{sec:analytics}, are given in Appendix \ref{sec:physical_scalings}.

\subsection{Overall Properties of Model Grid}

In Fig. \ref{fig:dlps} we show the duration-luminosity phase space of the model grid breakout properties. We additionally show the secondary shock-cooling peak for the subset of edge breakout events that have a clear double-peaked light curve. The resulting light curves will have timescales ranging from very rapid ($\sim$ minutes) to long-lasting ($\sim$ months); and peak luminosities spanning the sub-luminous $ 10^{41}$ erg s$^{-1}$ all the way to highly superluminous $\sim 10^{45}$ erg s$^{-1}$ events. The peak luminosities correlate inversely with duration, with a spread in the trend due to the diversity of CSM and ejecta parameters.  Note that more extreme events in terms of peak properties may be possible for an expanded parameter space broader than the range considered here.

The flashes from edge breakout events ($\xi\gg 1$) tend to occupy the high-luminosity and short-duration portion of phase space.
For CSM radii of $R_{\rm csm}\sim 10^2-10^3R_\odot$, the edge breakout flash resembles expected stellar surface shock breakout luminosities and durations, lasting on the order of a few minutes to hours. Larger CSM radii tend to produce  longer-lasting edge breakouts, as seen in Fig. \ref{fig:dlps}. Typical edge breakout luminosities range from a few times $10^{43}$ erg s$^{-1}$ on the lower end, reaching up to highly superluminous events $\gtrsim 10^{45}$ erg s$^{-1}$ for the most energetic interactions. 

The flashes from interior breakouts  ($\xi \lesssim 1$) bifurcate  into different regions of phase space depending on the value of $\eta = M_{\rm CSM}/M_{\rm ej}$. Heavy CSMs ($\eta \gtrsim 1$) occupy the brighter and longer-duration of the interior breakouts, spanning days to months in duration and peaks of $\sim 10^{44}$ erg s$^{-1}$. Interior breakouts from light CSMs ($\eta \ll 1$) are comparatively shorter (days to weeks) and dimmer, with a wider range in peak luminosities from $\sim 10^{42}$ to $10^{44}$ erg s$^{-1}$.

Compared to shock breakout, the post-breakout shock cooling emission generally produces lower peak luminosities and longer durations, comparable to those observed in typical radioactive nickel-powered transients. If the  breakout flash from these events is missed due to its rapid timescale, it may in practice be hard to distinguish between interaction and radioactive decay light curves using photometry alone.

\subsection{Dependence on Circumstellar Mass}

To numerically examine the effect of $M_{\rm csm}$ on the light curve, we hold constant the ejecta properties ($M_{\rm ej}=M_\odot$, $E_{\rm sn}=10^{51}$~erg) and vary the CSM mass in the range $10^{-2}M_\odot <M_{\rm csm}<10^2 M_\odot$. This covers the CSM regime from light $(\eta \ll 1)$ to heavy ($\eta \gtrsim 1$). Additionally, we adopt two fiducial values of $R_{\rm csm}=10^3R_\odot$ and $10^5 R_\odot$, which result in an edge and interior breakout, respectively.

The top row of Fig.~\ref{fig:csm_lightcurves} shows the resulting light curves as a function of $M_{\rm csm}$  while Figure~\ref{fig:csm_scalings} plots the dependence of $L_{\rm peak}$ and $\Delta t$ on $M_{\rm csm}$. The dependencies  are  non-monotonic and --  as expected from the analytic relations of \S\ref{sec:configuration} -- scale differently in the regimes of edge breakout ($\xi > 1$) and interior breakout $(\xi< 1$) and for $\eta \ll 1$ and $\eta \gtrsim 1$.

In Fig. \ref{fig:csm_scalings}, the breakout scalings of $L_{\rm bo}$ and $\Delta t_{\rm bo}$ with $R_{\rm csm}$ and $M_{\rm csm}$ are shown compared to the analytic scalings derived in Sec. \ref{sec:analytics}. Overall, the analytics agree well with the numerical results, correctly predicting different scalings depending on a light/heavy and edge/interior breakout scenario. In particular, the turnover in the dependence of $L_{\rm bo}$ on $M_{\rm csm}$ around $\eta\sim 1$  is reproduced for both edge and interior breakouts. The numerical results show a turnover in the edge breakout case at a lower CSM mass than the analytics predict, as the shock reaches the shallow inner portion of the ejecta and the self-similar scalings break down. Specifically, the ratio of outer to inner ejecta mass is equal to $(3-\delta)/(n-3)\approx 0.3$ for $\delta = 1$ and $n=10$ \citep{1989ApJ...341..867C}, and so for $\eta\gtrsim 0.3$ the shock behavior changes. Note that the exact behavior of the transition in the $\eta\approx 1$ range is not well-sampled in our numerical simulations, which limits the applicability of our analysis for interaction events in this parameter space.

Consider first the case of light CSM ($\eta \ll 1$). 
If we are in the regime of edge breakout ($\xi \gtrsim 1$) the luminosity of the light curve peak depends only weakly on $M_{\rm csm}$, since breakout happens at effectively the same radius $r_{\rm bo}\sim R_{\rm csm}$ and velocity $v_{\rm bo}\sim v_{\rm ej}$ (since there is not much deceleration for $\eta \ll 1$). If breakout occurs in the CSM interior ($\xi \lesssim 1$), the light curve is slightly brighter   for higher values of $M_{\rm csm}$, since the breakout location $\tau\sim c/v_{\rm bo}$  is reached later during the shock evolution (i.e. at a larger breakout radius $r_{\rm bo}$).

The breakout duration also scales differently depending on whether we are in the interior or edge breakout regime. For interior breakouts $(\xi \lesssim 1$), the duration is set primarily by shock crossing and radiative diffusion, giving a longer duration for larger $M_{\rm csm}$. However, for edge breakouts ($\xi \gtrsim 1$) from light CSM, the duration actually \textit{decreases} with increasing $M_{\rm csm}$. This can be understood by examining Eq. \ref{eq:breakout_dtbo}, where the edge breakout duration depends on the shock crossing of the breakout layer. For a light CSM, the width of the breakout layer decreases with increasing $M_{\rm csm}$ while the ejecta is not much decelerated $v_{\rm bo}\sim v_{\rm ej}$, and so the shock crossing time $\delta r_{\rm bo}/v_{\rm bo}$ (i.e. breakout duration) decreases. 

\begin{figure}
\includegraphics[width=0.5\textwidth]{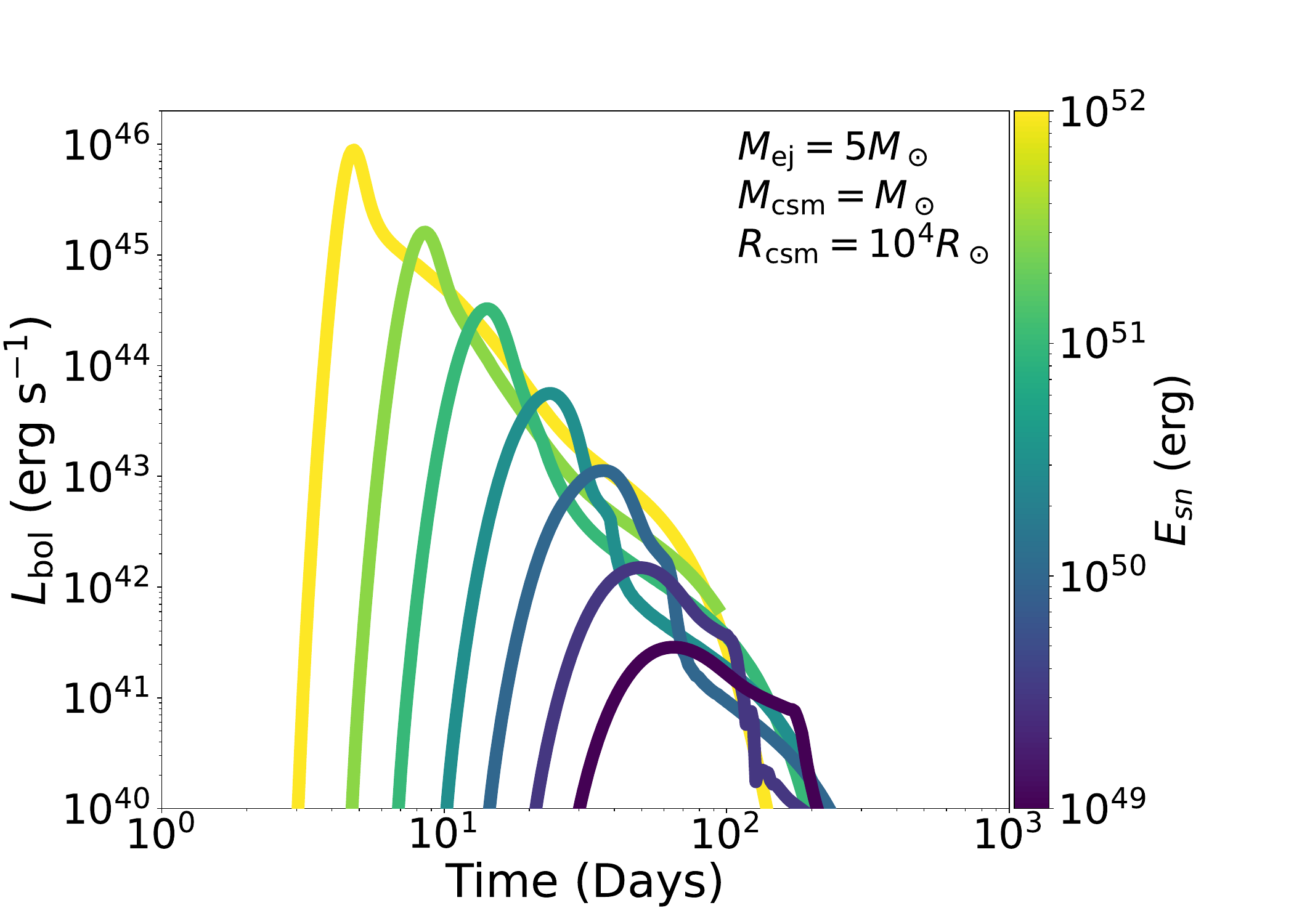}
\caption{Numerical light curves for fixed CSM properties and ejecta mass $M_{\rm ej}=5M_\odot$, varying the kinetic energy $E_{\rm sn}$.}
\label{fig:ejecta_energy_lightcurves.pdf}
\end{figure}

As we continue increasing $M_{\rm csm}$, we enter the heavy CSM regime, $\eta>1$. For this regime, the entirety of the ejecta is decelerated, and the maximum shock energy of $\sim E_{\rm sn}$ is reached at $\eta=1$. Any additional CSM mass beyond $M_{\rm ej}$ only acts to decrease the shock velocity $\sim v_{\rm ej}\eta^{-\alpha}$, resulting in a dimmer light curve. As a result, the heavy CSM $\eta \gtrsim 1$ breakout luminosity \textit{decreases} with $M_{\rm csm}$, with similar scalings for both edge and interior breakouts, as shown in Fig. \ref{fig:csm_scalings}. Furthermore, the breakout duration increases with $M_{\rm csm}$, with a steeper dependence for interior breakouts.

In addition to the breakout properties, $M_{\rm csm}$ will also impact the the post-breakout emission, i.e. shock cooling and continued interaction for edge and interior breakouts, respectively. For the edge breakout case in Fig. \ref{fig:csm_lightcurves}, the shock cooling emission becomes dimmer and longer-lasting with $M_{\rm csm}$. A larger $M_{\rm csm}$ results in a longer diffusion timescale, which keeps the radiation trapped for longer and exacerbates adiabatic losses in the cooling phase. This effect is most pronounced for the heavy CSM, which have a longer-lasting ``plateau'' of shock cooling emission, seen in the upper left panel of Fig. \ref{fig:csm_lightcurves}. For light CSM, the shock cooling appears more as a tail immediately following the breakout emission, while for heavy CSMs the shock cooling is more distinctly separated from breakout, appearing as a secondary feature in the light curve well after the breakout has subsided. 

The post-breakout emission in interior breakout events is produced through continued interaction, which powers a tail in the light curve before a sharp drop in luminosity at shock emergence.  The continued interaction luminosity is more luminous with increasing $M_{\rm csm}$ in the light CSM regime $\eta \ll 1$, since $v_{\rm sh}$ is only minimally decelerated while the CSM density increases with $M_{\rm csm}$. For $\eta \gtrsim 1$, due to the significant shock deceleration, the continued interaction luminosity instead decreases with $M_{\rm csm}$. This also leads to a much later shock emergence time once we reach heavy CSM masses.

The continued interaction tail reaches a maximum luminosity for masses $\eta = 1$, whereby any additional $M_{\rm csm}>M_{\rm ej}$ instead results in a less luminous light curve. Furthermore, the light curve slope becomes steeper as we enter the heavy CSM $\eta \gtrsim 1$ regime, as the shock begins to behave more as a snowplow blastwave whose exponents are given by Eq. \ref{eq:snowplow_exp} instead of the $\eta \ll 1$ exponents in Eq. \ref{eq:shock_lambda}.

\begin{figure}
\includegraphics[width=0.5\textwidth]{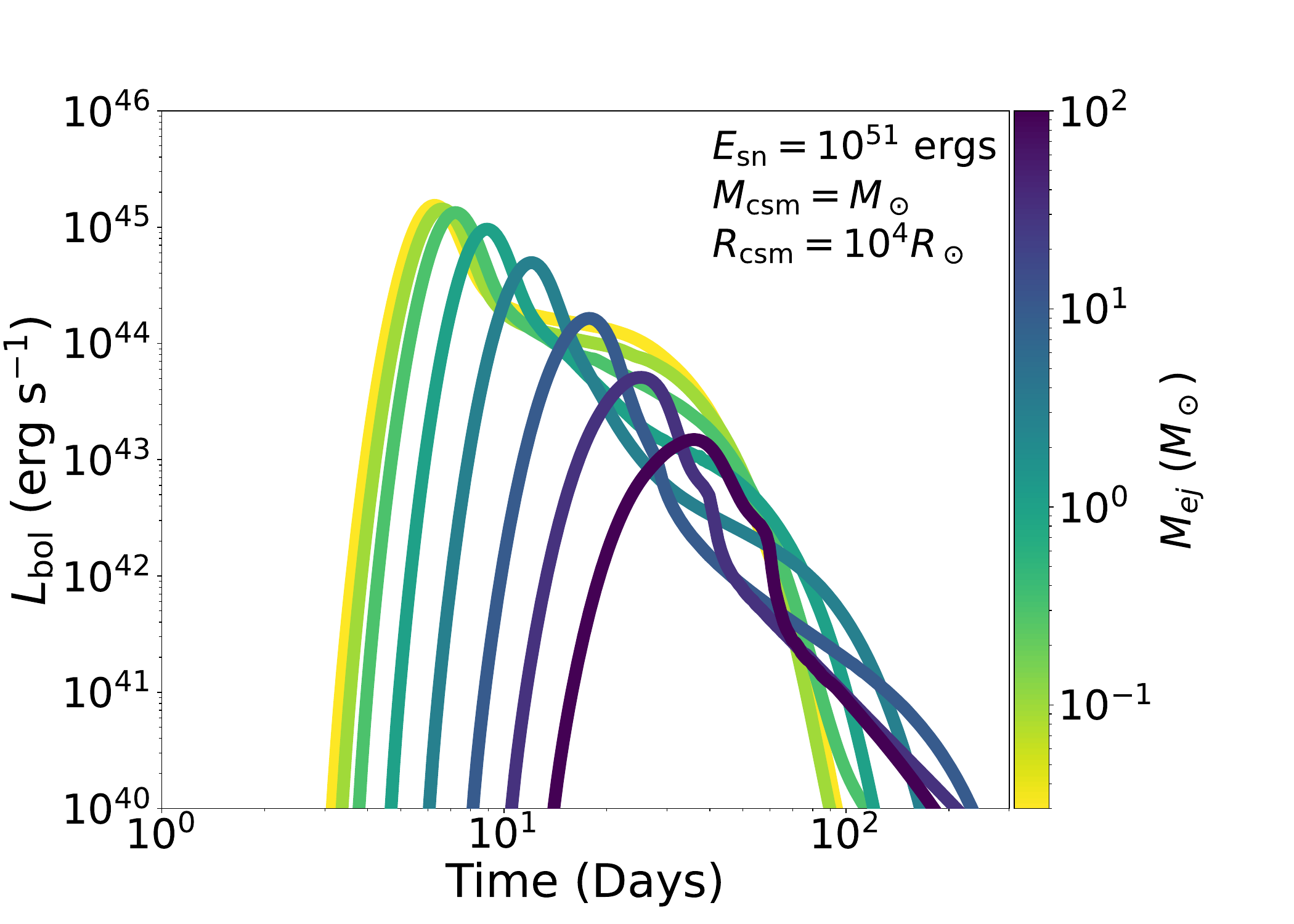}
\caption{Same as Fig. \ref{fig:ejecta_energy_lightcurves.pdf}, but varying the ejecta mass with a fixed $E_{\rm sn}=10^{51}$ ergs.}
\label{fig:ejecta_mass_lightcurves.pdf}
\end{figure}

\subsection{Dependence on Circumstellar Radius}

Here we use the same ejecta properties as in the previous section ($M_{\rm ej}=M_\odot$ and $E_{\rm sn}=10^{51}$ ergs), but instead vary $R_{\rm csm}$ while keeping $M_{\rm csm}$ fixed. We consider CSM radii in the range $10^2 R_\odot<R_{\rm csm}<10^6 R_\odot$ for two fiducial masses corresponding to a light $M_{\r csm}=0.1M_\odot$ and heavy $M_{\rm csm}=10M_\odot$ CSM.

As the models in the bottom row of Fig. \ref{fig:csm_lightcurves} have fixed $M_{\rm csm}$ and ejecta properties, the criteria $\xi > 1$ (i.e. breakout occuring at the CSM edge) is reached for $R_{\rm csm}\lesssim 10^4 R_\odot$. The light curves display a qualitative change in behavior in the two regimes of edge and interior breakout, transitioning from a double-peaked breakout with shock cooling for $R_{\rm csm}\lesssim 10^4 R_\odot$ $(\xi \gtrsim 1)$  to a single peak with a continued interaction tail for $R_{\rm csm}\gtrsim 10^4 R_\odot$ ($\xi \lesssim 1$).

In Fig. \ref{fig:csm_scalings} the analytic scalings of $L_{\rm bo}$ and $\Delta t_{\rm bo}$ on $R_{\rm csm}$ are compared with the numerical results. In the limit of an edge or interior breakout, the analytic scalings match reasonably well with the numerical simulations. In the intermediate regime around $10^4 R_\odot$, the scalings are less robust, given the assumption of a fixed $k_0$ as introduced in \S \ref{sec:analytics} for interior breakouts.
Note also that the analytics presented predict an independence of $L_{\rm bo}$ on $R_{\rm csm}$ for the edge breakout scenario. The numerical simulations qualitatively agree with this prediction, albeit with a slight positive correlation with $R_{\rm csm}$.

In general, larger CSM radii produce later and longer-lasting light curves, as the shock takes more time to reach the outer edge of the CSM. Furthermore, with increasing $R_{\rm csm}$ and fixed $M_{\rm csm}$, we are spreading the mass out over a larger volume, which decreases $\tau_{\rm sh}$ during interaction. Eventually, $R_{\rm csm}$ becomes large enough that we enter the interior breakout regime $\xi\lesssim 1$, whose light curve is marked by continued interaction rather than a shock cooling tail. This behavior holds for both a light and heavy CSM, with the primary difference of the two CSM mass regimes being the relative prominence of the post-breakout emission. 

For small CSM radii such that we are in the edge breakout regime, increasing $R_{\rm csm}$ results in a longer-lasting dark phase, brighter breakout peaks, and a slower breakout rise. This behavior holds for both light $\eta \ll 1$ and heavy $\eta \gtrsim 1$ CSM masses which also have similar scalings, shown in Figs. \ref{fig:csm_scalings}.
The post-breakout shock cooling luminosity increases strongly as $L_{\rm sc}\propto R_{\rm csm}$, although the cooling timescale appears nearly independent of the radius.

As we continue to increase the CSM radius we eventually enter the interior breakout regime $\xi \lesssim 1$. In this case, the breakout duration and luminosity turn over and begin decreasing slightly with increasing $R_{\rm csm}$. This break is more pronounced for light CSM masses, shown in the bottom-left panel of Fig. \ref{fig:csm_lightcurves}. Furthermore, the post-breakout emission changes from shock cooling to continued interaction at these larger radii. While the shock cooling luminosity increases with $R_{\rm csm}$, the continued interaction tail becomes less luminous for larger $R_{\rm csm}$; the decrease is more pronounced for light CSM masses. Finally, while the shock cooling duration is independent of $R_{\rm csm}$, the continued interaction tail scales directly with the shock emergence timescale $t_{\rm se}\propto R_{\rm csm}$.

\begin{figure}
    \centering
    \includegraphics[width=0.5\textwidth]{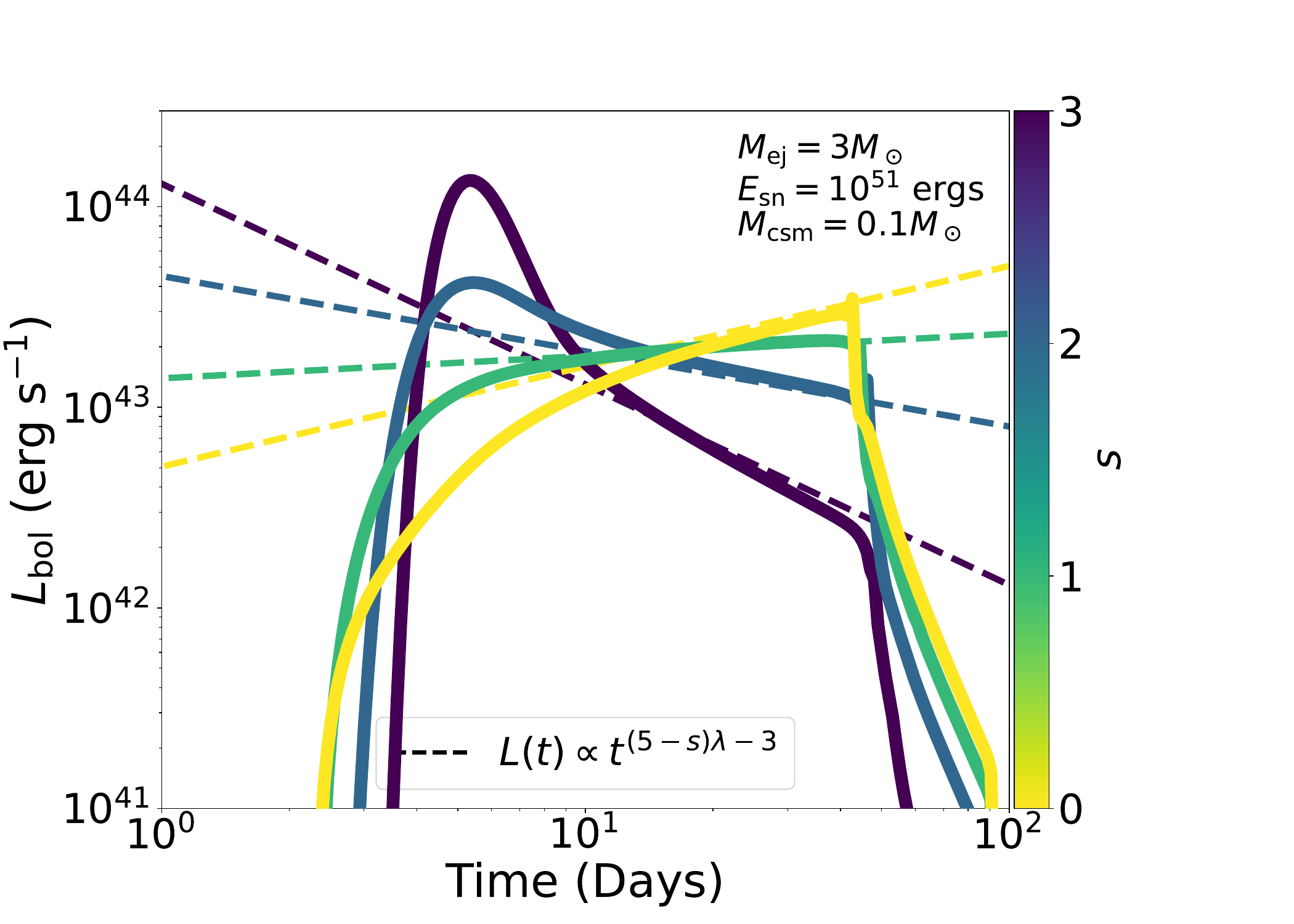}
    \caption{Numerical light curves for a light interior breakout ($\eta \ll 1$, $\xi < 1$), with different assumed CSM density profile $\rho\propto r^{-s}$. Dashed lines correspond to the power-law expression for the continued interaction phase Eq. \ref{eq:continued_Lsh}.}
    \label{fig:density_profile_lightcurves}
\end{figure}

\subsection{Dependence on Ejecta Mass and Energy}

Next, we consider the case of a fixed CSM, $M_{\rm csm}=M_\odot$ and $R_{\rm csm}=10^{14}$ cm (i.e. fixed $\tau_0\sim 10^3$), while holding $M_{\rm ej}=5M_\odot$ constant $(\eta=0.2)$. We vary the ejecta kinetic energy across the range $10^{49}\leq E_{\rm sn}\leq 10^{52}$ ergs, which is equivalent to a characteristic velocities between $10^{-3}\leq \beta_0\leq 0.05$.

 We show the resulting light curves for the different $E_{\rm sn}$ in Fig. \ref{fig:ejecta_energy_lightcurves.pdf}. We find higher-energy explosions produce earlier, faster, and brighter light curves, with $L_{\rm bo}\propto E_{\rm sn}^{5/2}$; this is due to higher kinetic energies producing faster and stronger shocks. The scale of $E_{\rm sn}$ does not just affect the characteristic timescale and luminosity of the light curve; it can also affect the type of interaction. As we go to lower energies in Fig. \ref{fig:ejecta_energy_lightcurves.pdf}, eventually we enter the $\xi\lesssim 1$ regime and the shock breaks out within the CSM rather than at the edge. In this case, the post-breakout emission will change from a shock cooling to a continued interaction phase.

We also examine the effect of a fixed ejecta kinetic energy $E_{\rm sn}=10^{51}$ ergs with the same CSM properties as above, but vary the ejecta mass across $0.03M_\odot \leq M_{\rm ej}\leq 100M_\odot$. We show the resulting light curves of the $M_{\rm ej}$ range in Fig. \ref{fig:ejecta_mass_lightcurves.pdf}. As we increase $M_{\rm ej}>M_\odot$, which corresponds to the $\eta>1$ case, the light curves become longer and dimmer since $v_{\rm sh}\approx v_{\rm ej}\propto M_{\rm ej}^{-1/2}$ for fixed $E_{\rm sn}$. Furthermore, as the amount of energy tapped in the $\eta<1$ case is $\sim M_{\rm csm}v_{\rm sh}^2$, we are also \textit{fractionally} converting less kinetic energy as we increase $M_{\rm ej}$. For large enough $M_{\rm ej}$, the shock velocity drops low enough that the post-breakout emission transitions from shock cooling to continued interaction, similar to the case of the lower-energy explosions.

On the other hand, as we decrease $M_{\rm ej}$ below $M_\odot$, we enter the $\eta \gtrsim 1$ regime where the light curve becomes nearly independent of $M_{\rm ej}$. This corresponds to the limit of a point explosion inside the CSM, and the only ejecta parameter that sets the light curve behavior will be $E_{\rm sn}$. Furthermore, the exact density structure of the ejecta is irrelevant, unlike the $\eta \ll 1$ case where the continued interaction tail is set directly by the outer density profile. In practice, this can make constraining the ejecta mass challenging in this limit due to the $M_{\rm ej}$ degeneracy.

\begin{figure}
    \centering
    \includegraphics[width=0.45\textwidth]{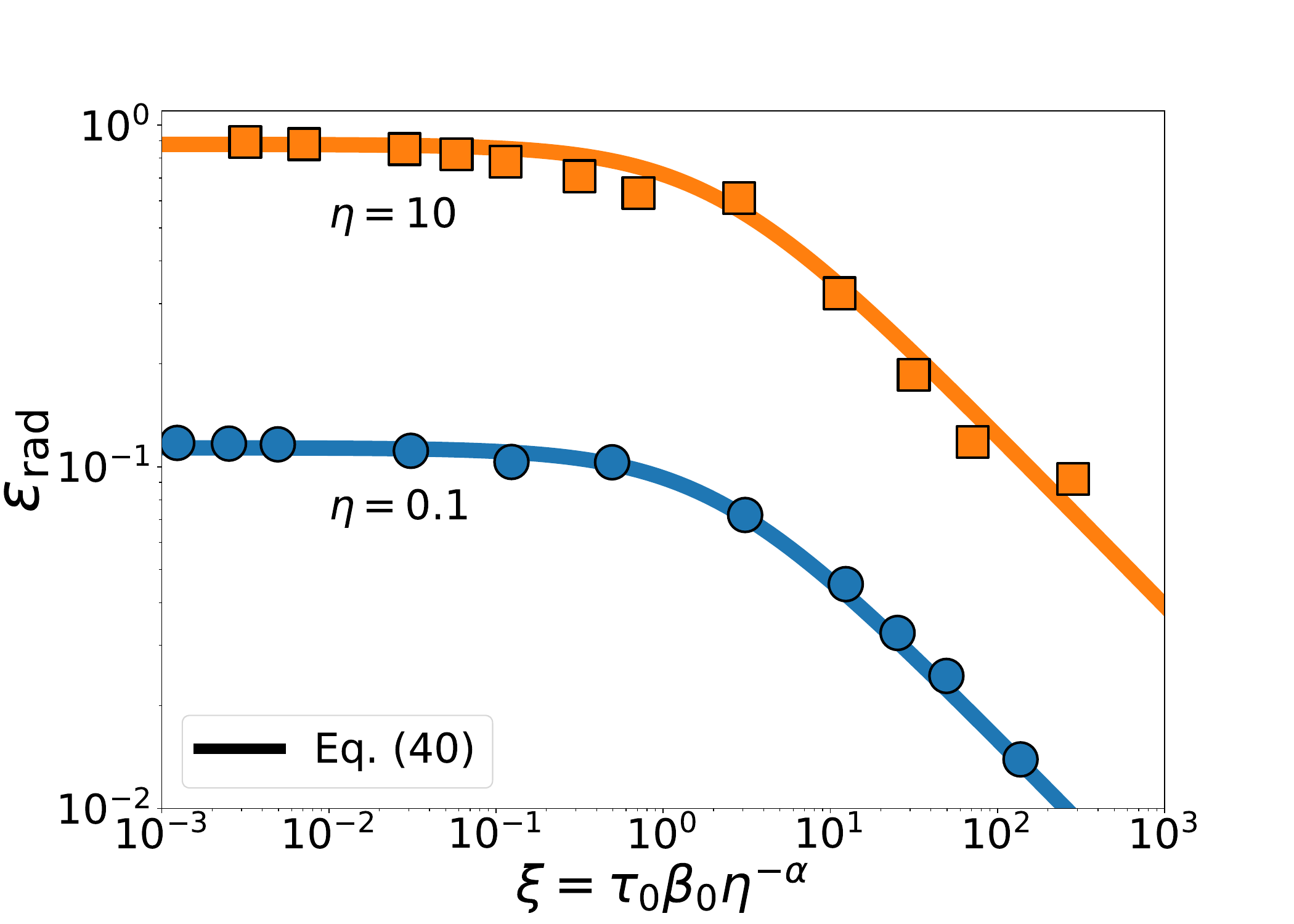}
    \caption{Radiated efficiency $\varepsilon_{\rm rad}$ as a function of the breakout parameter $\xi$. Points correspond to numerical simulations for the case of $\eta=0.1$ (blue circles) and $\eta=10$ (orange squares).
    Solid lines give the analytic expression in Eq. \ref{eq:epsrad} for each choice of $\eta$.}
    \label{fig:epsrad_betatau}
\end{figure}

\subsection{Radiated Efficiency}
\label{subsec:epsrad}

The efficiency with which interaction converts kinetic energy into observable radiation is an important consideration when physically interpreting transients. In particular,  for interaction to explain long-lasting luminous supernovae with integrated radiated energy reaching $10^{51}$~erg likely requires efficiencies not far below unity. Here we quantify the achievable efficiency throughout the parameter space.  

We can determine the radiated efficiency of our numerical models by integrating the light curve and comparing it to the initial ejecta kinetic energy,
\begin{align}\label{eq:epsrad_lc}
\varepsilon_{\rm rad} = \frac{1}{E_{\rm sn}}\int L(t)\,dt
\end{align}
For $\eta\ll 1$, from energy conservation, the interaction will convert a fraction $E_{\rm sh}\sim E_{\rm sn}\left(v_{\rm sh}/v_{\rm ej}\right)^{5-n}$ of the amount of kinetic energy contained in the steep outermost layer of the ejecta into internal gas/radiation energy, where $v_{\rm sh}\sim v_{\rm ej}\eta^{-\alpha}$ and $\alpha=1/(n-3)$ (see Appendix \ref{sec:shock_derivation} for a full derivation). Thus, the  radiated efficiency for $\eta\ll 1$, assuming no adiabatic losses and efficient conversion of shock energy into radiation, is roughly $\varepsilon_{\rm rad}\sim \eta^{a_0}$, where $a_0 = (n-5)/(n-3)$. For $\eta\gtrsim 1$, this reaches a potential maximum of unity, as the shock will tap into the entirety of the ejecta kinetic energy.

While the radiation remains trapped in the shocked region, it will suffer adiabatic losses that acts to degrade the efficiency by converting the radiation back into kinetic energy. For $\xi < 1$, adiabatic losses are minimal as the shock radiation is able to efficiently escape. For $\xi > 1$, the radiation will be adiabatically degraded by a factor proportional to $(t_d/t_0)\sim\left(\kappa M_{\rm csm} v/R_{\rm csm}^2 c\right)^{1/2}\sim \xi^{-1/2}$, where $t_d=\sqrt{\kappa M_{\rm csm}/v_{\rm ej}c}$ and $t_0 = R_{\rm csm}/v_{\rm ej}$.

We can interpolate between the regimes of light/heavy CSM and the effect of adiabatic losses with the analytic expression
\begin{align}\label{eq:epsrad}
\varepsilon_{\rm rad} \approx \frac{1}{\left(1+2/\eta\right)^{a_0}}\left(1+\frac{\xi}{2}\right)^{-1/2}
\end{align}
where $a_0= (n-5)/(n-3)=5/7$ for $n=10$.
In Fig. \ref{fig:epsrad_betatau} we show the efficiency compared to Eq. \ref{eq:epsrad} for a series of light ($\eta=0.1$) and heavy ($\eta=10$) CSM interactions, where we vary $R_{\rm csm}$ to produce a range of breakout parameters $\xi$. The numerical simulations agree well with Eq. (\ref{eq:epsrad}) across the different interaciton regimes. We see that the efficiency reaches a maximum in the regime of $\xi<1$, i.e the interior breakout regime. In this case, the photons from the breakout and continued interaction tail are able to escape before incurring much adiabatic losses, and hence are the more efficient class of interaction. The \textit{most} efficient case corresponds to $\eta=1$ and $\xi< 1$, where we tap almost all of the kinetic energy and quickly radiate away the shock photons. 

\begin{figure}
    \centering
    \includegraphics[width=0.5\textwidth]{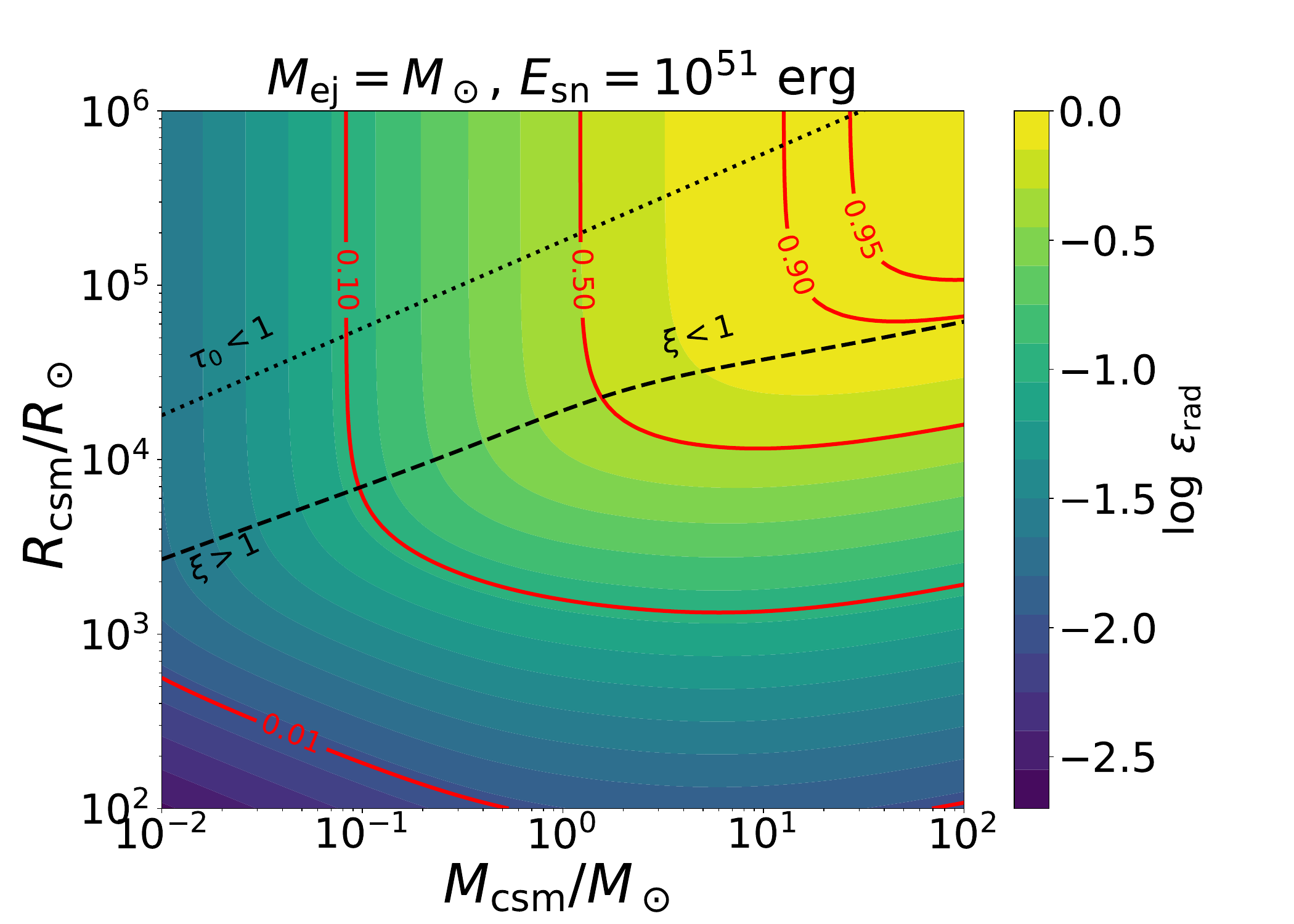}
    \caption{Radiated efficiencies from the interaction of a solar mass ejecta with $E_{\rm sn}=10^{51}$ ergs of kinetic energy, for a range of CSM masses and radii. Lighter regions denote higher efficiencies. }
    \label{fig:epsrad_csm}
\end{figure}

In contrast, once we enter the regime of $\xi > 1$, the photons can no longer quickly escape, coming out during the shock cooling phase after being adiabatically degraded. This corresponds to the edge breakout case, and $\varepsilon_{\rm rad}\propto \xi^{-1/2}$. Thus, although edge breakouts produce some of the more luminous transients expected from interaction, they are also reduced in their net radiative throughput due to the large optical depths of the CSM. In Fig. \ref{fig:epsrad_csm} we show the radiated efficiency for the case of interaction of a solar mass ejecta with kinetic energy $E_{\rm sn}=10^{51}$ ergs, in terms of the $M_{\rm csm}$-$R_{\rm csm}$ space. We see that massive, extended CSMs are the most efficient interactions, while a compact low-mass CSM only converts a small fraction of $E_{\rm sn}$.  

There is one other effect that will reduce the radiated efficiency of the interaction, which occurs when the shock is unable to cool efficiently, as described in \S \ref{sec:configuration}. Specifically, if the CSM is so optically thin $\tau_0\lesssim 0.3v_9$, then we enter the adiabatic shock regime, and few photons are produced. Thus, although low CSM optical depths improve efficiency by reducing adiabatic losses, it cannot be too low such that the shock is unable to cool. At the $\tau_0\lesssim 0.3v_9$ limit, we therefore expect the efficiency to turn over again.

\section{Discussion}
\label{sec:discussion}

When invoking circumstellar interaction to interpret an observed transient, care must be taken to ensure that the analysis is self-consistent. That is, we must first decompose the light curve and identify the separate phases outlined in \S \ref{sec:configuration}, as each  phase has a different dependence on the underlying physical parameters (\S \ref{sec:analytics}). The relative prominence of each phase and the corresponding scaling are determined by the dimensionless parameters $\eta$ and $\xi$. For example, as shown in Fig.~\ref{fig:csm_scalings}, the light curve duration and luminosity depend on $M_{\rm csm}$ and $R_{\rm csm}$ in a non-monotonic way as we transition from an edge $(\xi > 1$) to an interior breakout ($\xi < 1$), or from a light $(\eta \ll 1$) to heavy CSM ($\eta \gtrsim 1$). Applying an edge breakout scaling relation to e.g. an interior breakout would result in an incorrect estimate of the physical parameters of the system.

It may be  challenging to observe a light curve at a  high enough cadence to see all of the interaction phases.
The flash originating from an edge breakout is a particularly hard phase to capture, given its fast rise and immediate decline. If we are unable to catch the transient early enough, only the post-breakout shock cooling emission may be observed.  In contrast, an interior breakout from a heavy CSM will be much easier to observe given its gradual rise and fairly luminous peak, but the light curve may need to be followed up for a fairly long time to capture the continued interaction tail and shock emergence drop, which can take more than a year in certain cases. When all phases of the interaction  are not observed, there are typically degenerate solutions that fit the same  light curve photometry with very different CSM masses and radii (see Figure~\ref{fig:dlps}). Invoking significant heating by radioactive nickel or a central engine further increases this degeneracy.  In such cases, spectral information can be valuable for refining the interpretation. 

Another issue that can arise when interpreting observed light curve concerns the dark phase which is by definition unobservable. This phase obfuscates the exact time of progenitor explosion. In certain cases, the interaction may completely overshadow the stellar breakout burst and radioactive heating. In other cases, the transient light curve will be explained by a combination of early interaction emission followed by heating from additional sources e.g. radioactive decay \citep{2018Sci...362..201D,2019ApJ...887..169H}.

In what follows, we give four case studies of observed transients which apply the theoretical framework introduced in this work. We suggest a connection between each observed transient class and one of the four theoretical interaction classes delineated in \S \ref{sec:configuration}. While the true mapping may be more multi-faceted than this, we intend only to illustrate how the general framework can be useful in organizing data samples, as well as to point out the degeneracies that may arise when trying to fit observed events with interaction models. Due to these degeneracies, the model parameters may contain significant uncertainty when fitting to specific events, such as is shown in Fig. (\ref{fig:at2018cow_comp}).

\begin{figure*}
    \centering
    \includegraphics[width=0.95\textwidth]{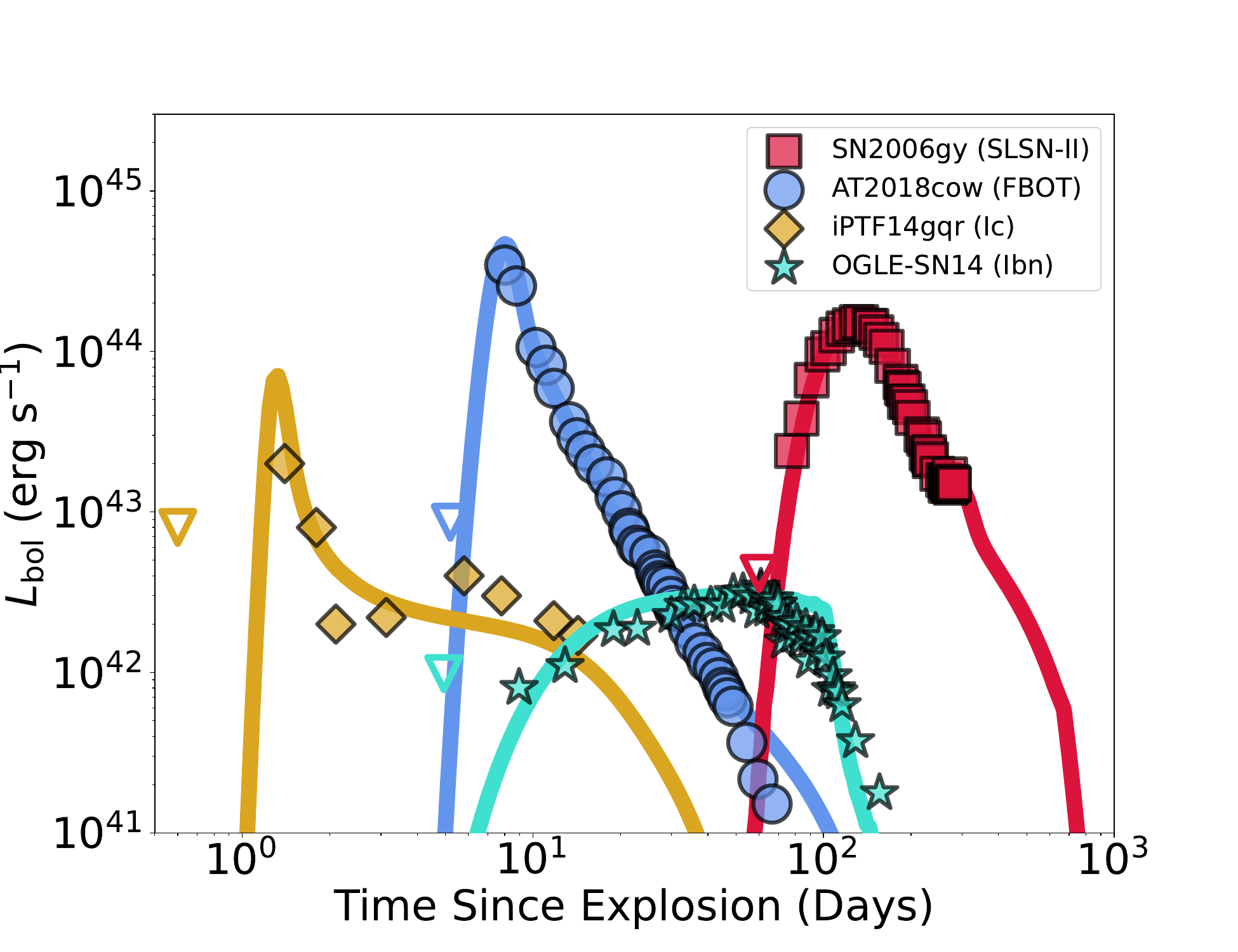}
    \caption{Fits of numerical simulations of each interaction type to several observed transients spanning a range of durations and luminosities. Open triangles denote upper limits on the light curve during the dark phase. Data taken from \cite{2017A&A...602A..93K} for OGLE-2014-SN-131; \cite{2007ApJ...666.1116S} for SN2006gy; \cite{2019MNRAS.484.1031P} for AT2018cow; and \cite{2018Sci...362..201D} for iPTF14gqr.} 
    \label{fig:data_fits}
\end{figure*}

\begin{deluxetable*}{lcccccccccc}
\tablehead{
\colhead{Transient (Type)} & \colhead{$M_{\rm ej}\,\,(M_\odot)$
} & \colhead{$E_{\rm sn}\,\,(10^{51}\,\,{\rm erg})$} & \colhead{$M_{\rm csm}\,\,(M_\odot)$} & \colhead{$R_{\rm csm}\,\,(10^3R_\odot)$} & \colhead{$s$} & \colhead{$\eta$} & \colhead{$\xi$} & $\varepsilon_{\rm rad}$ & \colhead{Interaction Type}  }
\startdata
SN2006gy (SLSN-IIn) & 12 & 2.5 & 40 & 80  & 2.5 & 3.3 & 0.6 & 0.56 &Heavy Interior \\
AT2018cow (FBOT)  & 5 & 1.0 & 0.9 & 4  & 2 & 0.2 & 10 & 0.15 & Light Edge \\
iPTF14gqr (Ic) & 0.1 & 0.02 & 0.3 & 0.4  & 2 & 3 & 50 & 0.22 & Heavy Edge \\
OGLE-2014-SN-131 (Ibn) & 4 & 0.1 & 0.8 & 20  & 1 & 0.2 & 0.1 & 0.23 & Light Interior
\enddata
\caption{Numerical fit parameters to the transient light curves in Fig. \ref{fig:data_fits}, as well as their interaction classification based on $\eta$ and $\xi$. Also given are the light curve radiated efficiencies $\varepsilon_{\rm rad}$. }
\label{tab:data_fit_params}
\end{deluxetable*}



\subsection{Light Interior Interaction as Type Ibcn/IIn Supernovae}

CSM interaction has historically been used to explain narrow emission features in supernova spectra \citep{1997ARA&A..35..309F,2017hsn..book..195G,2022Natur.601..201G},  where the narrow lines reflect the slow-moving velocity of the unshocked CSM. Such transients (i.e. type Ibn, Icn, IIn, and related events) may be associated with the continued interaction phase that occurs in interior breakout events. Given the typical luminosity range and inferred velocities, they are likely the result of an $\eta \ll 1$ "light" CSM interaction which converts only a fraction of the ejecta kinetic energy into radiation (i.e. $\varepsilon_{\rm rad}\ll 1$). 

Once the breakout radiation has subsided, the light curve will enter the continued interaction phase and track the instantaneous shock luminosity $L(t)=L_{\rm sh}$, where $L_{\rm sh}$ is given by Eq. (\ref{eq:shock_lum}).  If we assume power-law ejecta and CSM density profiles $\rho_{\rm ej}\propto r^{-n}$ and $\rho_{\rm csm}\propto r^{-s}$, then the numerically-calibrated continued interaction phase can be analytically expressed from Eq. (\ref{eq:continued_Lsh}) as

\begin{align}
L(t) \approx 0.2L_0\eta^{-3\alpha}\left[\frac{t}{t_{\rm se}}\right]^{(5-s)\lambda-3}
\end{align}
where $L_0=M_{\rm csm}v_{\rm ej}^3/R_{\rm csm}$; $t_{\rm se}=\eta^{\alpha}R_{\rm csm}/v_{\rm ej}$ is the shock emergence time Eq. \ref{eq:t_se}; and the exponents for the $\eta<1$ regime are given by Eqs. \ref{eq:alpha_eta} and \ref{eq:shock_lambda}  as
\begin{align*}
\lambda = \frac{(n-3)}{(n-s)},\,\,\alpha = \frac{1}{(n-3)}
\end{align*}

For the specific case of a constant wind mass loss $\dot{M}$, the CSM density profile is $\rho_{\rm csm}(r) = \dot{M}/4\pi r^2 v_w$ where $v_w$ is the wind velocity. 
The continued interaction phase of a wind will therefore evolve as
\begin{align}
L(t)\approx \frac{\dot{M}}{v_w} v_{\rm sh}^{3(n-3)/(n-2)}R_{\rm w}^{3/(n-2)}t^{-3/(n-2)}
\end{align}
where $R_{\rm w}\approx v_w t_w$ is the outer radius of a wind moving at a constant velocity $v_w$ for a duration $t_w$.
For our fiducial case of $n=10$ and scaled to physical units, this becomes
\begin{align}
L(t)\approx 8.8\cdot 10^{43}\,\,\dot{M}_{{\rm yr}}v_{w,5}^{-5/8}t_{w,{\rm yr}}^{3/8}v_{\rm sh,8}^{21/8}\,\,t_{\rm 1d}^{-3/8}\,\,\,{\rm erg}\,\,\,{\rm s}^{-1}
\end{align}
where $M_{{\rm yr}}$ is the wind mass loss in units of $M_{\odot}/$ yr; $v_{w,5}=v_w/ 10^5$ cm s$^{-1}$;  $t_{w,{\rm yr}}=t_w / 1$ yr; $v_{\rm sh,8}=v_{\rm sh}/10^8$ cm s$^{-1}$; and $t_{{\rm 1d}}$ is the time in days. From this we see that continued interaction with a steady wind will always produce a declining light curve. Thus, a flat or rising continued interaction phase requires a flatter wind density profile created by a non-constant mass loss episode $\dot{M}_w(t)$.

This class of interaction will display a fairly wide diversity of light curve morphologies due to its sensitive dependence on the CSM density profile. In Fig. \ref{fig:density_profile_lightcurves}, we show how such continued interaction-dominated light curves vary with the density profile. For steeper CSM profiles, the breakout becomes more luminous and prominent, even though the time of breakout as well as shock emergence does not change. For $\eta > 1$, the shape is not affected as drastically by the CSM density profile. Given that these events track the instantaneous shock luminosity, any minute variation in the CSM density profile will show up in the light curve as a ``bump'' during the continued interaction phase  \citep{2017A&A...605A...6N}. 

Fig. \ref{fig:data_fits} shows an example of an observed
interacting SN, OGLE-2014-SN-131,  a type Ibn event whose light curve rose gradually then abruptly fell off \citep{2017A&A...602A..93K}. In our framework, this can be interpreted as an $\eta<1$ and $\xi< 1$ interaction event with a sustained continued interaction phase. To get a rising light curve in this phase requires a shallow CSM density profile exponent $s < 5-3/\lambda$, where $\lambda$ is given above and in Eq. \ref{eq:shock_lambda}. For an ejecta density profile of $n=10$, this gives $\lambda = 7/(10-s)$ i.e. requiring a CSM density profile shallower than $s<5/4$.

The best-fitting model for OGLE-2014-SN-131 estimates $s\approx 1$ to get the correct rise and peak luminosity, which would indicate an episode of unsteady mass loss compared to the wind-like $s=2$. In this model, the sharp decline in the light curve after peak is associated with shock emergence from the outer edge of the CSM layer, which leads to a sudden halt to the interaction power.

\subsection{Heavy Interior Interaction as Superluminous Supernovae}

Some superluminous supernovae have total radiated energies in excess of $\sim 10^{51}$ ergs \citep{2012Sci...337..927G}. To achieve this in an interaction model requires efficient conversion of the ejecta kinetic energy to radiation (i.e. $\epsilon_{\rm rad}\sim 1$). From Figure~\ref{fig:epsrad_csm}, this can occur for heavy CSM ($\eta\gtrsim 1$) and an interior breakout scenario $\xi< 1$ for which adiabatic expansion losses are minimized. 

For heavy interior interactions, we can combine the light curve expressions Eqs. \ref{eq:interior_tbo} and  \ref{eq:interior_Lbo} using the Sedov exponents $\alpha=1/2$ and $\lambda=2/(5-s)$ (from Eqs. \ref{eq:alpha_eta} and \ref{eq:shock_lambda}) to get a constraint on the CSM mass as
\begin{align}\label{eq:mcsm_bo}
M_{\rm csm}\approx 5\,\kappa^{-2/3}L_{\rm bo,44}^{1/3}\Delta t_{\rm bo,30d}^{5/3}\,\,M_\odot
\end{align}
where $\Delta t_{\rm bo,30d} = \Delta t_{\rm bo} / 30$ days. This equation only remains valid if the inferred CSM mass $M_{\rm csm}$ is indeed greater than the supernova ejecta mass $M_{\rm sn}$. 

For SN2006gy \citep{2007ApJ...666.1116S}, the observed breakout properties were $L_{\rm bo}\approx 1.8\cdot 10^{44}$ erg s$^{-1}$ and $\Delta t_{\rm bo}\approx 60$ days. Thus, assuming a solar composition $\kappa=0.34$ cm$^2$ g$^{-1}$, we get $M_{\rm csm}\approx 40M_\odot$.

Additionally, for $\eta>1$, the shock emergence timescale Eq. \ref{eq:t_se} can be rewritten using $\alpha=1/2$ to get $t_{\rm se}\approx R_{\rm csm}M_{\rm csm}^{1/2}E_{\rm sn}^{-1/2}$. For heavy interior breakouts we can then approximate $E_{\rm sn}\approx \int L(t)dt=E_{\rm rad}$ by assuming large radiated efficiencies $\varepsilon_{\rm rad}\approx 1$ from Eq. \ref{eq:epsrad_lc}. Thus, our additional CSM constraint based on light curve measurements becomes
\begin{align}
R_{\rm csm}\approx 10^3\,E_{\rm rad,51}^{1/2}M_{\rm csm,\odot}^{-1/2}t_{\rm se,d}\,\,R_\odot
\end{align}
For SN2006gy, the observed measurements were $E_{\rm rad}\approx 1.2\cdot 10^{51}$ ergs and $t_{\rm se}\gtrsim 300$ days, giving $R_{\rm csm}\approx 5\cdot 10^{4}R_\odot$ using $M_{\rm csm}\approx 40M_\odot$ from above.

In Fig. \ref{fig:data_fits} we show the best-fit model based on the parameters given in Table \ref{tab:data_fit_params}, showing that our estimate is fairly close to the above analysis compared to a full numerical simulation. Note that the model kinetic energy is $E_{\rm sn}=2.5\cdot 10^{51}$ ergs, implying a radiated efficiency closer to $\varepsilon_{\rm rad}\approx 0.5$. We have also found that a CSM density profile of $r^{-2.5}$ rather than the fiducial wind-like $r^{-2}$ better fits the late-time light curve evolution, which would be indicative of a non-constant mass-loss episode in producing SN2006gy's circumstellar environment.

Finally, note that the continued interaction phase is still present in this case, although it may be less pronounced as that for Ibn/IIn due to the longer duration and much more luminous breakout peak, since it takes longer to subside and reveal the underlying instantaneous shock luminosity Eq. \ref{eq:continued_Lsh} with the snowplow exponents Eq.\ref{eq:snowplow_exp}. Similar to the previous case, any variations in the CSM density profile will be imprinted on the light curve, resulting in light curve bumps. Such behavior has been seen in superluminous supernovae \citep{2022ApJ...933...14H}.

\subsection{Light Edge Interaction as Fast Blue Optical Transients}

Interaction may be relevant in explaining the light curves of so-called fast blue optical transients, or FBOTs \citep{2013ApJ...774...58D,2018NatAs...2..307R,2019MNRAS.484.1031P,2021arXiv210508811H}. These events  generally rise in less than a day to reach peak luminosities in excess of $10^{44}$ erg s$^{-1}$. Perhaps the most well-studied example is AT2018cow, also referred to as "The Cow" \citep{2019MNRAS.484.1031P}. 

The fast rise and decline of the light curve favors a breakout flash as the theoretical interpretation, followed by a shock cooling tail. In order to get a rapid breakout flash, we must be in the edge breakout regime, $\xi > 1$. Furthermore, the rapid timescale of the shock cooling implies lowish diffusion times, i.e. $\eta < 1$.

To constrain the properties of the CSM in this regime, we can combine the edge breakout expressions Eqs. \ref{eq:breakout_lbo} and \ref{eq:breakout_dtbo} to get an expression for the CSM radius as
\begin{align}\label{eq:rcsm_bo}
R_{\rm csm}\approx 2\cdot 10^3\,\,\kappa^{1/3}L_{\rm bo,44}^{1/3}\Delta t_{\rm bo,d}^{2/3}\,\,R_\odot
\end{align}
For The Cow which had $L_{\rm peak}\approx 3\cdot 10^{44}$ erg s$^{-1}$ and a rise to peak time of $\sim 1.5$ days, this gives a radius $R_{\rm csm}\approx 3\cdot 10^{3} R_\odot$ (assuming $\kappa=0.34$ cm$^2$ g$^{-1}$), in rough agreement with the numerical best fit model shown in Fig. \ref{fig:data_fits}.
\begin{figure}
    \centering
    \includegraphics[width=0.5\textwidth]{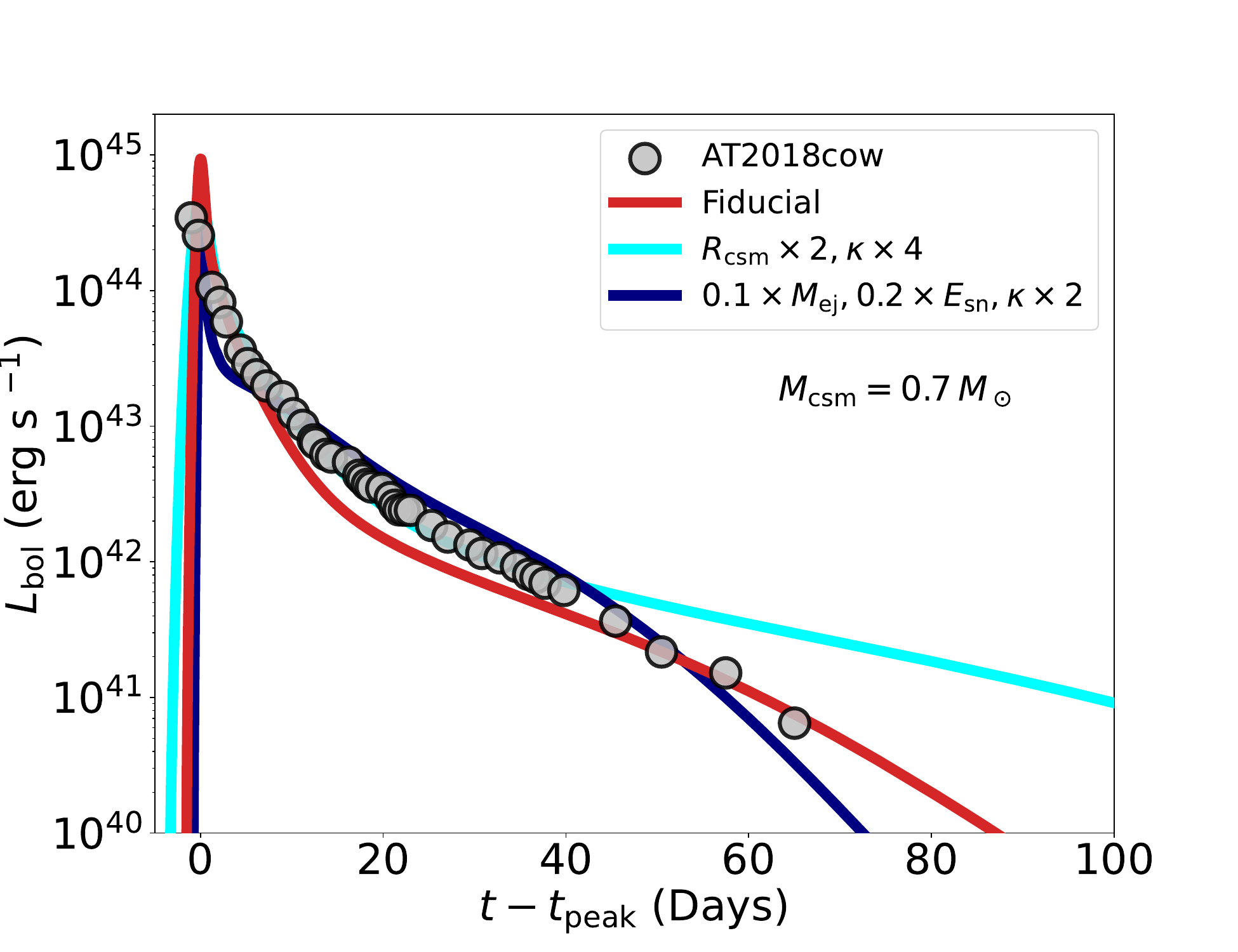}
    \caption{Numerical model fits to the light curve of AT2018cow \citep{2019MNRAS.484.1031P}, using slightly different ejecta and CSM parameters. The fiducial model corresponds to the one shown in Fig. \ref{fig:data_fits}.}
    \label{fig:at2018cow_comp}
\end{figure}

In the $\eta<1$ case, it may be difficult to distinguish between the breakout flash and subsequent shock cooling tail. In Fig. \ref{fig:at2018cow_comp} we show how the early part of The Cow can be fit with a variety of ejecta and CSM parameters, using a slightly different CSM mass than in Fig. \ref{fig:data_fits}. All three models give a reasonably good fit to the rise and peak luminosity of the light curve, i.e. the breakout. Their primary difference around peak is how much of the early emission comes from the breakout flash vs. shock cooling, each of which is described by different expressions Eqs. \ref{eq:breakout_lbo} and \ref{eq:L_cool}. It is only until much later that the models begin to reveal differences during the shock cooling phase, which may be harder to observe due to its lower luminosity and contamination from other effects such as recombination \citep{2020ApJ...899...56S,2021ApJ...915...80L} or radioactive decay \citep{2019ApJ...887..169H}.

\subsection{Heavy Edge Interaction as Double-peaked Transients}

\begin{figure}
    \centering
    \includegraphics[width=0.5\textwidth]{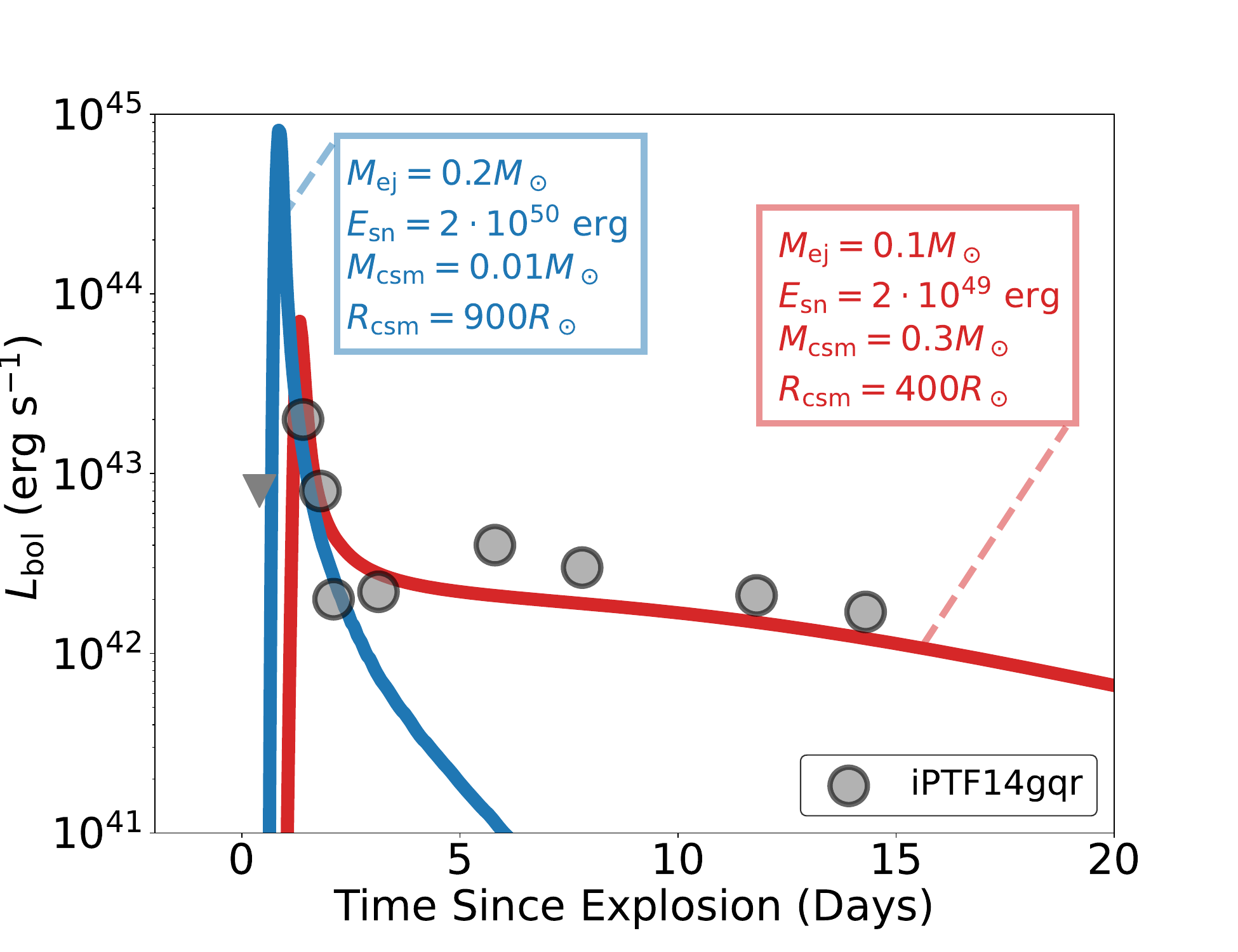}
    \caption{Numerical fits to the light curve of the Type Ic SN iPTF14gqr, with inset model parameters. The leftmost blue light curve fits only the first peak with a shock cooling tail, while the broader red light curve fits both the primary peak (shock breakout) and secondary bump (shock cooling). Light curve measurements taken from \cite{2018Sci...362..201D}.}
    \label{fig:doublepeak_fit}
\end{figure}

The observed Type Ic SN iPTF14gqr had a fast-rising ($\sim 1$ day) early luminosity excess, followed by a more extended primary light curve \citep{2018Sci...362..201D}. This light curve can be explained in multiple ways. In the original interpretation of \cite{2018Sci...362..201D}, shock cooling powered the early bump while radioactive decay powered the primary light curve. In Fig. \ref{fig:doublepeak_fit} we use the same parameters as described in \cite{2018Sci...362..201D}, showing that the early excess can indeed be fit by a shock cooling tail. However, this model also predicts a breakout flash that is two orders of magnitude more luminous than the brightest measurement. That such an observed breakout flash was unseen in iPTF14gqr may be a result of the cadence of the observations and the fact that this flash is primarily in very blue bands that might not have been easily captured by optical observations.

Alternatively, the double-peaked light curve can be explained entirely by interaction, \textit{without invoking multiple heating sources}. Namely, shock breakout produces the initial brief and luminous peak, while shock cooling produces the secondary longer-duration peak. The CSM mass must be sufficiently high, otherwise the shock cooling emission blends into the breakout emission, rather than forming a distinct double-peaked morphology (see Figure~\ref{fig:csm_lightcurves}). This can be quantified using  Eqs. \ref{eq:breakout_lbo} and \ref{eq:L_cool}, to write the ratio of the shock cooling peak to the breakout peak 
\begin{align}
\frac{L_{\rm sc}}{L_{\rm bo}}\sim \xi^{-1/2}
\end{align}
Similarly, the two timescales are, using Eqs. \ref{eq:breakout_dtbo} and \ref{eq:t_cool},
\begin{align}
\frac{\Delta t_{\rm bo}}{t_{\rm sc}}\sim \xi^{-3/4}
\end{align}
For scenarios in the regime $\xi \gg 1$ these equations imply $L_{\rm sc}< L_{\rm bo}$ and $\Delta t_{\rm bo}< t_{\rm sc}$, and hence distinct double peaks.

In Fig. \ref{fig:doublepeak_fit}, we additionally fit a numerical model to the entirety of iPTF14gqr's light curve assuming only interaction (i.e. no radioactive heating), where the first peak is produced by shock breakout rather than shock cooling. Although the CSM radius and ejecta mass are comparable in both interpretations, the simultaneous fit of both peaks requires an interaction consisting of ejecta an order of magnitude less energetic, and a larger CSM mass by a factor of about 30. In summary, both scenarios are plausible explanations for double-peaked events, depending on whether one invokes additional heating for the second peak, and whether the much bluer breakout flash is covered by the bands used.

\subsection{Caveats and Additional Physics}
\label{subsec:other_physics}

Several physical processes were neglected in constructing our analysis in order to provide a broadly applicable yet tractable theoretical framework. In \S \ref{sec:configuration} we briefly discussed the issue of inefficient radiative cooling of the shock. This will arise most likely in the $\eta \ll 1$ and $\xi<1$ regime, where the CSM is optically thin and the shock velocity is still sufficiently fast. The net effect of this is to reduce the radiative throughput and efficiency of the light curve (see \S \ref{subsec:epsrad}), and we must account for non-thermal emission processes of collisionless shocks that are typically encountered in the context of supernova remnants. 

\cite{2022ApJ...928..122M} have delineated the regimes in which circumstellar shocks behave, where the addition of inverse Compton scattering expands the space in which the shocks are radiative. They found that fast-moving (e.g. relativistic) shocks in an optically thin CSM are radiatively inefficient. Thus, our analysis is applicable only to non-relativistic shock velocities, particularly where the shock is moving slow enough and the CSM optically thick enough that free-free cooling is effective.

We have also neglected predictions of the color/spectra of the resulting interaction, which requires a careful treatment of all the relevant physical processes (e.g. photoionization, inverse Comptonization, and lines). Thus, our results are only applicable in a bolometric sense, which limits the predictive power if we do not have accurate bolometric corrections of observed events. Of particular interest is the photospheric behavior of the interaction as the shock progresses, and any reprocessing effects the CSM will have on the shock, such as ``thermalizing'' hot X-ray shock photons into optical wavelengths via large bound-free opacities \citep{2015MNRAS.449.4304D,2022ApJ...928..122M}.

While our models and analysis have used a simple constant opacity, the opacity may change significantly when temperatures become cool enough for atoms to recombine \citep{2020ApJ...899...56S}. This is of particular importance for interior breakouts from a heavy CSM, which substantially decelerates the ejecta velocity resulting in a lower-temperature shock. Additionally, for edge breakout events, the decreasing temperature of the expanding shock cooling region will drop low enough that the ejecta also recombines, which will affect the late-time light curves from these events. Our scalings and models only remain reliable when the temperatures remain greater than the recombination temperature of the matter.

The configuration of all of our models  consists of single a spherical CSM shell with a sharp outer edge, reminiscent of the CSM produced by an eruptive mass loss episode. When considering more general CSM configurations our expectations may need to be revised. If, for example, a long-duration wind produces a gradually declining CSM density profile without a sharp edge, then the situation will resemble one of our shell models with $R_{\rm csm}$ taken to be very large, such that breakout happens in the CSM interior and the phase of continued interaction  persists indefinitely. If on the other hand repeated episodes of eruptive mass occur, as for example in pulsational pair-instability supernovae \citep{2017ApJ...836..244W}, the CSM may consist of numerous spherical shells. If these  CSM shells are well separated, iterative application of the formalism presented here may be used to analyze each shell interaction individually. It is also possible that the CSM is non-spherical, with perhaps a disk-like configuration (see for example \cite{2018ApJ...856...29M,2019ApJ...887..249S}). In that case, the physical behavior of the escaping radiation differs from the spherically symmetric case, as only a sliver of the ejecta will participate in the interaction. 

We have assumed in the analytic scalings and numerical simulations that $R_*\ll R_{\rm csm}$, i.e. the inner edge of the CSM is much smaller than the outer CSM radius. However, it is possible to expect geometrically ``thin'' CSM shells due to e.g. brief episodes of mass loss, where $R_*\sim R_{\rm csm}$. This introduces an additional physical parameter to the dynamics that must be explicitly accounted for. In the case of an edge breakout, where radiation is only able to escape once the shock reaches $R_{\rm csm}$, the value of $R_*$ does not influence the resulting light curve or breakout duration, although it can alter the time of shock emergence relative to the supernova epxlosion, $t_{\rm se}\sim \eta^\alpha (R_{\rm csm}-R_*)/v_{\rm ej}$. The effect of a thin CSM shell will be most prominent for an interior breakout and resulting continued interaction phase, shown by the explicit dependence of $R_*$ in Eq. (\ref{eq:Lsh_s}).

Finally,  we have also assumed in our analytical treatment that the forward shock dominates the light curve at all phases, and the shock luminosity terminates at shock emergence. In reality, a reverse shock will form at the interface of the shocked ejecta region, illustrated in Fig. \ref{fig:shock_structure}, which will provide an additional luminosity source for the light curve \citep{2012ApJ...746..121C,2017hsn..book..875C}. The strength of the reverse shock depends on the CSM mass, where a larger contribution is expected in the $\eta>1$ heavy CSM regime. \cite{2019ApJ...884...87T} have shown that the reverse shock can be an important source of emission during the continued interaction phase. The reverse shock will also persist for some time after shock emergence, i.e. during the shock cooling phase. We have confirmed the existence of a luminous reverse shock in our numerical simulations, which is especially prominent for the $\eta>1$ models.  Further analytic study and numerical investigation is necessary to fully characterize its behavior.


\section{Conclusions}

CSM interaction significantly expands the light curve duration and luminosity phase space that normal supernovae may otherwise occupy. This is due to the efficient conversion of the large store of ejecta kinetic energy from the preceding supernova, into radiation at the shock front. Here, we articulated a conceptual framework to interpret interaction light curves. We decompose the interaction light curve into five distinct phases, each of which may produce distinct features in the light curve morphology. We separate interaction light curves into four distinct classes, which depend on a combination of (1) where shock breakout occurs; and (2) the relative masses of the ejecta and CSM.

In \S \ref{sec:analytics} we derived quantitative relations for the qualitative picture given in \S \ref{sec:configuration}. We provide scaling relations for each of the light curve phases, using a simplified model for the shock evolution. We then confirmed in \S \ref{sec:numerical} the analytical model by running a grid of one-dimensional radiation hydrodynamics simulations across a broad parameter space. Finally, we provided four case studies of observed transients in \S \ref{sec:discussion} to demonstrate how the framework can be used in practice.

Our results should be useful to study stellar mass loss through observations of supernova light curves. In particular, different physical mass loss mechanisms will have distinct predictions regarding the progenitor and structure of the CSM \citep{2017MNRAS.470.1642F,2021ApJ...906....3W}. While light interior breakouts can be explained by interaction with a typical stellar wind, heavy interior breakouts will require prodigious mass loss from a supermassive progenitor \citep{2012MNRAS.423L..92Q}. In other cases, the small radii necessary to produce edge breakouts will require episodes of significant mass loss near the end of the star's life \citep{2019ApJ...887..169H}. 

We have limited our analysis to the bolometric properties of the interaction light curve. The broadband spectra and observed colors also likely provide important information pertaining to the interaction, particularly given the wide range in shock temperatures that can result depending on how efficiently the shock can cool. We might therefore find that each light curve phase has a distinct color evolution and photospheric behavior. An accurate bolometric correction of the light curve may require detailed coverage from X-ray to optical wavelengths \citep{2022ApJ...928..122M}, although these capabilities are recently becoming attainable (see for example \cite{2017ApJ...835..140M,2019ApJ...872...18M}). This poses a unique observational challenge in several ways, particularly for the edge breakout flash due to its brief timescale and likely rapid color evolution. We have also neglected any non-thermal emission that may be produced by the shock, which is of particular importance in understanding radio observations of interacting supernovae.

In addition to the broadband colors, it would be interesting to connect the spectral evolution of the interaction to the different light curve phases and classes. Narrow emission lines have been the hallmark signature of interaction as it implies slow-moving material above the heating at the shock front. While the presence of narrow lines favors interaction, the absence of such features does not preclude CSM interaction as the mechanism behind the light curve. Interior breakouts are the natural interaction type to expect such features, while edge breakouts may have little to no narrow lines in their spectra as the bulk of the CSM has already been swept up. An accurate investigation of interaction spectra requires running expensive non-LTE radiation hydrodynamics simulations, such as is done in \cite{2015MNRAS.449.4304D} for the case of a heavy interior breakout.

In constructing a broadly applicable light curve framework, we have neglected several important physical effects that will influence the results presented in this paper. Potentially important effects are briefly discussed in \S \ref{subsec:other_physics} and warrant further investigation. Of particular interest include how asymmetric CSM configurations affect the observed phase properties, since the shock will only occupy a fraction of the full $4\pi$ solid angle of the ejecta. Indeed, several mass loss mechanisms such as binary interaction may produce a more disk-like geometry. The shock region is also prone to hydrodynamical instabilities which require high-resolution multidimensional radiation hydrodynamics simulations to fully investigate. It is unclear how such effects impact the resulting phases and classes discussed in this work.

\medskip
\begin{acknowledgments}
We thank Lars Bildsten, Anna Ho, Ben Margalit, Yuhan Yao, and Sivan Ginzburg for insightful discussions; Brenna Mockler and Benny Tsang for assistance in implementing the numerical code. We thank the ZTF Theory Network, funded in part by the Gordon and Betty Moore Foundation through Grant GBMF5076, for cultivating a fruitful scientific environment. DKK is supported by the National Science Foundation Graduate Research Fellowship. 
DNK research is supported in part by the U.S. Department of Energy, Office of Science, Office of Nuclear Physics, under contract number DE-AC02-05CH11231 and DE-SC0004658. This research is partly supported by a grant from the Simons Foundation (622817DK).
This research used resources of the National Energy Research Scientific Computing Center (NERSC), a U.S. Department of Energy Office of Science User Facility located at Lawrence Berkeley National Laboratory, operated under Contract No. DE-AC02-05CH11231 using NERSC award NP-ERCAP-0025048.

\end{acknowledgments}

\software{Sedona \citep{2006ApJ...651..366K,2015ApJS..217....9R}, NumPy \citep{harris2020array}, SciPy \citep{2020SciPy-NMeth}, matplotlib \citep{thomas_a_caswell_2023_7570264}, Adobe Illustrator}

\bibliographystyle{aasjournal}
\bibliography{references}
\appendix

\section{Numerical Scalings}
\label{sec:physical_scalings}

In \S \ref{sec:numerical} we performed numerical simulations of CSM interaction to determine the scaling behavior of the different light curve phases, and compare against the analytics presented in \S \ref{sec:analytics}. Namely, we proposed fitting formula of the $i$-th phase luminosity and timescale of

\begin{align}
L_{\rm i} &= a_i \eta^{-3\alpha_i}\xi^{k_i}L_0 \\
t_i &= b_i \eta^{\alpha_i}\xi^{c_i}t_0
\end{align}

where $L_0=M_{\rm csm}v_{\rm ej}^3/R_{\rm csm}$, $t_0=R_{\rm csm}/v_{\rm ej}$, $\eta=M_{\rm csm}/M_{\rm ej}$, and the breakout parameter is
\begin{align}
\xi = \beta_0\tau_0\eta^{-\alpha}\approx 10\,\kappa M_{\rm csm,\odot}v_9 R_4^{-2}\eta^{-\alpha}
\end{align}
Here, $\beta_0=v_{\rm ej}/c$ (where $v_{\rm ej}=\sqrt{2 E_{\rm sn}/M_{\rm ej}}$), $\tau_0=\kappa M_{\rm csm}/4\pi R_{\rm csm}^2$,  $v_9=v_{\rm ej}/10^9$ cm s$^{-1}$, $M_{\rm csm,\odot}=M_{\rm csm}/M_\odot$, and $R_4=R_{\rm csm}/10^4 R_\odot$. The value of $\alpha$ in the breakout parameter can be estimated from analytic arguments (see Appendix \ref{sec:shock_derivation}) as
\begin{align}\
\alpha = \left\{\begin{array}{lr}
    1/2, & (\eta\gtrsim 1) \\
    1/(n-3), & (\eta\ll 1)
\end{array}\right.
\end{align}
where $\rho_{\rm ej}\propto r^{-n}$ is the outer ejecta density profile, and $n\approx 7-10$ \citep{1989ApJ...341..867C}.Note that the scaling of $\alpha=1/(n-3)$ for $\eta\ll 1$ breaks down when the shock reaches the flatter inner portion of the ejecta. This occurs when the amount of swept up CSM mass exceeds the mass contained in the steep outer ejecta. The ratio of outer to inner ejecta mass is equal to $(3-\delta)/(n-3)= 2/7$ for the fiducial power-laws. Thus, for CSM masses $M_{\rm csm}\gtrsim 0.3M_{\rm ej}$, the behavior of the shockwave will change. 

\begin{figure}
    \centering
    \includegraphics[width=0.8\textwidth]{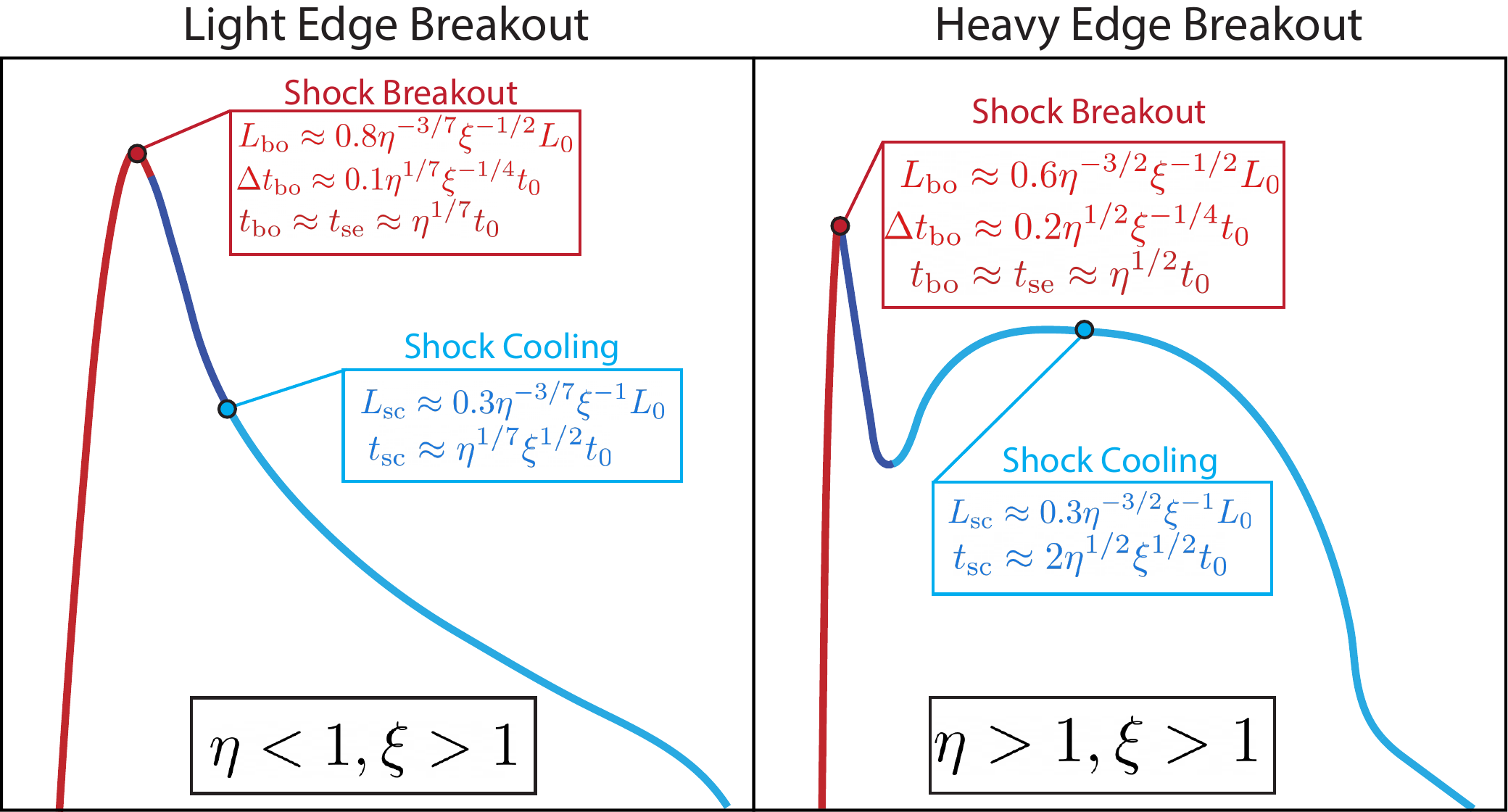}
    \caption{Schematic diagram with numerically calibrated scalings for edge breakout light curves, in the light \textit{(left)} and heavy \textit{(right)} CSM regimes. Assumes ejecta density profile $\rho_{\rm ej}\propto r^{-10}$ and CSM density profile $\rho_{\rm csm}\propto r^{-2}$. More general scalings for other profiles given in \S \ref{sec:analytics}.}
    \label{fig:edge_schematic}
\end{figure}

\begin{figure}
    \centering
    \includegraphics[width=0.8\textwidth]{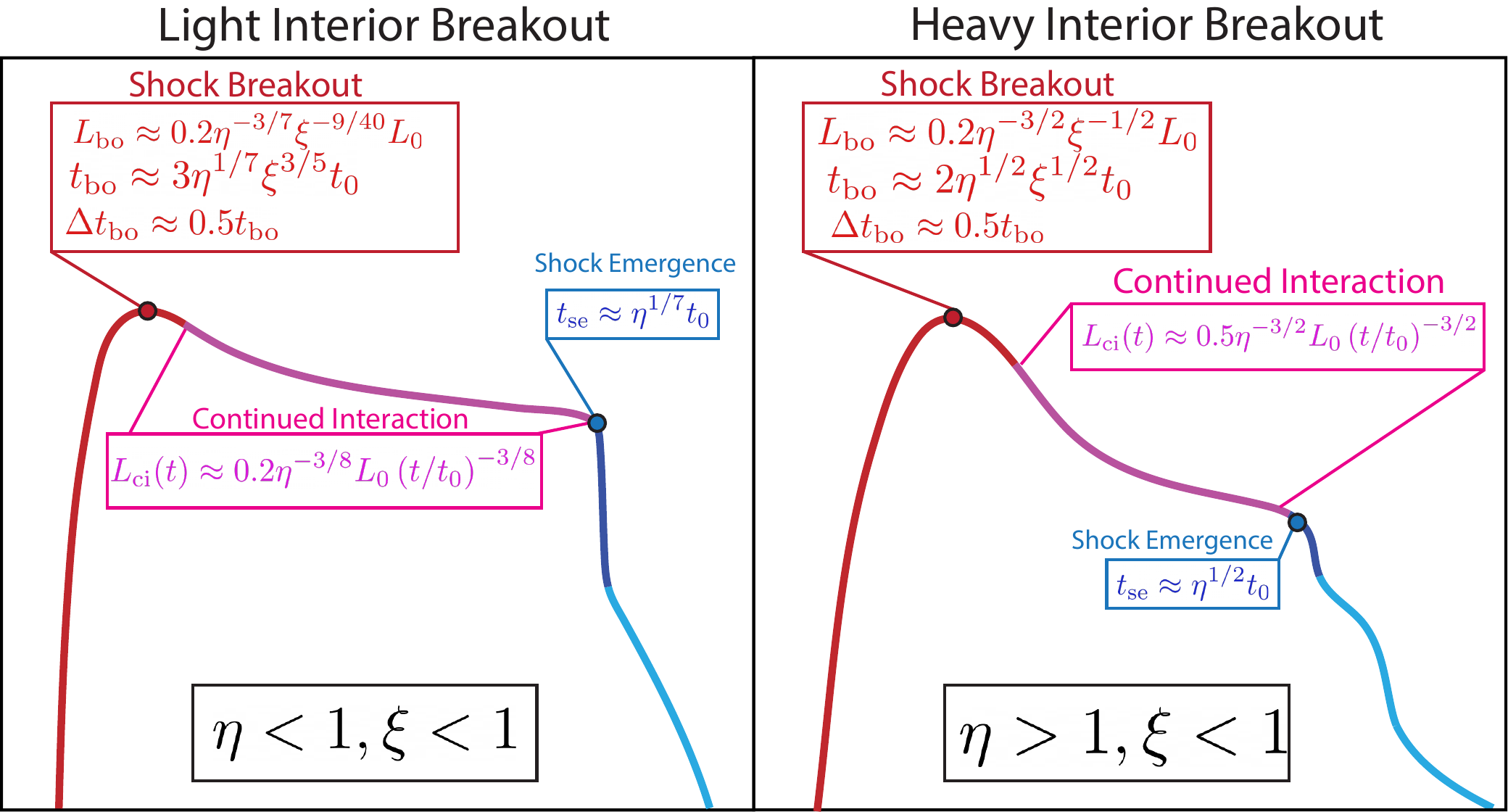}
    \caption{Same as Fig. \ref{fig:edge_schematic} but for interior breakout events, for light (\textit{left}) and heavy (\textit{right}) CSM regimes. Assumes ejecta density profile $\rho_{\rm ej}\propto r^{-10}$ and CSM density profile $\rho_{\rm csm}\propto r^{-2}$. More general scalings for other profiles given in \S \ref{sec:analytics}.}
    \label{fig:int_schematic}
\end{figure}

The scaling exponents $(a_i,k_i)$, $(b_i,c_i)$, and $\alpha_i$ are fit to the numerical simulations described in \S \ref{sec:numerical}. The numerical fits are given for each interaction class and phase in Figs. \ref{fig:edge_schematic} and \ref{fig:int_schematic}, which we then convert into physical scalings below. Note that the models assume an $n=10$ ejecta density profile, and an $s=2$ CSM density profile. For more general density profiles, refer to the analytical scalings in \S \ref{sec:analytics}.

\subsection{Light Edge Breakout ($\eta<1,\xi>1$)}

\begin{align}
L_{\rm bo}&\sim 7\,M_{\rm csm}^{1/7}M_{\rm ej}^{-25/28}E_{\rm sn}^{5/4}\kappa^{-1/2}c^{1/2} \\
\Delta t_{\rm bo}&\sim 0.1\,\,M_{\rm csm}^{-1/14}R_{\rm csm}^{3/2}M_{\rm ej}^{25/56}E_{\rm sn}^{-5/8}\kappa^{-1/4}c^{1/4} \\
t_{\rm bo}\approx t_{\rm se}&\sim 0.7\,\, M_{\rm csm}^{1/7}R_{\rm csm}^1 M_{\rm ej}^{5/14}E_{\rm sn}^{-1/2} \\
L_{\rm sc}&\sim 8\,M_{\rm csm}^{-2/7}R_{\rm csm}^{1}M_{\rm ej}^{-5/7}E_{\rm sn}^{1}\kappa^{-1}c \\
t_{\rm sc}&\sim 0.2\,M_{\rm csm}^{4/7}M_{\rm ej}^{5/28}E_{\rm sn}^{-1/4}\kappa^{1/2}c^{-1/2}
\end{align}


\subsection{Heavy Edge Breakout ($\eta>1,\xi>1$)}

\begin{align}
L_{\rm bo}&\sim 5\,M_{\rm csm}^{-3/4}E_{\rm sn}^{5/4}\kappa^{-1/2}c^{1/2} \\
\Delta t_{\rm bo} &\sim 0.2\,M_{\rm csm}^{3/8}R_{\rm csm}^{3/2}E_{\rm sn}^{-5/8}\kappa^{-1/4}c^{1/4} \\
t_{\rm bo}\approx t_{\rm se}&\sim 0.7\,M_{\rm csm}^{1/2}R_{\rm csm}^{1}E_{\rm sn}^{-1/2} \\
L_{\rm sc}&\sim 8\,M_{\rm csm}^{-1}R_{\rm csm}^{1}E_{\rm sn}^{1}\kappa^{-1}c \\
t_{\rm sc}&\sim 0.5\,M_{\rm csm}^{3/4}E_{\rm sn}^{-1/4}\kappa^{1/2}c^{-1/2}
\end{align}


\subsection{Light Interior Breakout ($\eta<1,\xi<1$)}

\begin{align}
L_{\rm bo}&\sim 0.9\,M_{\rm csm}^{53/140}R_{\rm csm}^{-11/20}M_{\rm ej}^{-111/112}E_{\rm sn}^{111/80}\kappa^{-9/40}c^{9/40} \\
t_{\rm bo} \approx 2\Delta t_{\rm bo} &\sim 0.6\,M_{\rm csm}^{23/35}R_{\rm csm}^{-1/5}E_{\rm sn}^{-1/5}M_{\rm ej}^{1/7}\kappa^{3/5}c^{-3/5} \\
t_{\rm se}&\sim 0.7\, M_{\rm csm}^{1/7}R_{\rm csm}^{1}M_{\rm ej}^{5/14}E_{\rm sn}^{-1/2} \\
L_{\rm ci}(t)&\sim 0.5\, M_{\rm csm}^{5/8}R_{\rm csm}^{-5/8}M_{\rm ej}^{-15/16}E_{\rm sn}^{-21/16}t^{-3/8}
\end{align}


\subsection{Heavy Interior Breakout ($\eta>1,\xi<1$)}

\begin{align}
L_{\rm bo}&\sim 2\,M_{\rm csm}^{-3/4} E_{\rm sn}^{5/4}\kappa^{-1/2}c^{1/2}\\
t_{\rm bo}\approx 2\Delta t_{\rm bo}&\sim 0.5\,M_{\rm csm}^{3/4}E_{\rm sn}^{-1/4}\kappa^{1/2}c^{-1/2} \\
t_{\rm se}&\sim 0.7\,M_{\rm csm}^{1/2}R_{\rm csm}^{1}E_{\rm sn}^{-1/2} \\
L_{\rm ci}(t)&\sim 0.8\,M_{\rm csm}^{-1/2}R_{\rm csm}^{1/2}M_{\rm ej}^{3/4}E_{\rm sn}^{3/4}\,t^{-3/2}
\end{align}


\section{Derivation of Shock Similarity Exponents}
\label{sec:shock_derivation}
In \S \ref{sec:analytics}, we introduced power-law forms of the radius with time as
\begin{align}
r_{\rm sh}(t) \sim R_{\rm csm}\left(\frac{t}{\eta^\alpha  t_0}\right)^{\lambda}
\end{align}
where $t_0 = R_{\rm csm}/v_{\rm ej}$,and $\eta=M_{\rm csm}/M_{\rm ej}$. The shock velocity is similarly expressed as
\begin{align}
v_{\rm sh}(t) = \frac{dr_{\rm sh}}{dt} \sim v_{\rm ej}\eta^{-\alpha}\left(\frac{t}{\eta^\alpha t_0}\right)^{\lambda-1}
\end{align}
These two expressions can be combined to get the shock velocity in terms of the shock radius as
\begin{align}
v_{\rm sh}(r_{\rm sh}) = v_{\rm ej}\eta^{-\alpha}\left(\frac{r_{\rm sh}}{R_{\rm csm}}\right)^{(\lambda-1)/\lambda}
\end{align}
For a power-law CSM density $\rho_{\rm csm}(r)\propto r^{-s}$ with inner radius $R_*$ and outer radius $R_{\rm csm}$, the amount of mass swept up by the shock will go as
\begin{align}\label{eq:dm_shock}
\delta M(r_{\rm sh}) &= \int_{R_*}^{r_{\rm sh}} 4\pi r^2 \rho_{\rm csm}(r)\,dr \\
&\approx M_{\rm csm}\left(\frac{r_{\rm sh}}{R_{\rm csm}}\right)^{3-s}
\end{align}
where we have assumed in the last step that $R_*\ll R_{\rm csm}$.

The shock radius and velocity depends on the two similarity exponents $\left(\alpha,\lambda\right)$, which in turn will depend on $\eta$. We now derive the shock equations and exponents in the different regimes.

\subsection{Light CSM Regime $(M_{\rm csm}<M_{\rm ej})$}
If $\eta<1$, then the steep outermost layer of the ejecta will dominate the bulk of the shock evolution. Only a portion of the ejecta mass of order $\sim M_{\rm csm}$ will participate in the interaction.

The amount of momentum contained in the ejecta above a radius $r_0$ is
\begin{align}
\delta (Mv) = \int_{r_0}^\infty 4\pi r^2\rho_{\rm ej}(r) v(r)\,dr
\end{align}
If we assume the ejecta expands homologously, $v=r/t$, we can write the ejecta density profile as a power-law in velocity coordinates,
\begin{align}
\rho_{\rm ej}(v_0) = \frac{f_\rho M_{\rm ej}}{v_{\rm ej}^3t^3}\left(\frac{v_0}{v_{\rm ej}}\right)^{-n}
\end{align}
where $\rho_0(t)\sim M_{\rm ej}/(v_{\rm ej} t)^3$ and $f_\rho$ is a constant of order unity (see Appendix \ref{sec:numeric_setup}). The momentum above a velocity coordinate $v_0$ is therefore
\begin{align}\label{eq:dmv_shock}
\delta(Mv)(v_0) = \frac{4\pi}{(n-4)}f_\rho M_{\rm ej}v_{\rm ej}\left(\frac{v_0}{v_{\rm ej}}\right)^{4-n}
\end{align}
As the shock runs through the CSM, it sweeps up as mass $\delta M_{\rm sh}$ at a velocity $v_{\rm sh}$. From conservation of momentum,
\begin{align}
2\delta M_{\rm sh}v_{\rm sh} = \delta(Mv)(v_{\rm sh})
\end{align}
Using Eq. \ref{eq:dm_shock} for the swept up mass and Eq. \ref{eq:dmv_shock} for the ejecta momentum, we get
\begin{align}
M_{\rm csm}\left(\frac{r_{\rm sh}}{R_{\rm csm}}\right)^{3-s}v_{\rm sh} \approx M_{\rm ej}v_{\rm ej}\left(\frac{v_{\rm sh}}{v_{\rm ej}}\right)^{4-n}
\end{align}
where we have dropped order-unity constants. Using $v_{\rm sh}=dr_{\rm sh}/dt$ and rearranging to solve for $r_{\rm sh}$, we get
\begin{align}
r_{\rm sh}(t) \sim R_{\rm csm}\left(\frac{t}{\eta^{1/(n-3)}t_0}\right)^{(n-3)/(n-s)}
\end{align}
 From this, we see that the similarity exponents in the light CSM regime $\eta<1$ are
\begin{align}
\lambda = \frac{(n-3)}{(n-s)},\,\,\, \alpha = \frac{1}{(n-3)}\,\,\,\,\,\,\, (\eta<1)
\end{align}
These similarity exponents hold for both adiabatic and radiative shocks in this regime.

\subsection{Heavy CSM Regime $(M_{\rm csm}>M_{\rm ej})$}
If the CSM mass exceeds the ejecta mass, $\eta>1$, then the interaction will tap the entirety of the ejecta kinetic energy, and will obey a blastwave evolution. The behavior of the blastwave depends on whether it is adiabatic or radiative.

\subsubsection{Adiabatic Blastwave (Sedov)}

Once the shock sweeps up of order $\delta M_{\rm sh}\sim M_{\rm ej}$, the shock transitions into a blastwave. If the blastwave is adiabatic, then energy is conserved and so
\begin{align}
M_{\rm ej}v_{\rm ej}^2 \approx \delta M_{\rm sh}v_{\rm sh}^2
\end{align}
where we have assumed the bulk of the ejecta kinetic energy is located near the shock front. 

Using Eq.\ref{eq:dm_shock} for $\delta M_{\rm sh}$ and $v_{\rm sh}= dr_{\rm sh}/dt$, we get
\begin{align}
r_{\rm sh}(t)\sim R_{\rm csm}\left(\frac{t}{\eta^{1/2}t_0}\right)^{2/(5-s)}
\end{align}
which is the usual Sedov-Taylor blastwave solution for a power-law medium $\rho\propto r^{-s}$. Thus, the similarity exponents for an energy-conserving blastwave are
\begin{align}
\lambda = \frac{2}{(5-s)},\,\,\,\alpha = \frac{1}{2}\,\,\,\,\,\,\,\,(\eta>1)
\end{align}

\subsubsection{Radiative Blastwave (Snowplow)}
If radiation is able to escape ahead of the shock, then energy is no longer conserved. Instead, from conservation of momentum, we have
\begin{align}
M_{\rm ej}v_{\rm ej}\approx \delta M_{\rm sh}v_{\rm sh}
\end{align}
Using Eq.\ref{eq:dm_shock} for $\delta M_{\rm sh}$ and $v_{\rm sh}= dr_{\rm sh}/dt$, we therefore get that a radiative blastwave will evolve in time as
\begin{align}
r_{\rm sh}(t)\sim R_{\rm csm}\left(\frac{t}{\eta t_0}\right)^{1/(4-s)}
\end{align}
which gives the evolution for a momentum-conserving ``snowplow'' blastwave in a power-law medium. Thus, the similarity exponents for a radiative blastwave are
\begin{align}
\lambda = \frac{1}{(4-s)},\,\,\alpha = 1\,\,\,\,(\eta>1)
\end{align}

\subsection{Shock Breakout Radius}
Shock breakout occurs when the shock optical depth $\tau_{\rm sh}$ equals $c/v_{\rm bo}$, where $v_{\rm bo}$ is the breakout velocity that depends on time. To find this point, we integrate the shock optical depth to the breakout radius
\begin{align}
\tau_{\rm bo} = -\int_{R_{\rm csm}}^{r_{\rm bo}}\kappa\rho_{\rm csm}(r)\,dr \approx -\frac{\tau_0}{(1-s)}\left[x_{\rm bo}^{1-s}-1\right]
\end{align}
where $\tau_0=\kappa M_{\rm csm}/4\pi R_{\rm csm}^2$ and $x_{\rm bo}=r_{\rm bo}/R_{\rm csm}$. For a shock radius that evolves in time as a power-law $r_{\rm sh}\propto t^\lambda$ and using the fact that $v_{\rm sh}=dr_{\rm sh}/dt$, the breakout velocity can be expressed in terms of $x_{\rm bo}$ as
\begin{align}
v_{\rm bo} \approx v_{\rm ej}\eta^{-\alpha}x_{\rm bo}^{1-1/\lambda}
\end{align}
Setting $\tau_{\rm sh}=c/v_{\rm bo}$ we get a non-linear equation for $x_{\rm bo}$
\begin{align}\label{eq:breakout_eq}
(1-s)x_{\rm bo}^{1/\lambda-1} = -\frac{\beta_0\tau_0}{\eta^{\alpha}}\left[x_{\rm bo}^{1-s}-1\right] = -\xi\left[x_{\rm bo}^{1-s}-1\right]
\end{align}
where $\beta_0=v_{\rm ej}/c$ and $\xi=\beta_0\tau_0\eta^{-\alpha}$. In general Eq. \ref{eq:breakout_eq} must be solved numerically for $x_{\rm bo}$, given $\beta_0,\tau_0,\eta^\alpha$, and the density profile $s$.
For the case of $s=2$ we can write this as
\begin{align}
x_{\rm bo}^{1/\lambda} = \left(1-x_{\rm bo}\right)\xi
\end{align}
Note that in the limit of $x_{\rm bo}\ll 1$, the breakout location becomes
\begin{align}\label{eq:xbo_less}
x_{\rm bo} \approx\xi^{\lambda} = \left[\beta_0\tau_0\eta^{-\alpha}\right]^{\lambda}
\end{align} Furthermore, for the case of $\xi\gg 1$ ($\beta_0\tau_0\gg \eta^{\alpha}$) Eq. \ref{eq:breakout_eq} is simply $x_{\rm bo}\approx 1$. We can interpolate between these two regimes with a free parameter $k_0$ as
\begin{align}\label{eq:xbo_approx}
x_{\rm bo}\approx \left(\frac{\beta_0\tau_0}{\eta^\alpha}\right)^{\lambda k_0} \approx \xi^{\lambda k_0}
\end{align}
where $0\leq k_0 \leq 1$, which is our proposed interior breakout expression Eq. \ref{eq:breakout_xbo} used in Sec. \ref{sec:analytics}. Note that Eq. \ref{eq:xbo_approx} is equivalent to Eq. \ref{eq:xbo_less} for the choice of $k_0=1$. Using the fact that $x_{\rm bo}=r_{\rm bo}/R_{\rm csm}$ from the earlier shock derivation, this corresponds to a breakout time of
\begin{align}
t_{\rm bo} \approx \frac{\kappa M_{\rm csm}}{4\pi R_{\rm csm}c}
\end{align}
which is the static diffusion time. For the case of $k_0\approx 0$, we instead have $x_{\rm bo}\approx 1$ and so $t_{\rm bo}\approx t_{\rm se}\approx \eta^{\alpha}t_0$, the shock emergence time. In general, $k_0$ will take on an intermediate value between these two regimes, and comparison with numerical simulations discussed in Sec. \ref{sec:numerical} and Appendix \ref{sec:physical_scalings} show that $k_0\approx 0.6$ works reasonably well for a range of interior breakout interactions.

\section{Numerical Simulations}
\label{sec:numeric_setup}

We have implemented 1D spherically-symmetric radiation hydrodynamics in Sedona \citep{2006ApJ...651..366K,2015ApJS..217....9R}, using the finite-volume moving-mesh method described in \cite{2016ApJ...821...76D}. We modify the hydrodynamics to include radiation using a comoving-frame grey flux-limited diffusion treatment based on \cite{2011ApJS..196...20Z,2013ApJS..204....7Z} and  \cite{2003JCoPh.184...53H}. We use an operator split approach where the hydrodynamics and radiation advection is solved explicitly with a second-order Runge Kutta timestepping with Courant condition $C_{\rm CFL}=0.1$; the nonlinear diffusion and matter-radiation coupling is solved implicitly using a Newton-Raphson method with relative error threshold of $\epsilon=10^{-10}$. The resulting linear system for the radiation energy density is directly solved using Thomas' tridiagonal matrix algorithm \citep{1986nras.book.....P}. We use an absorbing outer boundary condition for the radiation. 

We use a uniform gray opacity $\kappa$ for the ejecta and CSM that is constant in time, with a fiducial value of solar electron scattering, $\kappa=0.34$ cm$^2$ g$^{-1}$. We take a fraction of the scattering opacity to be absorptive, i.e. $\kappa_{\rm abs}=\epsilon_{\rm abs}\kappa$,
where $\epsilon_{\rm abs}=10^{-3}$ \citep{Lovegrove2017}. Note that we ignore recombination, which may be important in certain cases during lower-temperature shock interaction (see e.g. \cite{2020ApJ...899...56S}).

We initially evolve the simulations with a fixed Eulerian grid of $1024$ and $2048$ cells for the ejecta and CSM, respectively. Note that this resolution may be insufficient to completely resolve the shock structure for cases where the shock radiation efficiently escapes. However, completely resolving the shock structure is not necessary as we are concerned primarily with the bulk conversion of kinetic energy into escaping radiation. We have confirmed with additional simulations that increasing the resolution to resolve the shock structure does not impact the resulting light curve.

Once the shock nears the breakout layer, we turn on Lagrangian mesh motion to follow the shock breakout and subsequent expansion, and evolve the system out to late times to capture shock cooling. We extract bolometric light curves by taking the comoving radiative flux at the outermost boundary of the domain.

\subsection{Problem Setup}

We assume a spherically symmetric ejecta of mass $M_{\rm ej}$ and energy $E_{\rm sn}$ with initial density profile given by a broken power law \citep{1989ApJ...341..867C,2016ApJ...821...36K}
\begin{align}
\rho_{\rm ej}(r) = f_{\rho}\frac{M_{\rm ej}}{r_t^3}\left[\frac{r}{r_t}\right]^{-(\delta,n)}
\end{align}

where the inner and outer density profiles are $\delta$ and $n$, respectively. The normalization factor is given by
\begin{align}
f_\rho = \frac{1}{4\pi}\frac{(n-3)(3-\delta)}{(n-\delta)}
\end{align}
The transition radius of the broken power law is given by
\begin{align}
r_t = v_t t_0
\end{align}
where
\begin{align}
v_t = \sqrt{\frac{f_v E_{\rm sn}}{M_{\rm ej}}}
\end{align}
and
\begin{align}
f_v = \frac{2(5-\delta)(n-5)}{(n-3)(3-\delta)}
\end{align}
We choose an initial time $t_0$ such that $r_t \ll R_{\rm csm}$, the outer CSM radius, with a fiducial value of $t_0= 10^{3}$ seconds.

The ejecta velocity is assumed to be homologous,
\begin{align}
v_{\rm ej}(r) = v_t \left[\frac{r}{r_t}\right]
\end{align}
Finally, we initialize the ejecta to be initially cold, with a uniform temperature $T_{\rm ej}=10^{2}$ K.

\subsubsection{CSM Setup}

We assume a shell of CSM of mass $M_{\rm csm}$ and radius $R_{\rm csm}$. The CSM density profile is described by a power-law

\begin{align}
\rho_{\rm csm}(r) = f_{\rm csm}\frac{M_{\rm csm}}{R_{\rm csm}^3}\left[\frac{r}{R_{\rm csm}}\right]^{-s}~~~~~~~(R_*<r<R_{\rm csm})
\end{align}
where
\begin{align}
f_{\rm csm} = \frac{(3-s)}{4\pi}\left[1-\left(\frac{R_*}{R_{\rm csm}}\right)^{3-s}\right]^{-1}
\end{align}
Here, $R_*$ is the inner radius of the CSM, and set to a fixed value of $10^{-2}R_{\rm csm}$ (i.e. $R_*\ll R_{\rm csm}$).

At the outer CSM edge $r=R_{\rm csm}$, we stitch on a steep cutoff layer with density profile
\begin{align}
\rho(r) = f_{\rm csm}\frac{M_{\rm csm}}{R_{\rm csm}^3}\left[\frac{r}{R_{\rm csm}}\right]^{-p}~~~~~~(R_{\rm csm}<r<(1+f_{\rm bo})R_{\rm csm})
\end{align}
where $p\gg 1$, and $(1+f_{\rm bo})R_{\rm csm}$ is the outer radius of the breakout layer. We adopt a fiducial width for the breakout layer of $f_{\rm bo}=1/3$, and a density profile $p=30$. The exact numerical choice of $p$ does not affect the solution (see \S \ref{sec:analytics}), as long as $p\gg 1$.  Finally, the CSM is initially stationary (i.e. $v=0$) and cold, $T_{\rm csm}=10^2$ K.

\subsection{Comparison with Implicit Monte Carlo Radiation Hydrodynamics}
The flux-limited diffusion (FLD) approximation has the advantage of being computationally inexpensive compared to other more accurate methods for radiation hydrodynamics, such as moment-based \citep{2011JQSRT.112.1323V}, discrete ordinates \citep{2021ApJS..253...49J}, variable Eddington tensor \citep{1992ApJ...393..742E}, and Monte Carlo methods \citep{2015ApJS..217....9R}. FLD is particularly well-suited for optically thick problems, where the diffusion approximation is valid. However, the approximation breaks down once we enter the optically thin regions, which FLD addresses in an ad-hoc manner with a flux limiter. This situation is of particular concern in the $\xi<1$ interaction models.

To test the validity of our FLD results, we run the same interaction problem using the implicit Monte Carlo method of \cite{2015ApJS..217....9R}, which is a much more accurate but also costly approach to solving the equations of radiation hydrodynamics. We use the same finite-volume moving mesh hydrodynamics method of \cite{2016ApJ...821...76D} for the implicit Monte Carlo simulations, and adopt identical model parameters as the FLD runs. 

\begin{figure}
\centering
    \includegraphics[width=0.45\textwidth]{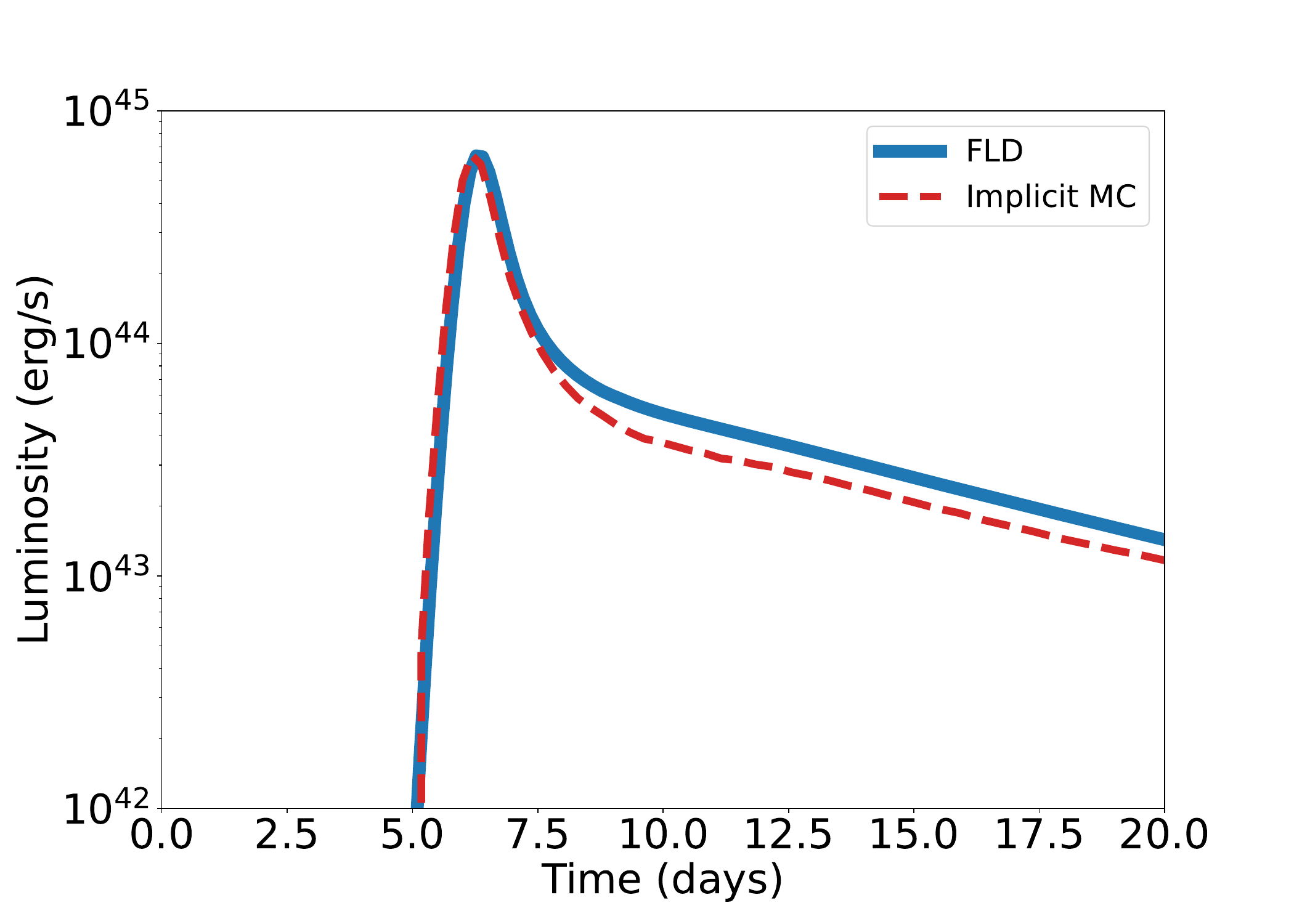}
    \includegraphics[width=0.45\textwidth]{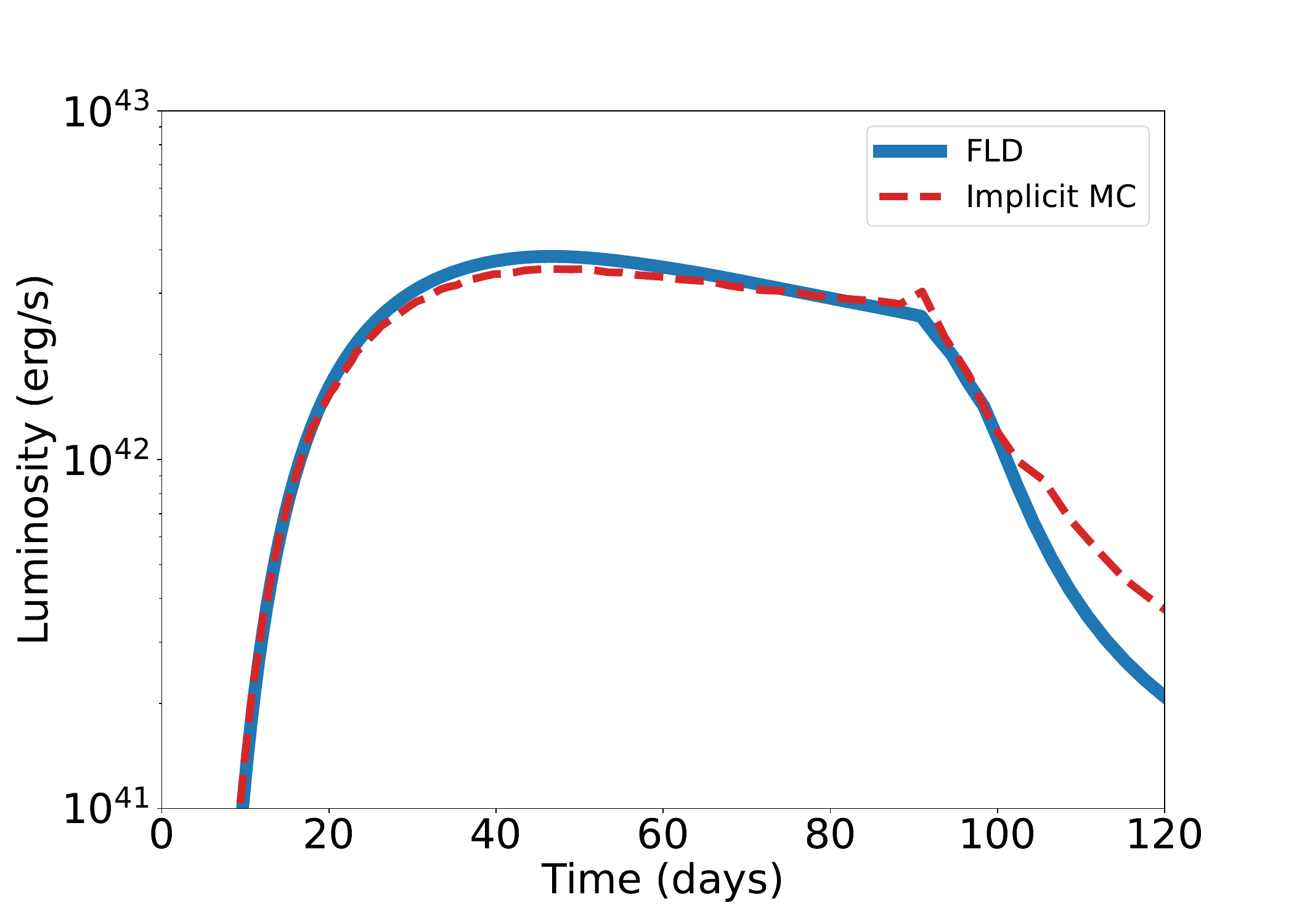}
    \caption{Numerical light curve comparison using the AT2018cow (left) and OGLE-2014-SN-131 (right) parameters in Table \ref{tab:data_fit_params}, using flux-limited diffusion (solid blue line) and implicit Monte Carlo radiation hydrodynamics (dashed red line).}
\end{figure}

In Fig. 18 we show the numerical light curves of the AT2018cow  ($\xi> 1$) and OGLE-2014-SN-131 ($\xi <1$) models (parameters listed in Table 1) for the two different transport methods. Overall we find excellent agreement across the different light curve phases in both cases, although FLD overpredicts the shock cooling tail in the Cow model by about $\sim 10-20\%$ at early times.

\end{document}